\documentclass[aps,twocolumn,amsmath,amssymb,superscriptaddress]{revtex4-2}
\pdfoutput=1
\usepackage[colorlinks=true,linkcolor=blue,citecolor=blue,urlcolor=blue]{hyperref}

\usepackage{graphicx}
\usepackage{graphics}
\usepackage[export]{adjustbox}
\usepackage{bm}
\usepackage{amsfonts}
\usepackage{amssymb}
\usepackage{amsmath}
\usepackage{wasysym}
\usepackage{bbold}
\usepackage{stmaryrd}
\usepackage[usenames]{color}
\usepackage{colordvi}
\usepackage{units}
\usepackage{bbm}
\usepackage{booktabs}
\usepackage{array}
\usepackage{placeins}
\usepackage{epsfig}
\usepackage{changes}
\usepackage{braket}
\usepackage{tabularx}
\usepackage[normalem]{ulem}
\newcolumntype{L}[1]{>{\raggedright\arraybackslash}p{#1}} 
\newcolumntype{C}[1]{>{\centering\arraybackslash}p{#1}} 
\newcolumntype{R}[1]{>{\raggedleft\arraybackslash}p{#1}} 

\newcommand{\be}{\begin{equation}}
\newcommand{\ee}{\end{equation}}
\newcommand{\beqn}{\begin{eqnarray}}
\newcommand{\eeqn}{\end{eqnarray}}

\usepackage{orcidlink} 

\begin{document}

\title{Entanglement witnesses in the $XY$ chain:\\ Thermal equilibrium and postquench nonequilibrium states}
\author{Ferenc Igl{\'o}i\,\orcidlink{0000-0001-6600-7713}}
\email{igloi.ferenc@wigner.hu}
\affiliation{Wigner Research Centre for Physics, Institute for Solid State Physics and Optics, H-1525 Budapest,  Hungary}
\affiliation{Institute of Theoretical Physics, University of Szeged, H-6720 Szeged, Hungary}
\author{G\'eza T\'oth\,\orcidlink{0000-0002-9602-751X}}
\email{toth@alumni.nd.edu}
\homepage{http://www.gtoth.eu}
\affiliation{Department of Theoretical Physics,
University of the Basque Country
UPV/EHU, E-48080 Bilbao, Spain}
\affiliation{EHU Quantum Center, University of the Basque Country UPV/EHU, 48940 Leioa, Biscay, Spain}
\affiliation{Donostia International Physics Center (DIPC), 
P.O. Box 1072,
E-20080 San Sebasti\'an, Spain}
\affiliation{IKERBASQUE, Basque Foundation for Science,
E-48013 Bilbao, Spain}
\affiliation{Wigner Research Centre for Physics, Institute for Solid State Physics and Optics, H-1525 Budapest, Hungary}
\date{\today}

\begin{abstract}
We use entanglement witnesses to detect entanglement in the $XY$ chain in thermal equilibrium and determine the temperature bound below which the state is detected as entangled. We consider the entanglement witness based on the Hamiltonian. Such a witness detects a state as entangled if its energy is smaller than the energy of separable states. {We also consider a family of entanglement witnesses related to the entanglement negativity of the state}. We test the witnesses in infinite and finite systems. We study how the temperature bounds obtained are influenced by a quantum phase-transition or a disorder line in the ground state. Very strong finite-size corrections are observed in the ordered phase due to the presence of a quasi-degenerate excitation. We also study the  postquench states in the thermodynamic limit after a quench when the parameters of the Hamiltonian are changed suddenly. {In the case of the Ising model, we find that the mixed postquench state is detected as entangled by the two methods if the parameters of the Hamiltonian before and after the quench are close to each other. We find that the two witnesses give qualitatively similar results, showing that energy-based entanglement witnesses are efficient in detecting the nearest-neighbor entanglement in spin chains in various circumstances.} {For other $XY$ models, we find that the negativity based witnesses also detect states in some parameter regions where the energy-based witness does not, in particular, if the quench is performed from the paramagnetic phase 
to the ferromagnetic phase and vice versa.} The domains in parameter space corresponding to postquench states detected as entangled by the energy-based witness have been determined analytically, which stresses further the utility of our method.

\vspace{1em}
\noindent DOI: \href{https://doi.org/10.1103/PhysRevResearch.5.013158}{10.1103/PhysRevResearch.5.013158}
\end{abstract}


\maketitle

\section{Introduction}
\label{sec:introduction}

Entanglement lies at the heart of quantum mechanics and also plays an important role in quantum information theory (QIT) \cite{Horodecki2009Quantum,Guhne2009Entanglement,Friis2019}. For pure states it is equivalent to correlations, while for mixed states the two notions differ. A quantum state is entangled, if its density matrix cannot be written as a mixture of product states. Based on this definition, several sufficient conditions have been developed. In special cases, e.g. for $2\times 2$ (two-qubit) and $2\times 3$ bipartite systems \cite{Peres1996Separability,Horodecki1997Separability} and for multi-mode Gaussian states \cite{Giedke2001Entanglement} even necessary and sufficient conditions are known.

However, in an experimental situation usually only limited information about the quantum state is available. This is true even for theoretical calculations for very large systems. Only those approaches for entanglement detection can be applied that require the measurement of a few observables. One of such approaches is using entanglement witnesses. They are observables that have a positive expectation value for all separable states. Thus, a negative expectation value signals the presence of entanglement. The theory of entanglement witnesses has recently been rapidly developing \cite{Horodecki1996Separability,Terhal2000Bell,Lewenstein2000Optimization}. It has been shown how to construct entanglement witnesses that detect entanglement close to a given quantum state, even if it is mixed or a bound entangled state \cite{Acin2001Classification}. It is also known how to optimize a witness operator in order to detect the most entangled states \cite{Lewenstein2000Optimization}.

Apart from determining optimal entanglement witnesses, it is also important to find witnesses that are easy to measure in an experiment or possible to evaluate in a theoretical calculation.  From both point of views, witnesses based on spin chain Hamiltonians attracted considerable attention \cite{Toth2005EntanglementWitnesses,Toth2006Detection,Guhne2006Energy,Guhne2005Multipartite,Brukner2004MacroscopicB,Dowling2004Energy,Wu2005Entanglement}.  There have been already calculations for infinite chains \cite{Toth2005EntanglementWitnesses,Wu2005Entanglement,Dowling2004Energy}. It has been shown that the optimal witness for the thermal state of the chain is not necessarily the Hamiltonian \cite{Wu2005Entanglement}. Besides entanglement in general, witnesses based on energy can be used to detect multiparticle entanglement  \cite{Guhne2005Multipartite,Guhne2006Energy}. Note that even a direct relationship between entanglement measures and the energy of the thermal state has been observed in isotropic Heisenberg chains \cite{Wang2002Threshold}. The energy-based witnesses have been used in various physical systems such as nanotubular systems \cite{Vertesi2006Thermal}, in molecular nanomagnets \cite{Sioli2012Towards}, in heterometallic wheels \cite{Siloi2013Quantum}, and also for theoretical calculations in theoretical spin models  \cite{Troiani2012Energy,Homayoun2019Energy,Troiani2013Detection}. 

In this paper, we extend the approach to the $XY$ model. {This model is exactly solvable and several entanglement based properties have been studied recently \cite{Osterloh2002,PhysRevA.66.032110,Patan__2007,PhysRevB.89.134101,PhysRevA.88.052305}.} We also consider another approach, based on {a family of witnesses that detect} entanglement whenever the entanglement negativity of the nearest-neighbor two-spin density matrix is nonzero \cite{Vidal2002Computable}, i.e., when the state violates the entanglement criterion based on the positivity of the partial transpose (PPT) \cite{Peres1996Separability,Horodecki1997Separability}.  We consider finite and infinite systems in thermal equilibrium and compare the temperature bounds for separability obtained from the energy-based witness and from the negativity-based {witnesses}.

Then, we test the entanglement witnesses in mixed states based on the following idea. We place the system in the ground state of a given $XY$ Hamiltonian. Then, considering a quench, we change the parameters of the Hamiltonian \cite{PhysRevA.2.1075,PhysRevA.3.2137,PhysRevLett.85.3233,PhysRevA.69.053616,RevModPhys.83.863}. Since the state is not an eigenstate of the new Hamiltonian, dynamics start. In the infinite time limit, the system approaches a stationary state, which is some mixture of the states appearing during the dynamics. If the Hamiltonian is non-integrable, the system is expected to be thermalized and the stationary state is described by a Gibbs ensemble with an effective temperature  \cite{Sotiriadis_2012,PhysRevA.79.021608,Sotiriadis_2009,PhysRevA.78.013626,PhysRevLett.100.100601,PhysRevLett.101.063001,PhysRevLett.100.030602,PhysRevLett.98.210405,PhysRevLett.97.156403,PhysRevLett.96.136801,PhysRevLett.98.050405}, see however Refs.~\cite{PhysRevE.93.032116,Larson_2013,PhysRevLett.106.025303,Olshanii2012}. For integrable systems, such as the $XY$ chain, the stationary state is assumed to be described by a so-called Generalized Gibbs Ensemble (GGE)  \cite{Vidmar_2016,Ilievski_2016,PhysRevLett.115.157201,PhysRevA.91.051602,Pozsgay_2014a,Pozsgay_2014,PhysRevA.90.043625,PhysRevLett.113.117203,PhysRevLett.113.117202}, for which different effective temperatures are assigned to each conserved quantities. This type of description has been exactly calculated for the quantum Ising chain \cite{Calabrese_2012}, and a similar formalism is conjectured for the $XY$ chain \cite{Blass_2012}. 

In this article, we show for the $XY$ model that the postquench state can still be handled efficiently for large systems and the expectation value of the witness operators mentioned above can also be computed. We analyze, in which cases the mixed state is detected by the energy-based witness and by the {witnesses} based on entanglement negativity. We find that the energy-based witness is efficient in detecting entanglement in these systems.

Our paper is organized as follows. In Sec.~\ref{sec:model}, we introduce the $XY$ model, present its free-fermion representation, calculate thermal averages and present its conjectured GGE after a global quench. In Sec.~\ref{sec:witness}, the entanglement witnesses are described. In Sec.~\ref{sec:thermal}, the temperature bounds are calculated both by the energy- and the entanglement negativity-methods and also finite-size corrections are studied. In Sec.~\ref{sec:post_quench}, the bounds for postquench states are calculated. In Secs.~\ref{sec:discussion} and ~\ref{sec:conclusions}, we close our paper with a discussion and conclusions, respectively. In the Appendix, we present the calculation of the thermal average of the energy in finite periodic chains.

\begin{figure}[t!]
\includegraphics[width=1\columnwidth]{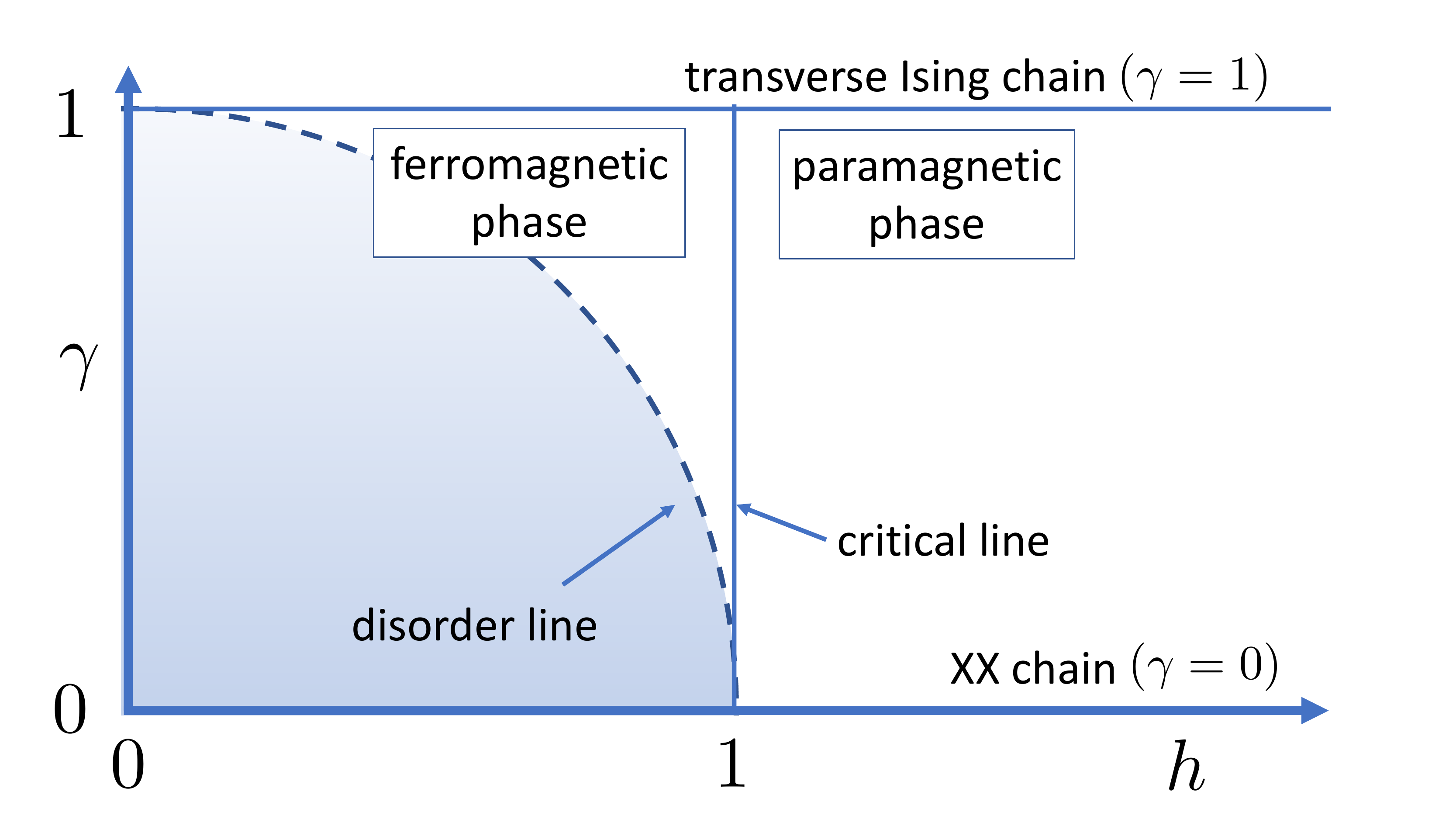}
	\vskip 0cm
\caption{Phase diagram of the $XY$ chain with the Hamiltonian Eq.~(\ref{hamilton}). The equation of the disorder line is  given by Eq.~(\ref{eq:disorderline}). For a qualitative description of the regions on the two sides of the disorder line, see Sec.~\ref{sec:disorder}.
\label{fig_1}}	
\end{figure} 

\section{Model and methods}
\label{sec:model}

In this section, we describe the model, present the $XY$ spin-chain Hamiltonian and show how to calculate important quantities for it in a free-fermion representation, including averages for finite temperatures. Finally, we discuss how to handle nonequilibrium stationary states after a quench.

\subsection{The $XY$ chain}
\label{sec:XY-chain}
The $XY$ chain is defined by the Hamiltonian
\be
\scalebox{0.925}
{$\displaystyle{\cal H}=-\sum_{l=1}^L \left[\dfrac{1+\gamma}{2} \sigma^x_l \sigma^x_{l+1}+\dfrac{1-\gamma}{2} \sigma^y_l \sigma^y_{l+1}\right]
-h \sum_{l=1}^L \sigma^z_l$},
\label{hamilton}
\ee
where $\sigma_l^{x},$ $\sigma_l^{y},$ and $\sigma_l^{z}$  are Pauli spin operators acting on the spin at site $l,$ and  $\sigma_{L+1}^{\alpha}\equiv\sigma_{1}^{\alpha}$ for $\alpha=x,y,z.$ We consider chains with periodic boundary conditions and generally calculate quantities in the thermodynamic limit, $L \to \infty$. In a few cases we study also finite chains up to $L=12$ through numerical techniques. The parameters $0 \le \gamma \le 1$ and $h \ge0$ denote the strength of the anisotropy and the transverse field, respectively. The special case $\gamma=1$ represents the transverse Ising model, and for $h=0$, $\gamma=0$ the Hamiltonian reduces to the $XX$ chain. The equilibrium phase diagram is shown in Fig.~\ref{fig_1}.

\subsection{Free-fermion representation}

Using standard techniques \cite{Lieb1961Two,Pfeuty1970The}, the Hamiltonian in Eq.~(\ref{hamilton}) is expressed in terms of fermion creation and annihilation operators $\eta^{\dag}_p$ and $\eta_p$ as
\be
{\cal H}=\sum_p \varepsilon\left(p\right)\left(\eta^{\dag}_p \eta_p-\frac{1}{2}\right),
\ee
where the sum runs over $L$ quasi-momenta,  which are equidistant in $[-\pi,\pi]$ for periodic  boundary conditions and almost equidistant in $[0,\pi]$ for free boundary conditions and a finite $L$, but again equidistant in the limit $L\to\infty$. The energy of the modes is given by \cite{PhysRevA.3.2137,PhysRevA.2.1075,Blass_2012}
\be
\varepsilon\left(p\right)=2\sqrt{\gamma^2 \sin^2 p+\left(h-\cos p\right)^2}
\label{eps}
\ee
and the Bogoliubov angle $\varTheta_p$ diagonalizing the Hamiltonian is given by 
\be\tan\varTheta_p=-\gamma\sin p/\left(h-\cos
p\right).\ee
The energy of the ground state equals
\be
E_0=-\sum_p \frac{\varepsilon\left(p\right)}{2},
\ee
since the state is the fermionic vacuum.

\subsubsection{Disorder line}
\label{sec:disorder}

The disorder line, also shown in Fig.~\ref{fig_1}, is given by
\be
h^2+\gamma^2=1. \label{eq:disorderline}
\ee
Alternatively, we can express the $h$ values as a function of $\gamma$ corresponding to the disorder line as 
\be
h_d=\sqrt{1-\gamma^2}.\label{eq:hd}
\ee
As we will see, $h_d$ will turn out to be a central quantity in our paper.

The system has a special behavior at the disorder line given by Eq.~(\ref{eq:disorderline}). Here the energy of modes has the simple form
\be
\varepsilon\left(p\right)=2(1-h_d \cos p),
\label{disorder}
\ee
and the ground-state energy density in the thermodynamic limit is given by
\be
\frac{E_0}{L}=-\frac{1}{\pi}\int_0^{\pi}(1-h_d \cos p) \textrm{d} p=-1.\label{eq:E0_over_L_disorderline}
\ee
Below the disorder line 
\be
h^2+\gamma^2<1 
\ee
holds and the long-range two-point correlation functions have an oscillatory behavior, whereas at the disorder line they are constant \cite{PhysRevA.2.1075,PhysRevA.3.2137}. Above the disorder line we have 
\be
h^2+\gamma^2>1 
\ee
and the two-point correlation functions do not have an oscillatory behavior.

\subsection{Averages at finite temperature}
\label{section:av_temperature}

Next, we summarize results for the thermodynamic limit, $L \to \infty,$  used in the article. Calculations of the energy for finite periodic chains are presented in the Appendix~\ref{sec:finite_fermionic}.

\subsubsection{Energy}

At finite temperature, $T>0$, the partition function is expressed as
\be
Z=2^L\prod_p \cosh\left(\frac{\varepsilon\left(p\right)}{2T}\right),
\ee
and the average value of the energy is given by
\be
\langle {\cal H} \rangle_T=-\sum_p t(p,T)\frac{\varepsilon\left(p\right)}{2},
\label{E_T}
\ee
with the definition 
\be
t(p,T)=\tanh\left(\frac{\varepsilon\left(p\right)}{2T}\right),
\label{t(p)}
\ee
setting $k_B=1$.

\subsubsection{Correlation functions}

The correlation functions of the $XY$ model are calculated in Refs.~\cite{PhysRevA.3.2137,PhysRevA.2.1075} and the nearest-neighbor correlations can be expressed as
\begin{align}
\langle \sigma_l^x \sigma_{l+1}^x \rangle_T&=g_c-g_s,\nonumber\\
\langle \sigma_l^y \sigma_{l+1}^y \rangle_T&=g_c+g_s,\nonumber\\
\langle \sigma_l^z \sigma_{l+1}^z \rangle_T&=g_0^2-g_c^2+g_s^2,
\label{nn_corr}
\end{align}
where we define the sums
\begin{align}
g_c&=\frac{2}{L}\sum_p \cos p (\cos p -h)~t(p,T)\varepsilon^{-1}\left(p\right),\nonumber\\
g_s&=-\gamma \frac{2}{L}\sum_p \sin^2 p ~t(p,T)\varepsilon^{-1}\left(p\right),\nonumber\\
g_0&=\frac{2}{L}\sum_p (h-\cos p) ~t(p,T)\varepsilon^{-1}\left(p\right).
\label{par_corr}
\end{align}
We stress once more that the relations above are valid in the thermodynamic limit, $L \to \infty$. One can easily check that
\be
\frac{\langle H \rangle_T}{L}=-\dfrac{1+\gamma}{2}\langle \sigma_l^x \sigma_{l+1}^x \rangle_T-\dfrac{1-\gamma}{2}\langle \sigma_l^y \sigma_{l+1}^y \rangle_T-h\langle \sigma_l^z \rangle_T
\label{H_corr}
\ee
holds with $\langle \sigma_l^z \rangle_T=g_0,$ which further supports the correctness of our calculations. In particular, it helps to verify that constant factors have been taken into account correctly.

\subsection{Nonequilibrium stationary states after a quench}
\label{sec:quench}

We consider global quenches at zero temperature, which suddenly change the parameters of the Hamiltonian from $\gamma_0$, $h_0$ for $t<0$ to $\gamma$, $h$ for $t>0$. For $t<0$ the system is assumed to be in equilibrium, i.e., in the ground state $\left|\Phi_0\right\rangle$ of the Hamiltonian ${\cal H}_0$ with parameters $\gamma_0$ and $h_0.$ After the quench, for $t>0$, the state evolves coherently according to the new Hamiltonian ${\cal H}$ as
\be
\left|\Phi_0(t)\right\rangle=\exp(-i{\cal H}t)\left|\Phi_0\right\rangle.
\label{Phi(t)}  
\ee
Correspondingly, the time evolution of an operator in the Heisenberg picture is
\be
\sigma_l\left(t\right)=e^{i{\cal H}t} \sigma_l e^{-i{\cal H}t}.
\ee

The energy of the system after the quench is given as
\be
\langle \Phi_0| \displaystyle{\cal H} |\Phi_0 \rangle=\sum_p \varepsilon\left(p\right)\left(\langle \Phi_0|\eta^{\dag}_p \eta_p|\Phi_0 \rangle-\frac{1}{2}\right),
\label{E_Phi}
\ee
where  the occupation
probability of mode $p$ in the initial state $\left|\Phi_0\right\rangle$ is given as 
\be
f_p=\left\langle \Phi_0\right|\eta^{\dag}_p \eta_p \left|\Phi_0\right\rangle.
\ee 
For the $XY$ model it is expressed through the difference
$\Delta_p=\varTheta^0_p-\varTheta_p$ of the Bogoliubov angles as
\be 
f_p=\tfrac{1}{2}\left(1-\cos \Delta_p\right)
\label{f_p}
\ee 
with the cosine of the difference $\Delta_p$ given as 
\be
\scalebox{0.925}
{
$\cos \Delta_p=4\dfrac{\left(\cos p -h_0\right)\left(\cos p -h\right)+\gamma\gamma_0\sin^2 p}{\varepsilon\left(p\right)\varepsilon_0\left(p\right)},$
}
\label{Delta}
\ee
where the index $0$ refers to quantities before the quench \cite{Blass_2012}. In the thermodynamic limit, Eq.~(\ref{E_Phi}) can be rewritten as
\be
\frac{\langle \Phi_0| \displaystyle{\cal H} |\Phi_0 \rangle}{L}=-\frac{1}{4\pi}\int_{-\pi}^\pi \varepsilon(p)\cos \Delta_p {\rm d}p.\label{eq:quenchth}
\ee

After large enough time and in the thermodynamic limit the system is expected to reach a stationary state, 
\be
{\bm\rho}_{q}=\lim_{\tau\rightarrow\infty}\frac1 \tau \int_{0}^{\tau} e^{-i {\cal H}t} |\Phi_0\rangle\langle \Phi_0| e^{+i{\cal H}t}  {\rm d}t,
\ee
for which the energy equals the energy of the initial state
\be
{\rm Tr}({\bm\rho}_{q} {\cal H})=\langle \Phi_0| {\cal H} |\Phi_0 \rangle.\label{eq:pqstate}
\ee
{Similarly, for an observable ${\cal O}$ the stationary value is given by ${\rm Tr}({\bm\rho}_{q} {\cal O})$.}

In the stationary  state, due to conserved symmetries, averages of correlations are described by a Generalized Gibbs Ensemble (GGE)\cite{Vidmar_2016,Ilievski_2016,PhysRevLett.115.157201,PhysRevA.91.051602,Pozsgay_2014,PhysRevA.90.043625,PhysRevLett.113.117203,PhysRevLett.113.117202}. In this case to each fermionic mode an effective temperature, $T_{\rm eff}(p)$ is attributed through the relation \cite{Blass_2012}
\be
\tanh\left(\frac{\varepsilon\left(p\right)}{2T_{\rm eff}(p)}\right)=|2f_p-1|=|\cos \Delta_p|.
\label{T_eff}
\ee
This follows by comparing the relations in Eqs.~(\ref{E_T}), (\ref{t(p)}) and (\ref{E_Phi}). In this way the nearest-neighbor correlations in the stationary state can be obtained as in section \ref{section:av_temperature}, just replacing $t(p,T)$ defined in Eq.~(\ref{t(p)}) by $|\cos \Delta_p|$
\be
t(p,T) \to |\cos \Delta_p|.
\label{t_to_cos}
\ee
In particular, we have to apply Eq.~(\ref{t_to_cos}) for the correlation functions in Eqs.~(\ref{nn_corr}) and (\ref{par_corr}).

\section{Entanglement witnesses}
\label{sec:witness}

Generally an operator ${\cal W}$ is called an entanglement witness, if its expectation value, $\langle {\cal W} \rangle$ satisfies the following requirements \cite{Lewenstein2001Characterization,Terhal2002Detecting}.

(i) For all separable states  \be\langle {\cal W} \rangle \ge 0\ee  holds.

(ii) For some entangled state we have 
\be 
\langle {\cal W} \rangle < 0. 
\ee  
We say that such a state is detected by the witness as entangled. Entanglement witnesses have been used in various physical systems to verify the presence of entanglement \cite{Bourennane2004Experimental, Walther2005ExperimentalOneWay,Kiesel2005Experimental,Wieczorek2009Experimental,Prevedel2009Experimental,Gao2010Experimental,Genuine2019Gong,Haffner2005Scalable,14qubit2011Monz,Feldman2008,Garttner2018Relating}.

A single entanglement witness cannot detect all entangled states. On the other hand, if for some states ${\bm \rho}_k$ we have
\be
\langle {\cal W} \rangle_{{\bm \rho}_k}=\langle {\cal W} \rangle_{{\bm \rho}_{k'}}
\ee
for all $k,k',$ then for any mixture of such states 
\be
{\bm \rho}=\sum_k p_k {\bm \rho}_k\label{eq:rhok}
\ee
the expectation value of the witness remains the same, i.e.,
\be
\langle {\cal W} \rangle_{{\bm \rho}}=\langle {\cal W} \rangle_{{\bm \rho}_k}
\ee
holds for all $k.$ Thus, if ${\bm \rho}_k$ are detected as entangled, ${\bm \rho}$ given in Eq.~(\ref{eq:rhok}) is also detected as entangled, even if ${\bm \rho}$ is highly mixed.

Let us consider now the case, in which the $\bm{\rho}_k$ family is obtained via a unitary dynamics from an initial state. If for a state ${\bm \rho}$ we have $\langle {\cal W} \rangle_{{\bm \rho}}$ then for any state given as 
\be
{\bm \rho}'(K)=e^{-i K}{\bm \rho}e^{+i K}\label{eq:rhoK}
\ee
we have the same expectation value
\be
\langle {\cal W} \rangle_{{\bm \rho}'}=\langle {\cal W} \rangle_{{\bm \rho}},
\ee
if  $K$ commutes with $\cal W$ as 
\be
[K,{\cal W}]=0.
\ee
It is easy to see that the expectation value is equal to $\langle {\cal W} \rangle_{{\bm \rho}}$ even for an arbitrary mixture [like in Eq.~(\ref{eq:rhok})] of the states given in Eq.~(\ref{eq:rhoK}). Thus, if ${\bm \rho}$ is detected as entangled by the witness ${\cal W},$ an arbitrary mixture of the states given in Eq.~(\ref{eq:rhoK}) is also detected.  Based on these, we can see that entanglement witnesses might be especially useful in detecting entanglement in a mixture of states obtained from a unitary dynamics with a Hamiltonian that commutes with the witness. {In general, an entanglement witness that is conserved during the quench, might be especially useful in detecting entanglement.}

In this section, we present the energy-based witness \cite{Toth2005EntanglementWitnesses}, which is constructed with the Hamiltonian and thus, based on the arguments above, it is especially suited for entanglement detection in a postquench state given in Eq.~(\ref{eq:pqstate}). We also present the negativity-based {witnesses} \cite{Wu2005Entanglement} what we will use to analyze the entanglement properties of the $XY$ chain.

\subsection{Energy-based witness}
\label{sec:energybased}

In this section, we review the idea of detecting entanglement with energy measurement \cite{Toth2005EntanglementWitnesses,Brukner2004MacroscopicB,Dowling2004Energy}.

First we calculate the minimum of $\langle \displaystyle{\cal H} \rangle$ for product states of the form
\be
|\Psi\rangle=|\psi\rangle_1 \otimes |\psi\rangle_2 \otimes \dots \otimes|\psi\rangle_L,\label{eq:state1}
\ee
with the single particle states
\be
|\psi\rangle_k=\cos \phi e^{i\theta_{\uparrow}}|\uparrow\rangle_k+\sin \phi e^{i\theta_{\downarrow}}|\downarrow\rangle_k,\label{eq:state2}
\ee
where $k$ labels the site in the real space.
The energy per site for the state given in Eqs.~(\ref{eq:state1}) and (\ref{eq:state2}) is
\be
\frac{\langle \Psi|\displaystyle{\cal H}|\Psi \rangle}{L}=-\frac{1}{2} \sin^2 2\phi[1+\gamma \cos 2(\theta_{\uparrow}-\theta_{\downarrow})]-h \cos 2\phi,
\ee
which has a minimum at $\theta_{\uparrow}=\theta_{\downarrow}$ and for
\begin{align}
\cos 2\phi&=\frac{h}{1+\gamma}, &\textrm{for } {h} \le {1+\gamma},\nonumber\\
\phi&=0, &\textrm{for } {h}>{1+\gamma}.\label{eq:phigammah}
\end{align}
Thus, the minimum energy per site for product states is given by
\be
\frac{E_{\rm sep}}{L}=
\begin{cases} 
-\frac{(1+\gamma)^2+h^2}{2(1+\gamma)}, &\textrm{for } {h} \le {1+\gamma},\\
-h, &\textrm{for } {h}>{1+\gamma}.
\label{E_sep}
\end{cases}
\ee

Then, we consider separable states given as \cite{Werner1989Quantum}
\be
{\bm \rho}_{\rm sep}=\sum_m p_m {\bm \rho}_m^{(1)}\otimes {\bm \rho}_m^{(2)}\otimes \dots \otimes{\bm \rho}_m^{(L)},
\ee
where ${\bm \rho}_m^{(k)}$ are single particle pure states. The bound given in Eq.~(\ref{E_sep}) is also the bound for mixed separable quantum states,
since the expectation value 
\be
\langle \displaystyle{\cal H} \rangle={\rm Tr}({\bm \rho} H)
\ee
is linear in ${\bm \rho},$ and the set of separable states is convex. 

Then we can simply write the witness detecting entanglement based on the energy as
\be
{\cal W}_E ={\cal H} - E_{\rm sep}\openone.\label{eq:ewit}
\ee
We will compute $\langle {\cal W}_E \rangle$ for thermal states [see in Eq.~(\ref{E_T})] and for postquench states [see in Eq.~(\ref{E_Phi})]. 

Along the disorder line $h$ and $\gamma$ fulfil Eq.~(\ref{eq:disorderline}). Then,  based on Eq.~(\ref{eq:E0_over_L_disorderline}), we have for the ground state energy
\be
E_0=-L.
\ee
The ground state is a product state of the form given in Eqs.~(\ref{eq:state1}) and (\ref{eq:state2}) for $\theta_{\uparrow}=\theta_{\downarrow}$ and $\phi$ fulfilling Eq.~(\ref{eq:phigammah}). 

Note that based on the bound for separable states given in Eq.~(\ref{E_sep}) we have 
\be
E_{\rm sep}=-L,
\ee
which means that there is a separable state with energy $E_{\rm sep}$. However, since $E_0=E_{\rm sep}$ and the ground state is non-degenerate, there is only a single pure state having this energy and it must be a product state. Thus, only by knowing $E_0$ and $E_{\rm sep},$ and the fact that the ground state is non-degenerate, we can conclude that the ground state must be a product state along the disorder line.

\subsection{Negativity-based {witnesses}}
\label{sec:negativtybased}

In this section, we summarize the method presented in Ref.~\cite{Wu2005Entanglement} and show how to apply it to the $XY$ chain. It suggests to use not the Hamiltonian but another operator as an entanglement witness for spin chains in thermal equilibrium, which is shown to be connected to partial transpose of the density matrix \cite{Peres1996Separability,Horodecki1997Separability}, and to entanglement negativity \cite{Vidal2002Computable}.

Deciding whether a quantum state is entangled is a hard task in general. However, there are some necessary conditions for separability, that are easy to test. If these conditions are violated then the state is entangled. One of the most important conditions of this type is the PPT condition \cite{Peres1996Separability,Horodecki1997Separability}. For a bipartite density matrix given as
\be
{\bm \rho}=\sum_{kl,mn}{\bm \rho}_{kl,mn} \ket{k}\bra{l}\otimes \ket{m}\bra{n}
\ee
the partial transpose according to first subsystem is defined by exchanging subscripts $k$ and $l$ as 
\be
{\bm \rho}^{T_A}=\sum_{kl,mn}{\bm \rho}_{lk,mn} \ket{k}\bra{l}\otimes \ket{m}\bra{n}.
\ee
It has been shown that for separable quantum states  \cite{Peres1996Separability,Horodecki1996Separability}
\be
{\bm \rho}^{T_A}\ge 0
\ee
holds. Thus, if ${\bm \rho}^{T_A}$ has a negative eigenvalue then the quantum state is entangled. For $2\times2$ and $2\times3$ systems, the PPT condition detects all entangled states \cite{Horodecki1996Separability}. For systems of size $3\times3$ and larger, there are PPT entangled states \cite{Horodecki1997Separability,Horodecki1998Mixed-State}. One can even use the partial transpose to tell how much a quantum state is entangled. The entanglement negativity \cite{Vidal2002Computable} is defined as 
\be
{\cal N}(\bm{\rho})=2{\rm max}(0,-{\rm min}(\mu_{\nu})),
\label{negativity}
\ee
where $\mu_{\nu}$ are the eigenvalues of the partial transpose $\bm{\rho}^{T_A}.$

Let us turn now to $XY$ chains. Let us consider the nearest-neighbor reduced density matrix, $\bm{\rho}$, which will be defined in the $\sigma^z$ basis given by $|\uparrow\rangle$ and $|\downarrow\rangle.$ We use the convention 
\be |\uparrow\uparrow\rangle=|1\rangle, \quad |\uparrow\downarrow\rangle=| 2 \rangle, \quad  |\downarrow\uparrow\rangle=| 3 \rangle, \text{ and }   |\downarrow\downarrow\rangle=| 4 \rangle.\nonumber\ee Due to the symmetries of the problem, $\bm{\rho}$ is a direct sum of two $2\times 2$ matrices living in the space spanned by the states $|1\rangle,|4\rangle$ and $|2\rangle,|3\rangle$, respectively. Consequently, it is represented as
\be
\bm{\rho}=
\begin{bmatrix}
\rho_{11} & 0 & 0 & \rho_{14}\\
0 & \rho_{22} & \rho_{23} & 0\\
0 & \rho_{32} & \rho_{33} & 0\\
\rho_{41} & 0 & 0 & \rho_{44}\\
\end{bmatrix},
\ee
where we indicated the elements that are necessarily zero explicitly.
The matrix is real and symmetric, hence $\rho_{14}=\rho_{41}$ and $\rho_{23}=\rho_{32}$ hold, furthermore we have the constraint due to unit trace, $\rho_{11}+\rho_{22}+\rho_{33}+\rho_{44}=1$. Due to the permutational symmetry of the problem we also have
\be
\rho_{22}=\rho_{33}.\label{eq:rho2233}
\ee
{Due to Eqs.~(\ref{nn_corr}) and (\ref{par_corr}), $\rho_{14}=-2g_s,$ and 
\be
\rho_{14}\ge0 \label{eq:rho140}
\ee 
holds if $\gamma\ge0.$ In Appendix~\ref{ap:rho230} we also show that 
\be
\rho_{23}\ge0 \label{eq:rho230}
\ee
holds.}

For the partially transposed density matrix, by indicating the zero elements explicitly we obtain
\be
\bm{\rho}^{T_A}=
\begin{bmatrix}
\rho_{11} & 0 & 0 & \rho_{23}\\
0 & \rho_{22} & \rho_{14} & 0\\
0 & \rho_{41} & \rho_{33} & 0\\
\rho_{32} & 0 & 0 & \rho_{44}\\
\end{bmatrix}.
\label{ptranspose}
\ee
The lowest eigenvalue of $\bm{\rho}^{T_A}$, denoted by $\mu_{\rm min}$ is non-negative if and only if the state is separable \cite{Peres1996Separability,Horodecki1996Separability}. According to Eq.~(\ref{ptranspose}), the minimal eigenvalues of the $2\times2$ submatrices are
\begin{subequations}
\begin{align}
\mu_{\rm min}^{(1)}&=\frac{\rho_{11}+\rho_{44}-\sqrt{(\rho_{11}-\rho_{44})^2+4\rho_{23}\rho_{32}}}{2},\label{eq:mumin1}\\
\mu_{\rm min}^{(2)}&=\frac{\rho_{22}+\rho_{33}-\sqrt{(\rho_{22}-\rho_{33})^2+4\rho_{14}\rho_{41}}}{2},\label{eq:mumin2}
\end{align}\label{eq:mumin}
\end{subequations}
and the minimal eigenvalue is just 
\be
\mu_{\rm min}=\min(\mu_{\rm min}^{(1)},\mu_{\rm min}^{(2)}).
\ee

Taking into account all our knowledge of the density matrix elements, Eq.~(\ref{eq:mumin2}) can be simplified to 
\be
\mu_{\rm min}^{(2)}=\rho_{22}-\rho_{14}.\label{eq:mumin2_simple}
\ee
The matrix elements of $\bm{\rho}$ appearing in Eq.~(\ref{eq:mumin2_simple}) can be expressed through nearest-neighbor correlations~{\cite{PhysRevA.66.032110}. Here we use} the relations
\begin{align}
&\langle \sigma_l^z \sigma_{l+1}^z \rangle-1\nonumber\\
&\quad\quad=\rho_{11}-\rho_{22}-\rho_{33}+\rho_{44}-(\rho_{11}+\rho_{22}+\rho_{33}+\rho_{44})\nonumber \\
&\quad\quad=-2(\rho_{22}+\rho_{33})=-4\rho_{22}
\end{align}
and
\begin{align}
\langle \sigma_l^x \sigma_{l+1}^x \rangle-\langle \sigma_l^y \sigma_{l+1}^y \rangle 
&=2(\langle \sigma_l^+ \sigma_{l+1}^+ \rangle+\langle \sigma_l^- \sigma_{l+1}^- \rangle) \nonumber \\
&=2(\rho_{14}+\rho_{41})=4\rho_{14}.
\end{align}
For $\rho_{14}$ the relation Eq.~(\ref{eq:rho140}) holds. Thus, the eigenvalue is obtained as 
\be
\mu_{\rm min}^{(2)}=-\frac{1}{4}(\langle \sigma_l^x \sigma_{l+1}^x \rangle-\langle \sigma_l^y \sigma_{l+1}^y \rangle +\langle \sigma_l^z \sigma_{l+1}^z \rangle-1).
\label{witness}
\ee
{The other eigenvalue, $\mu_{\rm min}^{(1)}$, can also be expressed with nearest-neighbor correlations, since
\begin{align}
&\rho_{11}+\rho_{44}=(\langle \sigma_l^z \sigma_{l+1}^z \rangle+1)/2,\nonumber\\
&\rho_{11}-\rho_{44}=(\langle \sigma_l^z\rangle+\langle \sigma_{l+1}^z \rangle)/2,\nonumber\\
&\rho_{23}=\rho_{32}=(\langle \sigma_l^x \sigma_{l+1}^x \rangle+\langle \sigma_l^y \sigma_{l+1}^y \rangle)/4.
\end{align}
Then, we obtain the eigenvalue as 
\begin{align}
&\mu_{\rm min}^{(1)}=\frac{\langle \sigma_l^z \sigma_{l+1}^z \rangle+1}{4}\nonumber\\
&-\frac{1}{4}\sqrt{(\langle \sigma_l^z\rangle+\langle \sigma_{l+1}^z \rangle)^2+(\langle \sigma_l^x \sigma_{l+1}^x \rangle+\langle \sigma_l^y \sigma_{l+1}^y \rangle)^2}.\label{eq:mumin1_corr}
\end{align}}

In order to proceed, we need to know that the partial transposition of a two-qubit state has at most one negative eigenvalue and all the eigenvalues lie in $[-1/2,1]$\cite{PhysRevA.58.826,PhysRevA.87.054301}. Thus, only one of the $\mu_{\rm min}^{(1)}$ and $\mu_{\rm min}^{(2)}$ can be negative, and when they are equal to each other, they must be non-negative and the state must be separable.

Let us now examine the relation between the eigenvalues $\mu_{\rm min}^{(1)}$ and $\mu_{\rm min}^{(2)},$ and the parameters $h$ and $\gamma.$ At the disorder line where  Eq.~(\ref{eq:disorderline}) is satisfied, we have $\mu_{\rm min}^{(1)}=\mu_{\rm min}^{(2)}$ and here the state is separable even at $T=0$, which is consistent with what we mentioned about the eigenvalues of the partial transpose. We have checked numerically in finite systems, that in the oscillatory region where $h^2+\gamma^2\le 1$, the relation \be\mu_{\rm min}^{(1)}\le\mu_{\rm min}^{(2)}\label{eq:mumin12ineq}\ee holds. On the other hand, if $h^2+\gamma^2>1$ then  we have \be\mu_{\rm min}^{(1)}>\mu_{\rm min}^{(2)}.\label{eq:mumin12ineqB}\ee 

Let us now show that a simple entanglement witness can be obtained based on the expression for $\mu_{\rm min}^{(2)}$ given in Eq.~(\ref{witness}). In particular, we can define the operator
\be
{\cal W}_N=-\frac{1}{4}( \sigma_l^x \sigma_{l+1}^x -\sigma_l^y \sigma_{l+1}^y  + \sigma_l^z\sigma_{l+1}^z -\openone),
\label{neg_witness1}
\ee
and hence its expectation value gives one of the eigenvalues of the partial transpose of the two-qubit density matrix 
\be
\langle {\cal W}_N\rangle=\mu_{\rm min}^{(2)},
\ee
which also implies that ${\cal W}_N$ is an entanglement witness.
{It is instructive to rewrite the witness given in Eq.~(\ref{neg_witness1}) as 
\be
{\cal W}_N=\frac1 2\openone-\vert \Phi^+\rangle\langle \Phi^+\vert,
\label{neg_witness}
\ee
where
\be
\ket{\Phi^+}=\frac1 {\sqrt 2}(\ket{\uparrow\uparrow}+\ket{\downarrow\downarrow})\label{eq:maxent}
\ee}The form in Eq.~\eqref{neg_witness} expresses the fact that the witness ${\cal W}_N$ detects entangled states in the vicinity of the state given in Eq.~(\ref{eq:maxent}). If $\mu_{\rm min}^{(1)}>\mu_{\rm min}^{(2)}$ holds then ${\cal W}_N$ is an entanglement witness and its expectation value even gives us the minimal eigenvalue of the partial transpose of the nearest-neighbor two-qubit density matrix. The minimum of the expectation value of the witness is $\langle {\cal W}_N\rangle=-1/2$ for the maximally entangled state given in Eq.~(\ref{eq:maxent}). If $\mu_{\rm min}^{(1)}\le\mu_{\rm min}^{(2)}$ holds  then based on the previous arguments we have $\langle {\cal W}_N \rangle = \mu_{\rm min}^{(2)}\ge 0.$ 

{Let us now show that a simple entanglement witness can be obtained based on the expression for $\mu_{\rm min}^{(1)}$ given in Eq.~(\ref{eq:mumin1_corr}). For the special case when
\be
\langle \sigma_l^z\rangle+\langle \sigma_{l+1}^z \rangle=0, \label{eq:condz}
\ee
we can derive an entanglement witness
\be
{\cal W'}_N=-\frac{1}{4}( \sigma_l^x \sigma_{l+1}^x +\sigma_l^y \sigma_{l+1}^y  - \sigma_l^z\sigma_{l+1}^z -\openone).
\label{neg_witness2}
\ee
It is instructive to rewrite the witness given in Eq.~(\ref{neg_witness2}) as 
\be
{\cal W}_N'=\frac1 2\openone-\vert \Psi^+\rangle\langle \Psi^+\vert,
\label{neg_witness3}
\ee
where the state is defined as
\be
\ket{\Psi^+}=\frac1 {\sqrt 2}(\ket{\uparrow\downarrow}+\ket{\downarrow\uparrow}).\label{eq:singlet}
\ee
If the condition in Eq.~(\ref{eq:condz}) holds, then
\be
\langle {\cal W}_N' \rangle = \mu_{\rm min}^{(1)}.
\ee
However, in general,
\be
\langle {\cal W}_N' \rangle \ge \mu_{\rm min}^{(1)},
\ee
holds and the witness ${\cal W'}_N$ is not sufficient to detect all entangled states, if Eq.~(\ref{eq:mumin12ineqB}) does not hold.}

{We will show that a one-parameter family of entanglement witnesses is sufficient. Let us define the entanglement witness \cite{Horodecki2009Quantum,Guhne2009Entanglement}
\be
{\cal W}_{N,p}''=(\ket{\Psi_p}\bra{\Psi_p})^{T_A},\label{eq:wfamily}
\ee
where the state is given as
\be
\ket{\Psi_p}=\sqrt{p}\ket{\uparrow\uparrow}-\sqrt{1-p}\ket{\downarrow\downarrow}.\label{eq:Psi_p}
\ee
Since for $\rho_{23}$ the relation Eq.~\eqref{eq:rho230} holds, the two coefficients have an opposite sign. 
Based on the general relation 
\be
{\rm Tr}(XY)={\rm Tr}(X^{T_A}Y^{T_A}),
\ee
we can write that
\begin{align}
{\rm Tr}({\cal W}_{N,p}'' \bm{\rho})&={\rm Tr}[ ({\cal W}_{N,p}'')^{T_A} \bm{\rho}^{T_A} ]\nonumber\\
&= {\rm Tr}( \ket{\Psi_p}\bra{\Psi_p} \bm{\rho}^{T_A} ) \ge \mu_{\rm min}^{(1)}. \label{eq:WNp}
\end{align}
Based on Eq.~\eqref{ptranspose},  we can see that when  Eq.~(\ref{eq:mumin12ineqB}) does not hold, the eigenvector of $\bm{\rho}^{T_A}$ with the negative eigenvalue is of the form given in Eq.~\eqref{eq:Psi_p}. Hence, for some $p$ the inequality in  Eq.~\eqref{eq:WNp} is saturated. Thus, the eigenvalue can be obtain as a minimization over the expectation values of the witnesses as
\be
\min_p \langle {\cal W}_{N,p}''\rangle_{ \bm{\rho}}=\mu_{\rm min}^{(1)},
\ee
which is an approach somewhat different from that of Ref.~\cite{Wu2005Entanglement}, which presented a nonlinear witness operator. Note that ${\cal W}_N'$ is member of the family
\be
{\cal W}_N'={\cal W}_{N,1/2}''.
\ee}

In summary, the witness ${\cal W}_N$ given in Eq.~(\ref{neg_witness1}) can be used to detect entanglement in the nearest-neighbor state of an $XY$ system in thermal equilibrium if $h^2+\gamma^2>1$ holds. {Otherwise, ${\cal W}_N'$ given in Eq.~(\ref{neg_witness2}) will detect many entangled states. However, all entangled states are detected by the family of witnesses given in Eq.~(\ref{eq:wfamily}).}

In the case of a quench the parameters of the Hamiltonian are changed suddenly, as described in Sec.~\ref{sec:quench}. Using the quench protocol $(h_0,\gamma) \to (h,\gamma)$ we noticed numerically, that the condition Eq.~(\ref{eq:mumin12ineqB}) is only valid in a part of the phase diagram. We {find}  that ${\cal W}_N$ does not detect entanglement in the region $h_0<h_d$ and $h<h_d$.

\section{Temperature bounds for equilibrium thermal states}
\label{sec:thermal}

In this section, we consider thermal states, which are generally entangled at low temperature, but at a sufficiently high temperature they are separable. Using the energy-based witness given in Eq.~(\ref{eq:ewit}) and the {negativity-based witnesses given in Eqs.~(\ref{neg_witness1})} and (\ref{eq:wfamily}), we calculate temperature bounds below which the state is detected as entangled. We note that at specific points there have been calculations for infinite chains \cite{Toth2005EntanglementWitnesses,Wu2005Entanglement,Dowling2004Energy}. Here we consider several parts of the phase diagram and also study the finite size corrections that turn out to be very important in the ordered phase.

\begin{figure}[t!]
\vspace{0.5cm}
\includegraphics[width=1.\columnwidth,angle=0]{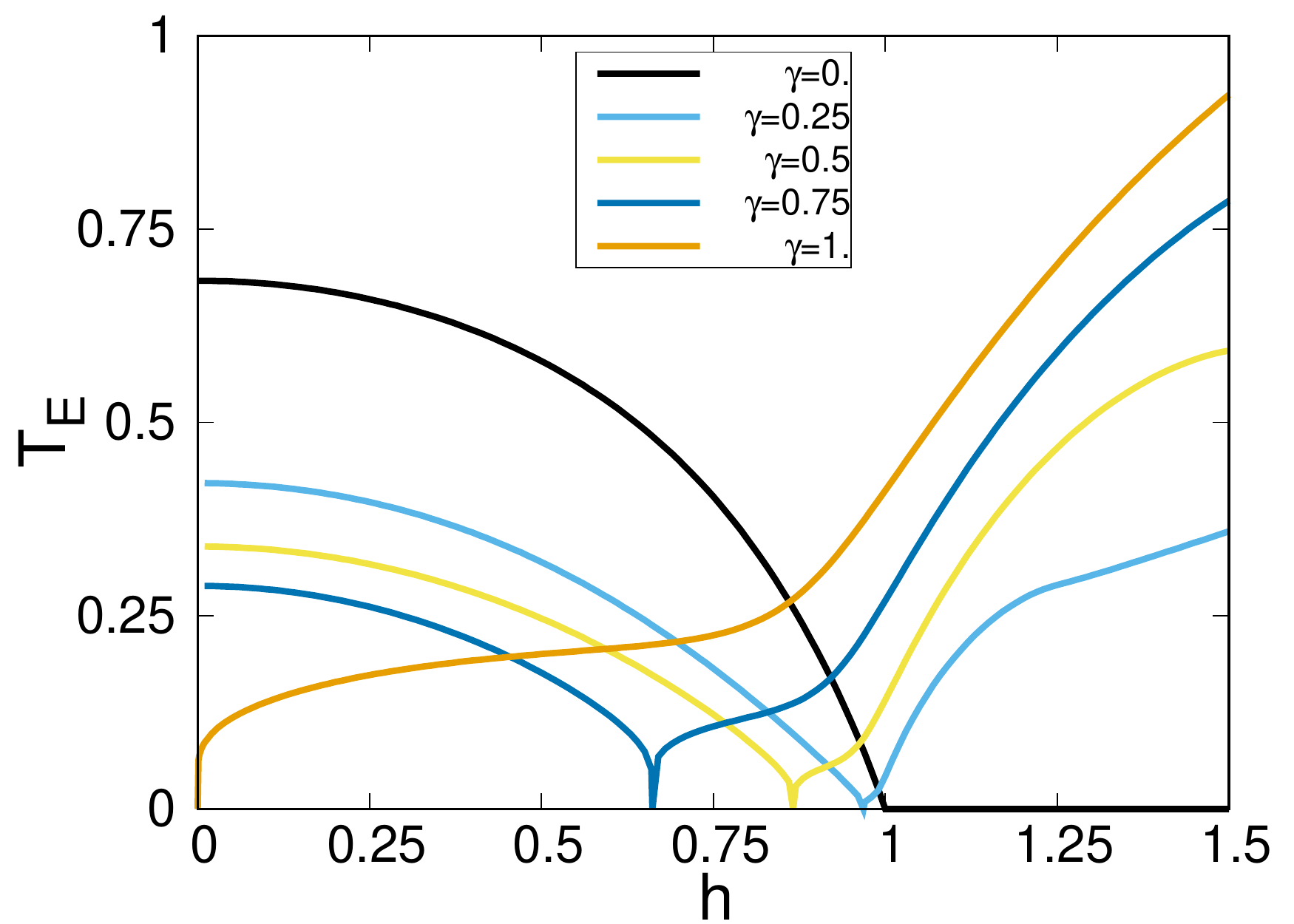}
\caption{Temperature bound of entangled equilibrium thermal states of an $XY$ chain in the thermodynamic limit calculated using the energy-based entanglement witness given in Eq.~(\ref{eq:ewit}). The Hamiltonian $\cal H$ is given in Eq.~(\ref{hamilton}). The witness detects a state entangled, if $\langle {\cal H}\rangle$ is smaller than the minimal energy for separable states, given in Eq.~(\ref{E_sep}). The value
$\gamma=1$ corresponds to the transverse Ising model.
At the disorder point, $h$ is given by Eq.~(\ref{eq:hd}).}
\label{fig_2}
\end{figure} 

\begin{figure}[t!]
\vspace{0.5cm}
\includegraphics[width=1.\columnwidth,angle=0]{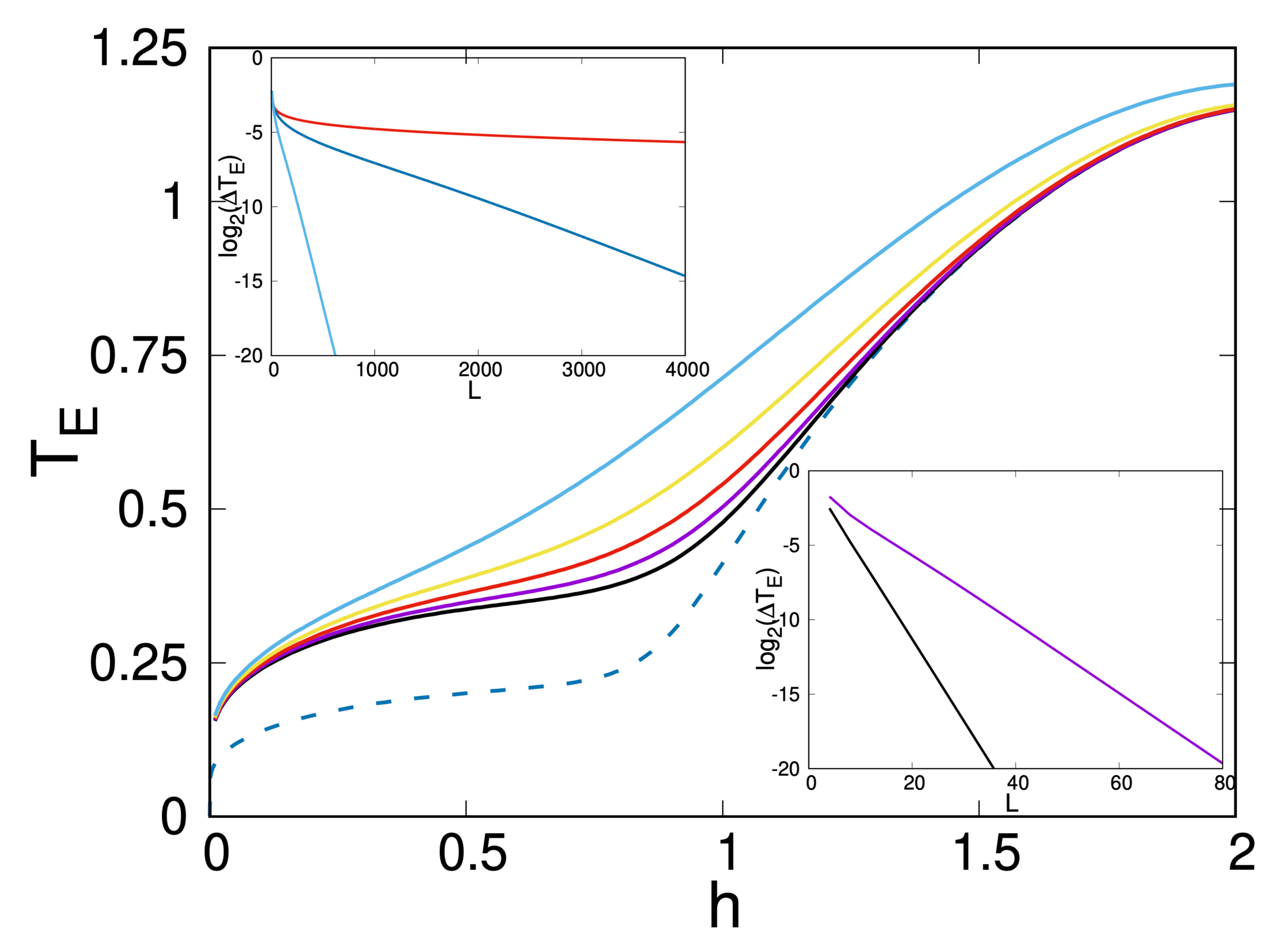}
\caption{Temperature bound of entangled equilibrium thermal states of an $XY$ chain for finite systems calculated using the energy-based entanglement witness given in Eq.~(\ref{eq:ewit}). Main panel: (solid) Temperature bound of entangled equilibrium thermal states in finite transverse Ising chains for (from top to bottom) $L=4,6,8,10,12.$ (dashed) The $L \to \infty$ case, which is the same as the $\gamma=1$ line in Fig.~\ref{fig_2}. Insets: finite-size corrections in a lin-log scale at different values of $h$. Upper inset: corrections in the ordered phase for (from top to bottom) $h=0.25,0.5$ and $0.75.$ Lower inset: (top) at the critical point, $h=1.0$ and (bottom) in the paramagnetic phase, $h=1.25.$ Note the different scales in the $x$ axis.
\label{fig_2a}}	
\end{figure} 

\subsection{Energy-based witness}

The energy-based witness ${\cal W}_E$ described in Sec.~\ref{sec:energybased} detects a state as entangled if $\langle H \rangle<E_{\rm sep}.$ For thermal states, there is a corresponding temperature bound $T_E$ such that the state is detected as entangled by the witness if $T<T_E.$ We will study thermal states in the $XY$ chain based on these ideas. The Hamiltonian $\cal H$ is given in Eq.~(\ref{hamilton}), and the minimal energy for separable states is given in Eq.~(\ref{E_sep}). 

\subsubsection{Thermodynamic limit}

First, we consider the thermodynamic limit. In Fig.~\ref{fig_2}, we show $T_E$ as a function of $h$, for different values of the anisotropy. We can make the following observations. 

Let us start with the paramagnetic phase where $h>1,$ as shown in Fig.~\ref{fig_1}. At a fixed $h$ value, $T_E$ monotonously decreases with $\gamma$, and we even have $T_E \to 0$ as $\gamma \to 0.$  At a fixed $0 < \gamma\le1$, $T_E$ also decreases with decreasing $h$ and passing through the critical point at $h=1$ in the ordered phase approaches zero at the disorder point $h=h_d,$ where $h_d$ is given in Eq.~(\ref{eq:hd}). Decreasing $h$ further at the other side of the disorder point, the temperature bound starts to increase monotonously. In Fig.~\ref{fig_2}, when going from bottom to top, the order of the $ T_E (h) $ curves at $ h = 0 $ is the opposite of that at $ h> 1.$ That is, if $\gamma_1>\gamma_2$ then we have 
\be
T_E(h,\gamma_1)>T_E(h,\gamma_2),
\ee
if $h>1.$ On the other hand, we have 
\be
T_E(h,\gamma_1)<T_E(h,\gamma_2),
\ee
at $h=0.$ Finally, at the disorder point $h=h_d,$ there is a singularity
 \be
 T_E \sim \ln^{-1}|h-h_d|. 
 \ee
Let us turn to the quantum critical point $h=1$. Close to it there is an inflection point of the $T_E(h)$ curve, the position of which approaches $h=1$ as $\gamma \to 0$.

{If $\gamma=0$ then we have the $XX$-model and it has a polarized ground state if $h\ge1,$ which is separable. Consequently, $T_E=0$ in this region.}

\subsubsection{Finite-size corrections}

We have repeated the calculation of the temperature bound on finite transverse Ising chains working in the Ising spin basis up to $L=12$. Results are shown in the main panel of Fig.~\ref{fig_2a}. The corrections are relatively small in the paramagnetic phase, while in the ordered phase they are quite large. To see more precisely the finite-size dependence of the correction term we repeated the calculation in the free-fermion basis as described in the Appendix~\ref{sec:finite_fermionic}. 

\begin{figure}[t!]
\vspace{0.5cm}
\includegraphics[width=1.\columnwidth,angle=0]{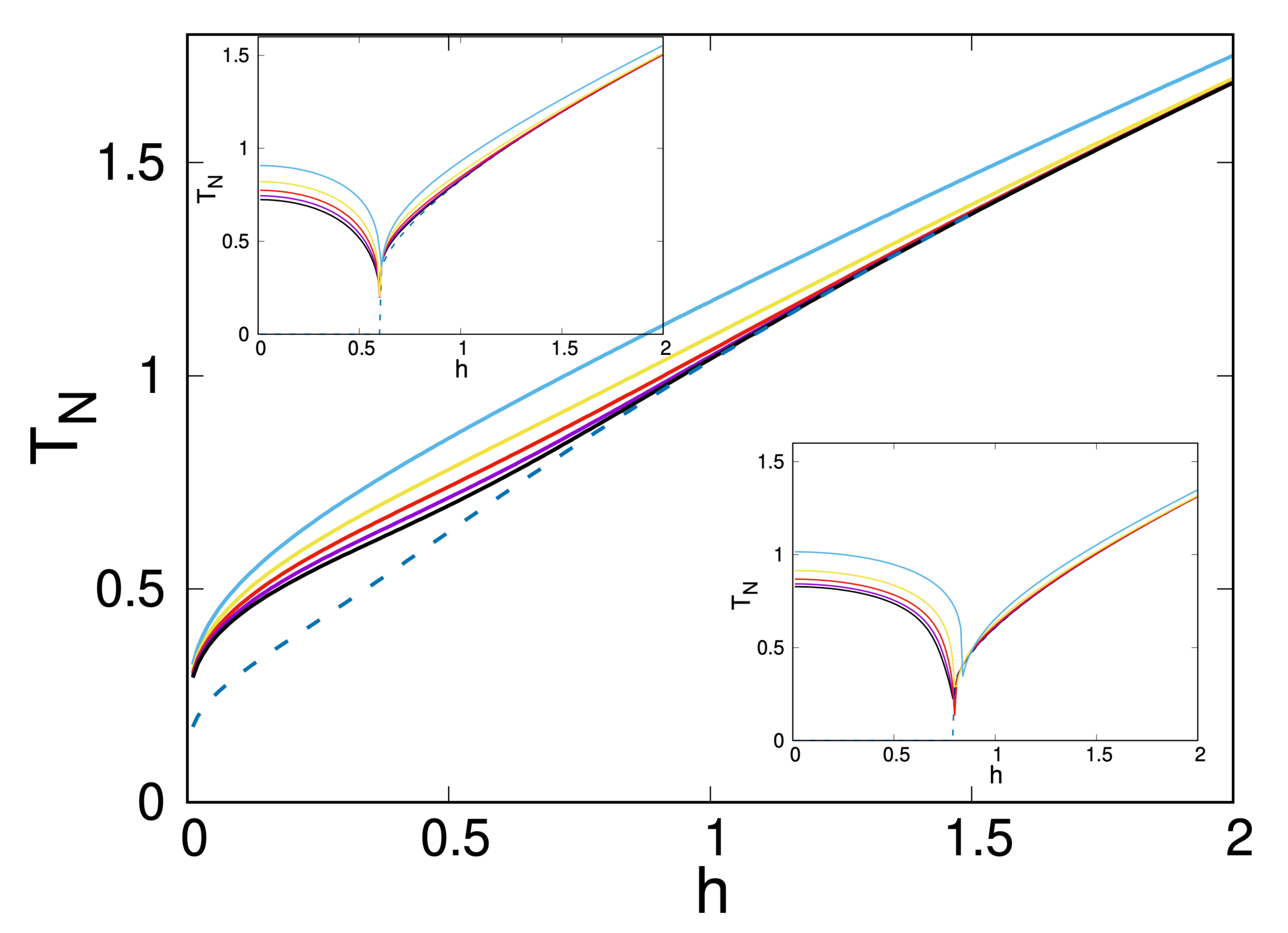}
\caption{Temperature bound of entangled equilibrium thermal states of an $XY$ chain for finite systems calculated using the negativity-based entanglement {witnesses}. (Solid) (from top to bottom) $L=4,6,8,10,12.$ (Dashed) $L \to \infty$ case. Main panel: $\gamma=1$, quantum Ising chain. Upper inset: $\gamma=0.8.$ Lower inset: $\gamma=0.6.$ 
\label{fig_2b}}	
\end{figure} 

We have studied the finite-size dependence of the correction term
\be
\Delta T_{E}(L)=T_{E}(L)-T_{E}
\ee
at different points of the phase diagram. We considered several values of $h$ corresponding to the ordered phase and to the paramagnetic phase, as well as to the critical point. Using lin-log scale in the figures the curves are asymptotically linear, and thus the size dependence is well described by the form
\be
\Delta T_{E}(L) \simeq A L^{-a}\exp(-L/L_0),
\ee
where the power-law correction term is relevant for short chains, $L < L_0$. In the paramagnetic phase and at the critical point $L_0$ is just a few lattice spacings ($L_0\approx 2.6$ and $6.2$, for $h=1.25$ and $1.0$, respectively), while in the ferromagnetic phase it is much longer ($L_0\approx 52$, $610$ and $5800$, for $h=0.75$, $0.5$ and $0.25,$ respectively) and tends to infinity at $h \to 0$. (In the ferromagnetic phase the exponent $a$ is $h$-dependent, it is approximately $0.25$ and $0.42$, for $h=0.25$ and $0.5$, respectively.) The slow finite-size convergence of the results is due to the presence of an exponentially small gap in the ordered phase, which has a large correction.

\subsection{Negativity-based {witnesses}}

The negativity-based {witnesses} ${\cal W}_N$ {and ${\cal W}_{N,p}''$} described in Sec.~\ref{sec:negativtybased} {detect} a state as entangled if its negativity is nonzero. For thermal states, there is a corresponding temperature bound $T_N$ such that the state is detected as entangled by the witness if $T<T_N.$ We will study thermal states in the $XY$ chain based on these ideas. 

Results for the transverse Ising chain are in the main panel of Fig.~\ref{fig_2b}. We carried out calculations for the finite chain in the Ising spin bases. We also made calculations in the thermodynamic limit using the formula with the nearest-neighbor correlations given in Eq.~(\ref{witness}). We find that the temperature bounds obtained using the entanglement negativity {witness} given in Eq.~(\ref{neg_witness1}) are considerably higher than those obtained using the energy-based witness; { the difference is around a factor of two for $h\approx 0,$ while the difference is less for $h>1.$} The finite-size corrections appear almost negligible in the paramagnetic phase ($h>1$), but are much larger in the ferromagnetic phase ($h<1$). The finite-size corrections observed for the negativity-based entanglement witness have the same origin, as those for the energy witness. Indeed, both the entanglement negativity [given via the minimum of the values in Eqs.~(\ref{witness}) and (\ref{eq:mumin1_corr})] and the energy [given  in Eq.~(\ref{H_corr})] can be obtained via nearest-neighbor correlation functions.

We have repeated the calculation for two values fulfilling $0<\gamma<1,$ and the results are shown in the insets in Fig.~\ref{fig_2b}. 
{The results are calculated using the negativity-based witness given in Eq.~(\ref{neg_witness1}) for $h>h_d$ and the family of negativity-based witnesses in Eq.~(\ref{eq:wfamily}) for $h<h_d.$ Equivalently, we used the formula giving the smallest eigenvalue of the partial transpose of the two-qubit state. In particular, we used the formula for $\mu^{(2)}_{\rm min}$ in  Eq.~(\ref{witness}) for $h>h_d,$ and the formula for $\mu^{(1)}_{\rm min}$ in  Eq.~(\ref{eq:mumin1_corr}) for $h<h_d.$ If the smallest eigenvalue is negative, then the two-qubit state is entangled.}
In these cases too, the convergence to the thermodynamic limit is very fast in the paramagnetic phase, while in the ferromagnetic phase the convergence is much slower. We find that the temperature bounds obtained using the entanglement negativity witness are considerably higher than those obtained using the energy-based witness.

\section{Entanglement in nonequilibrium postquench states}
\label{sec:post_quench}

In this section, we consider global quenches in the system, as described in Sec.~\ref{sec:quench} and study the entanglement properties of nonequilibrium stationary states, which are obtained in the large-time limit after the quench. To calculate averages we use the GGE protocol and assign different effective temperatures to each fermionic modes, as described in Eqs.~(\ref{T_eff}) and (\ref{t_to_cos}). First we apply the energy-based witness in Eq.~(\ref{eq:ewit}) and then the entanglement negativity-based {witnesses in Eqs.~(\ref{neg_witness1})  and (\ref{eq:wfamily}).}

\subsection{Energy-based witness}

The energy-based witness detects the postquench state as entangled if
\be
\langle \Phi_0| \displaystyle{\cal H} |\Phi_0 \rangle < E_{\rm sep}\label{eq:Esepcond}
\ee
holds, where $\cal H$ is the Hamiltonian after the quench. In the following, for simplicity we consider the case when $\gamma_0=\gamma$ holds and study the area in the $(h_0,h)$ plane corresponding to quenches in which entanglement has been detected. 

We will now obtain the boundaries analytically using the bound  $E_{\rm sep}$ given in Eq.~(\ref{E_sep}). We have to consider two cases.

(i) Let us consider first the case when 
\be
h \le 1+\gamma\label{eq:hcond1}
\ee
holds.
Then, from Eq.~(\ref{eq:Esepcond}), we obtain the relation
\be
-h I_1(h_0,\gamma)-I_2(h_0,\gamma)<-\frac{(1+\gamma)^2+h^2}{2(1+\gamma)}.
\label{h_h0}
\ee
The right-hand side of Eq.~(\ref{h_h0}) corresponds to the top line in the equation defining the bound for separable states in Eq.~(\ref{E_sep}). The left-hand side is equal to the energy of the state after quench given in Eq.~(\ref{eq:quenchth}). The integrals are defined as 
\begin{align}
I_1(h_0,\gamma)&=\frac{1}{\pi}\int_0^{\pi} \frac{h_0-\cos p}{\sqrt{\gamma^2 \sin^2 p+\left(h_0-\cos p\right)^2}} \textrm{d} p,\nonumber\\
I_2(h_0,\gamma)&=\frac{1}{\pi}\int_0^{\pi} \frac{-h_0\cos p+\cos^2 p+\gamma^2\sin^2p}{\sqrt{\gamma^2 \sin^2 p+\left(h_0-\cos p\right)^2}} \textrm{d} p. \label{eq:int}
\end{align}
The left-hand side and the right-hand side of Eq.~(\ref{h_h0}) are up to second-oder in $h$. Thus, the condition given in Eq.~(\ref{h_h0}) holds if
\be
h_- < h <h_+,\label{eq:hminusplus}
\ee
where the lower and upper bounds are defined as
\be
h_{\pm}=\left(I_1 \pm \sqrt{I_1^2+2I_2/(1+\gamma)-1} \right)(1+\gamma).
\label{h_pm}
\ee
Note that $h_-$ and $h_+$ depend on $h_0.$ Note also that we assumed that the condition given in Eq.~(\ref{eq:hcond1}) is satisfied, and thus the interval given by Eq.~(\ref{eq:hminusplus}) must be reduced taking into account Eq.~(\ref{eq:hcond1}).
 
(ii) Let us consider now the case when 
\be
h > 1+\gamma \label{eq:hcond2}
\ee
holds.
Then, from Eq.~(\ref{eq:Esepcond}), we obtain the relation
\be
-h I_1(h_0,\gamma)-I_2(h_0,\gamma)<-h.
\label{h_h0b}
\ee
The right-hand side of Eq.~(\ref{h_h0b}) corresponds to the bottom line in the equation defining the bound for separable states in Eq.~(\ref{E_sep}). The left-hand side is equal to the energy of the state after quench given in Eq.~(\ref{eq:quenchth}). The integrals are defined in Eq.~(\ref{eq:int}).
For a fixed $h_0,$ the left-hand side and the right-hand side of Eq.~(\ref{h_h0b}) are up to first-oder in $h$.
Thus, Eq.~(\ref{h_h0b}) holds if
\be
h <\tilde{h}_+,\label{eq:hhtilde}
\ee
where the upper bound is defined as
\be
\tilde h_+=\frac{I_2}{1-I_1},
\label{h_pm2}
\ee
and it depends on $h_0.$  Note that we assumed that the condition given in Eq.~(\ref{eq:hcond2}) is satisfied, thus the set of $h$ values satisfying Eq.~(\ref{eq:hhtilde}) must be reduced taking into account Eq.~(\ref{eq:hcond2}).

Let us determine now the set of $h$ values for which the quantum state is detected as entangled. Let us start from a small $h_0$ value. In this case, simple algebra yields
\be
h_-\le h_+<\tilde h_+<1+\gamma 
\ee 
and if the postquench state is detected as entangled, then $h$ must fulfill Eq.~(\ref{eq:hminusplus}). Let us increase $h_0.$ We arrive at a point when 
\be
h_-<h_+=\tilde h_+=1+\gamma.
\ee
Increasing $h_0$ further, we find that 
\be
h_-<1+\gamma< h_+ <\tilde h_+.
\ee 
If the postquench state is detected as entangled, then $h$ must fulfill the relation
\be
h_- < h <\tilde h_+.\label{eq:hminusplus2}
\ee
Based on these, for a given $h_0$ and $h$, the postquench state is entangled if 
\be
h_- < h < \begin{cases} h_+\text{  if }\tilde h_+\le 1+\gamma,\\ \tilde h_+\text{ if }\tilde h_+>1+\gamma, \end{cases}
\ee
where $h_+$  and $h_-$ are given in Eq.~(\ref{h_pm}), and  $\tilde h_+$ is given in Eq.~(\ref{h_pm2}). We stress that we obtained the boundaries analytically.

\begin{figure*}[t]
\centering
  \begin{tabular}{@{}cc@{}}
\includegraphics[width=1.\columnwidth,angle=0]{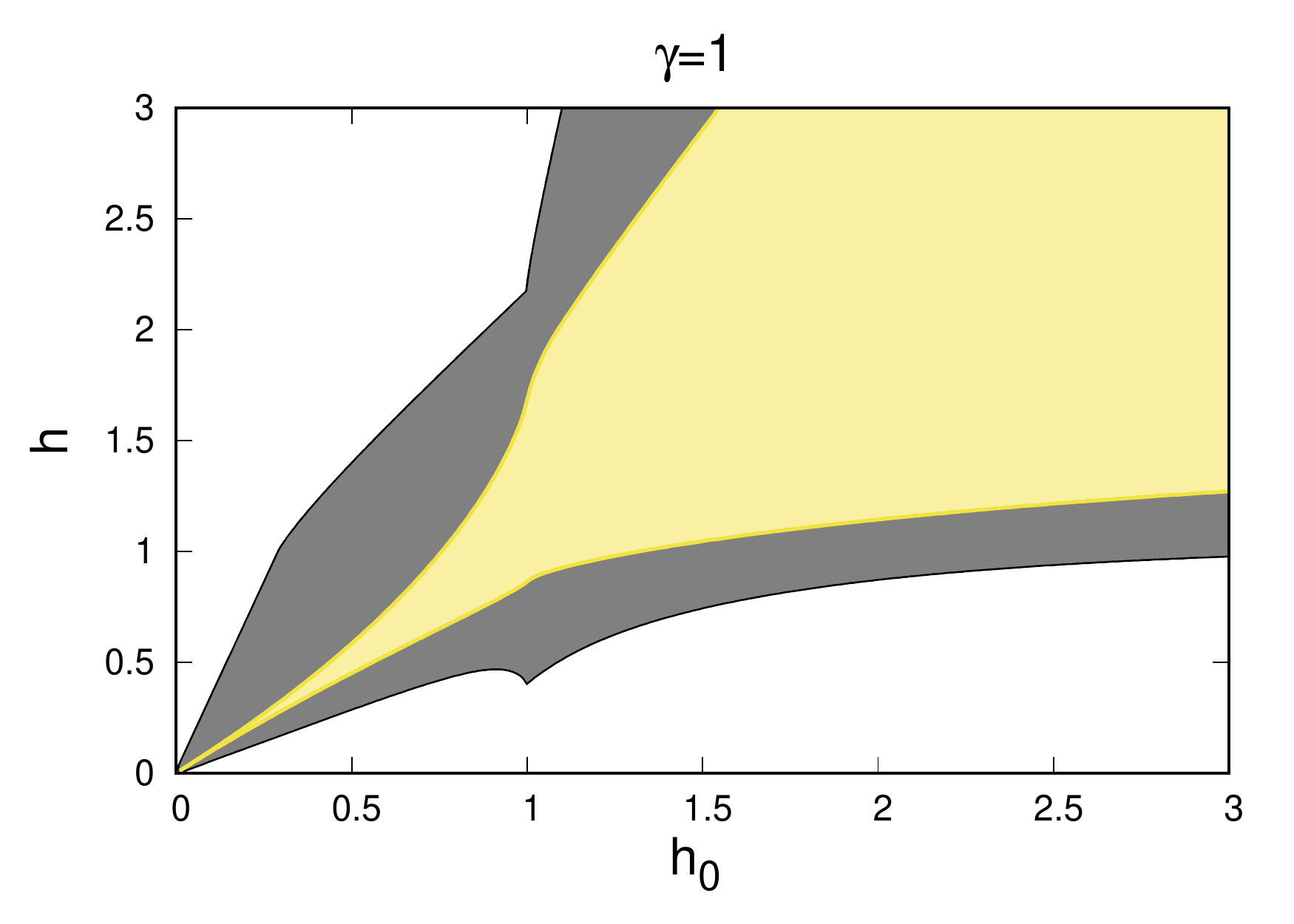} &
\includegraphics[width=1.\columnwidth,angle=0]{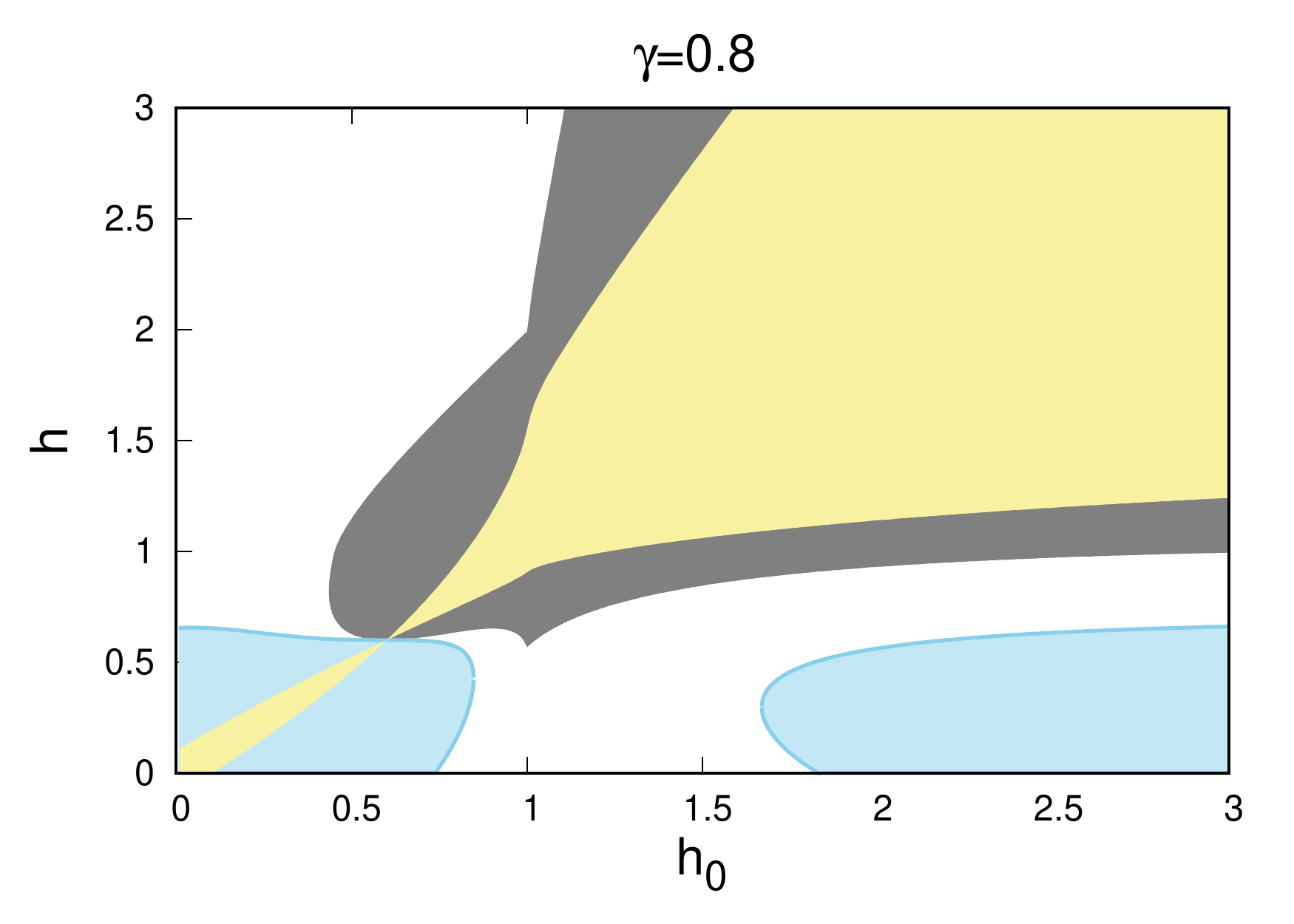} \\
\hspace{0.7cm}(a) & \hspace{1cm}(b)\\
&\\
\includegraphics[width=1.\columnwidth,angle=0]{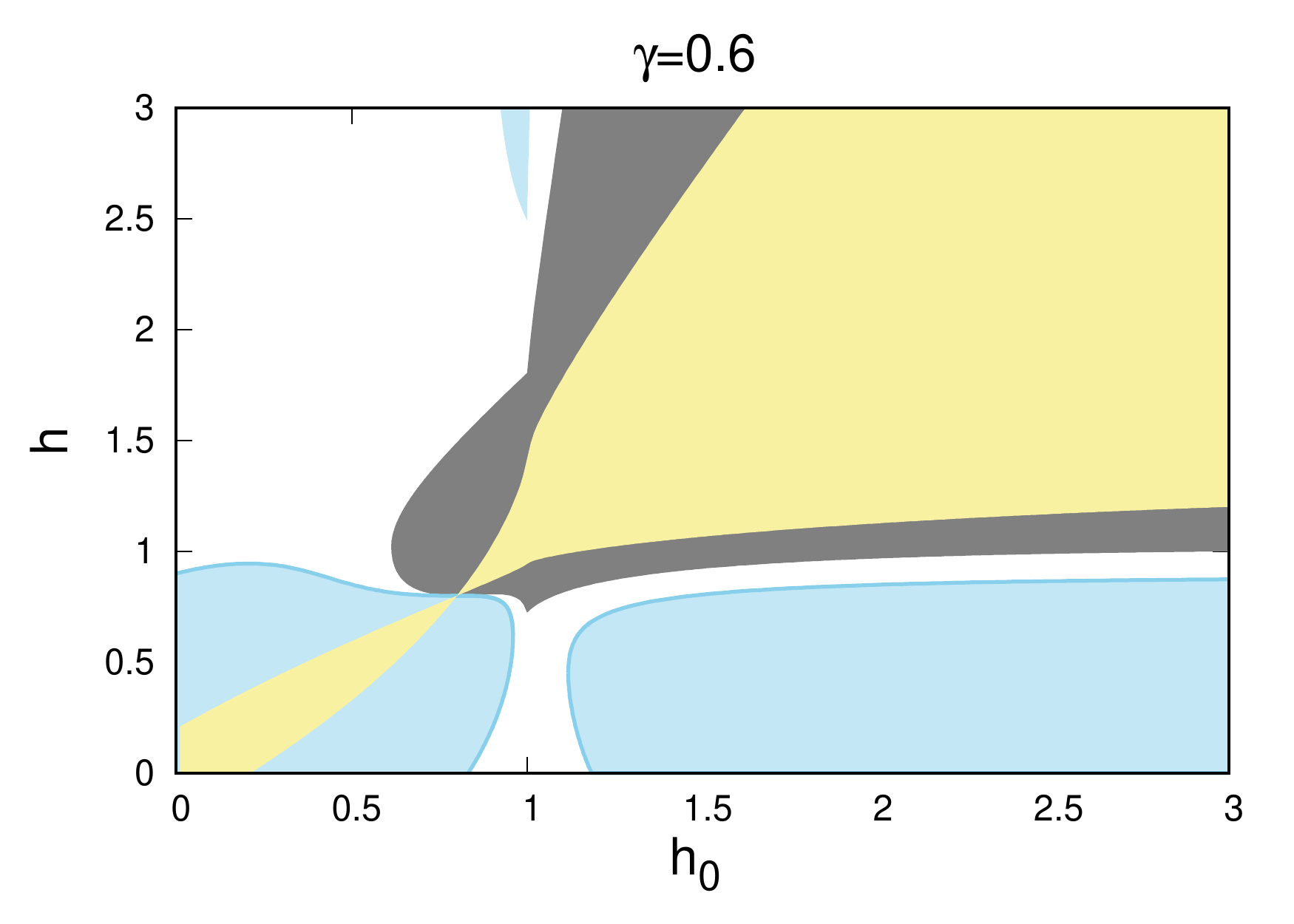} &
\includegraphics[width=1.035\columnwidth,angle=0]{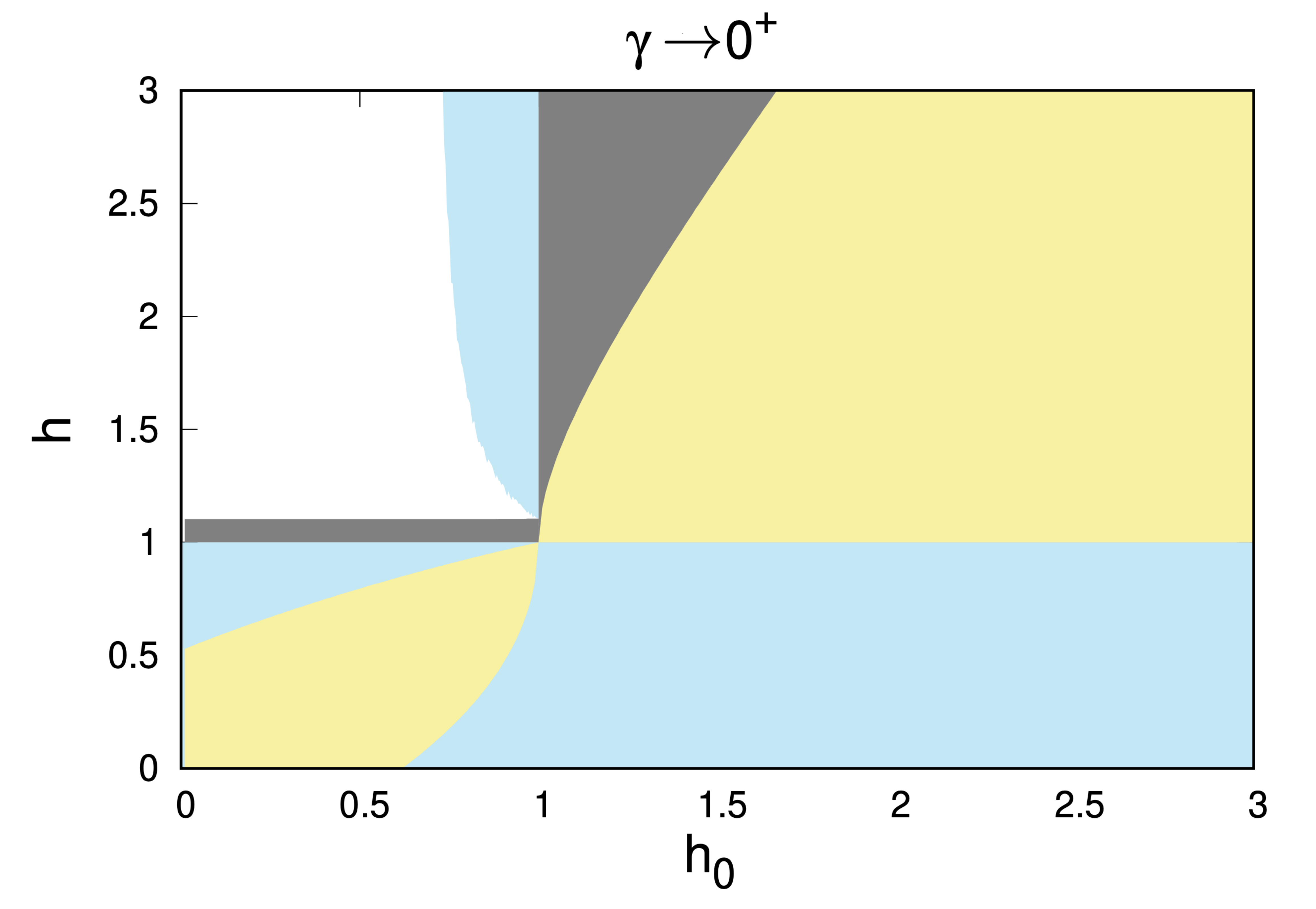} \\
\hspace{0.7cm}(c) & \hspace{1cm}(d)\\
  \end{tabular}
\caption{Postquench states after a sudden quench protocol $(h_0,\gamma) \to (h,\gamma)$ in the $XY$ chain. (Yellow) Entanglement is detected by the energy-based witness. {(Blue) Entanglement is detected in the postquench state by the negativity-based method using $\mu^{(1)}_{\rm min}$ in Eq.~(\ref{eq:mumin1_corr}). Equivalently, the entangled states can be detected by the family of negativity-based entanglement witnesses given in Eq.~(\ref{eq:wfamily}). (Gray)  Entanglement is detected in the postquench state by the negativity-based witness in Eq.~(\ref{neg_witness1}). The blue and the gray areas touch each other at the disorder point, where $h_0=h_d$ and $h=h_d,$ and $h_d$ is given in Eq.~(\ref{eq:hd}). The blue and the gray filled areas overlap with the yellow area such that all yellow-filled area is part of the blue and the gray filled areas. Thus, all postquench states detected by the energy-based witness are also detected by the negativity-based witness.} {In panel (d) the limit $\gamma \to 0^+$ is considered, which is different from the case when $\gamma=0$, see text.}
\label{fig_7}}	
\end{figure*} 

In Fig.~\ref{fig_7}, the region in which the postquench states are detected as entangled by the energy-based witness are indicated with a yellow filled area. In Fig.~\ref{fig_7}(a), we consider the case with $\gamma=1$ and hence we have a quantum Ising chain. The region corresponding to postquench states detected as entangled consists of a single connected part. In Fig.~\ref{fig_7}(b), (c), and (d),  we have $\gamma=0.8$, $\gamma=0.6$ and $\gamma \to 0^+$, respectively. The region corresponding to postquench states detected as entangled consists of two connected parts, which touch each other at 
\be
(h_0=h_d,h=h_d), \label{eq:point}
\ee
where $h_d$ is given in Eq.~(\ref{eq:hd}). The point given by Eq.~(\ref{eq:point}) corresponds to a quench in which the system is at the disorder line before and after the quench, and due to $h=h_0$ the parameter $h$ does not change. Thus, the thermal state is separable, as discussed in Sec.~\ref{sec:energybased}.

If the initial state is in the ferromagnetic domain, $h_0<1$, the region of postquench states detected as entangled is rather narrow, the values of $h$ corresponding to an entangled entangled postquench states are close to $h_0$. Starting the quench from the paramagnetic phase, $h_0>1,$ the the domain of postquench states detected as entangled is relatively wider. 

Starting the quench just at the critical point, $h_0=1$, the boundaries exhibit a singularity since
\be
\frac{\textrm{d} h_{\pm}}{\textrm{d} h_{0}} \sim -(h_{\pm}-1)\ln|h_0-1|\;
\ee
holds.

\subsection{Negativity-based {witnesses}}

We have calculated the region of postquench states detected as entangled by the entanglement negativity-based witness. {In Fig.~\ref{fig_7}, the results are shown indicating that the postquench states  are detected as entangled with blue (gray) filled areas, if the calculation is used $\mu^{(1)}_{\rm min}$ given in Eq.~(\ref{eq:mumin1_corr})  ($\mu^{(2)}_{\rm min}$ given in Eq.~(\ref{witness})). In other words, the blue (gray) area denotes states that are detected  by the family of negativity-based witnesses in Eq.~(\ref{eq:wfamily}) (by the negativity-based witness given in Eq.~(\ref{neg_witness1})). All postquench states that are detected by the energy-based witness are also detected by the negativity-based methods.} 

In Fig.~\ref{fig_7}(a), we consider the case with $\gamma=1$ and hence we have a quantum Ising chain. In this case, the negativity-based witness with {$\mu^{(2)}_{\rm min}$ is applicable in the whole phase diagram. In Fig.~\ref{fig_7}(b), (c), and (d),  we have $\gamma<1,$ and the condition $\mu_{\rm min}^{(1)}<\mu_{\rm min}^{(2)}$ is fulfilled in a part of the phase diagram. Thus, entangled postquench states are also detected based on $\mu^{(1)}_{\rm min}$}. 

{In Fig.~\ref{fig_7}(d) the limit $\gamma \to 0^+$ is considered, which is however not identical to the $\gamma=0$ case, which corresponds to the $XX$-model. In the latter model, for all $h_0$ and $h$ we have
\be
[{\cal H}_0,{\cal H}]=0,
\label{H0_H}
\ee
hence this case needs a special treatment. If the system is ideally isolated, than according to Eq.(\ref{Phi(t)}) the system remains in its original state, $|\Phi_0(t)\rangle=\exp(-i\tilde{E}_0 t) |\Phi_0\rangle$, with $\tilde{E}_0=\langle \Phi_0|{\cal H}|\Phi_0\rangle$. In this state the the correlations in Eqs.~(\ref{nn_corr}) and (\ref{par_corr}) are given by:
\begin{align}
g_c&=\begin{cases} 
0,& h_0\ge 1,\\
\frac{2}{\pi}\sqrt{1-h_0^2}, & h_0 < 1,\nonumber\\
\end{cases}\\
g_s&=0,\nonumber\\
g_0&=\begin{cases} 
1, & h_0\ge 1,\\
1-2\frac{\arccos (h_0)}{\pi}, & h_0 < 1.
\end{cases}
\label{par_corr_g0}
\end{align}
Thus, the postquench state is separable for $h_0 \ge 1$ and entangled for $h_0<1$, independently of the value of the postquench parameter $h$.}

We note, that $|\Phi_0(t)\rangle$ is generally not the ground state of ${\cal H}$, since the modes for $p_1<p<p_2$ are occupied, where $p_1={\rm min}[\arccos (h_0),\arccos (h)]$ and $p_2={\rm max}[\arccos (h_0),\arccos (h)]$. This follows from Eqs.~(\ref{f_p}) and (\ref{Delta}).
In the limit $\gamma \to 0^+,$ the system can be considered not ideally isolated, which means that an arbitrarily small, but non-zero interaction is present with the environment. Then the system for sufficiently long time will decay to its ground state, having no occupied modes. This is formally equivalent to the condition in Eq.~(\ref{T_eff}), which for having  $|\cos \Delta_p|=1$ predicts $T_{\rm eff}(p)=0$ for each modes. As a consequence the correlations in Eqs.~(\ref{nn_corr}) and (\ref{par_corr}) are the same as in the ground state of the postquench Hamiltonian, and the postquench state is separable for $h \ge 1$ and entangled for $h<1$, independently of the value of the initial parameter $h_0$.

We can see that the entangled region detected by the negativity method is larger, than that detected by the energy-based witness. Even if the initial state is separable, corresponding to $h_0=h_d,$ the postquench state can be detected as entangled by the negativity-based witness, if $h\ne h_d.$ In this case, the entanglement is obtained due to the entangling dynamics of the Hamiltonian $\cal H.$

Note that the border of the entangled domain is singular at $h_0 = 1,$ which corresponds to the case when the system is critical before the quench. Note also that for $\gamma<1$ the border of the entangled domain is horizontal where the two disconnected parts meet at the point given in Eq.~(\ref{eq:point}).

\section{Discussion}
\label{sec:discussion}

Entanglement in mixed quantum states is a difficult problem, in particular when the degrees of freedom is large and we approach the thermodynamic limit. The possible systems of investigation are usually quantum spin systems, most often quantum spin chains. Such type of quantum spin chains could have experimental realizations in condensed matter systems  \cite{dutta_aeppli_chakrabarti_divakaran_rosenbaum_sen_2015} or they could be engineered artificially through ultracold atomic gases in an optical lattice. Recently, this latter type of technique is very well developed and different intriguing questions could be studied experimentally \cite{Greiner2002,Paredes2004,Sadler2006,PhysRevLett.98.160404,Kinoshita2006,Hofferberth2007,RevModPhys.80.885,Trotzky2012,Cheneau2012}. 

{In this paper we consider the $XY$ chain, which is integrable through free-fermionic techniques and several exact results are available, mainly in the ground state but there are some known results even at finite temperature \cite{PhysRevA.2.1075,PhysRevA.3.2137}. We consider the entanglement properties of mixed states of the $XY$ chain. To detect entanglement we use different entanglement witnesses: an energy-based witness and a family of witnesses that detect states with a nonzero bipartite entanglement negativity. Using the former is technically simpler, but it does not detect all states that have nearest-neighbor entanglement. In contrast, the witnesses based on entanglement negativity can detect all states that have nearest-neighbor entanglement, but it is generally more complicated to calculate their expectation value.}

First we considered thermal states, for which some previous calculations are available at specific points for infinite chains  \cite{Toth2005EntanglementWitnesses,Wu2005Entanglement,Dowling2004Energy}. Here we performed the entanglement detection both for finite chains and in the thermodynamic limit. In the ordered phase of the system very strong finite-size corrections are detected, which are due to the presence of a quasidegenerate first excited state and the corresponding gap is exponentially small with the length of the chain.

One of the main novelties of the present paper is that we considered also mixed states which are due to a quench, when parameters of the Hamiltonian of the system are changed abruptly and the time evolution of the system is governed by the new Hamiltonian. After a sufficiently long time the system will approach a nonequilibrium stationary state, the properties of which are of vital importance. The postquench state is a mixed quantum state. For general, non-integrable systems it is expected to be a thermal state, which is described by an appropriate Gibbs ensemble. For integrable systems, for which examples are the $XXZ$ or the $XY$ chains, the postquench state is described by a so-called Generalized Gibbs Ensemble.

We observed, that the postquench state has an entangled two-qubit reduced state for the nearest neighbors, if the parameters involved during the quench (in our case the transverse fields $h$ and $h_0$) are sufficiently close to each others. If the parameters change significantly, then the nearest-neighbor two-qubit reduced state for the nonequilibrium stationary state becomes non-entangled. We expect that the latter statement is generally valid in various systems for the entanglement of nonequilibrium postquench states, at least for the entanglement of few-particle blocks. {Additionally, in the XY models different from the Ising model, we found entanglement if the quench is performed from the paramagnetic phase  $(h_0>1)$ to the ferromagnetic phase $(h<1)$  and vice versa. These cases are relevant especially for small $\gamma.$ } It would be interesting to check the entanglement properties of postquench states of other (Bethe-Ansatz) integrable models.

While we studied the nearest-neighbor entanglement of the postquench state, other properties uncovering hidden criticality of the initial system not detectable by local quantities have recently been considered \cite{Paul2022Hidden}. The method has been based on efficient lower bounds on the negativity in $XY$ chains \cite{Eisler_2015,PhysRevB.97.165123}. We have shown that the criticality of the initial state can still be seen in the boundaries of the regions with nearest-neighbor entanglement.

\section{Conclusions}
\label{sec:conclusions}
We used energy-based entanglement witnesses to detect entanglement in the thermal states of the infinite and finite $XY$ chain, as well as in the mixed states arising after quench. We compared their performance to the negativity-based entanglement witness. We find that they efficiently detect entanglement in the systems we considered.

\begin{acknowledgments}
We thank I.~Apellaniz, I.~L.~Egusquiza, C.~Klempt, J.~Ko\l ody\'nski, I.~A.~Kov\'acs, G.~Muga, J.~Siewert, Sz.~Szalay, R.~Tr\'enyi, G. Vitagliano, and Z.~Zimbor\'as for discussions. This work was supported by the Hungarian Scientific Research Fund under Grants No. K128989 and No. KKP-126749, No. 2019-2.1.7-ERA-NET-2020-00003, and by the National Research, Development and Innovation Office of Hungary (NKFIH) within the Quantum Information National Laboratory of Hungary.   We acknowledge the support of the  EU (COST Action CA15220, QuantERA CEBBEC, QuantERA MENTA, QuantERA QuSiED),  the Spanish MCIU (Grant No. PCI2018-092896, No. PCI2022-132947), the Spanish Ministry of Science, Innovation and Universities and the European Regional Development Fund FEDER through Grant No. PGC2018-101355-B-I00 (MCIU/AEI/FEDER, EU) and through Grant No. PID2021-126273NB-I00 funded by MCIN/AEI/10.13039/501100011033 and by "ERDF A way of making Europe", the Basque Government (Grant No. IT986-16, No. IT1470-22).  We thank the "Frontline" Research Excellence Programme of the NKFIH (Grant No. KKP133827). G.~T. acknowledges a  Bessel Research Award of the Humboldt Foundation. We acknowledge Project no. TKP2021-NVA-04, which has been implemented with the support  provided by the Ministry of Innovation and Technology of Hungary from  the National Research, Development and Innovation Office (NKFIH), financed under  the TKP2021-NVA funding scheme. 
\end{acknowledgments}


\appendix

\section{Thermal average of the energy for finite periodic chains}
\label{sec:finite_fermionic}

For periodic chains in the fermionic representation there are two sectors, depending on the parity of the number of fermions \cite{Burkhardt_1985,PhysRevB.35.7062}.

(i) For \textit{even number of fermions}, let us denote the Hamiltonian by ${\cal H}^{(+)}$, and the possible values of the momenta are as follows:
\be
p=\pm\frac{\pi}{L}(2m -1),\quad m=1,2,\dots,L/2.
\ee

(ii) For \textit{odd number of fermions}, let us denote the Hamiltonian by ${\cal H}^{(-)}$, and the possible values of the momenta are as follows:
\be
q=0,\pi,\pm\frac{2\pi}{L}n,\quad n=1,2,\dots,L/2-1.
\ee
In this sector the energy of the $q=0$ mode is
\be
\varepsilon(q=0)=2(h-1).
\ee
The partition function of the system is given as
\be
Z=Z^{(+)}+Z^{(-)},
\ee
where $Z^{(\pm)}$ are the partition functions calculated with even and odd number of fermions, respectively.

The even partition function for $L=2\ell$ sites is given as
\be
Z_{2\ell}^{(+)}=2^{2\ell-1}\left[\prod_{p>0} \cosh^2\left(\frac{\beta\varepsilon(p)}{2}\right)+\prod_{p>0} \sinh^2\left(\frac{\beta\varepsilon(p)}{2}\right)\right],
\ee
and the analogous odd partition function is expressed as
\begin{align} 
Z_{2\ell}^{(-)}&=2^{2\ell-2}\left[C^{(+)}\prod_{0<q<\pi} \cosh^2\left(\frac{\beta\varepsilon(q)}{2}\right)\right. \nonumber\\
&\left.+C^{(-)}\prod_{0<q<\pi} \sinh^2\left(\frac{\beta\varepsilon(q)}{2}\right)\right],
\end{align}
with the definitions
\be
C^{(\pm)}=\cosh(\beta)\pm \cosh(\beta h).
\ee
Here, as usual $\beta=1/T.$
The thermal average of the energy is given by
\be
\langle {\cal H} \rangle_T=\langle {\cal H}^{(+)} \rangle_T \frac{1}{1+Z^{(-)}/Z^{(+)}}+\langle {\cal H}^{(-)} \rangle_T \frac{Z^{(-)}/Z^{(+)}}{1+Z^{(-)}/Z^{(+)}},
\ee
with
\begin{align}
\langle {\cal H}^{(+)} \rangle_T=\frac{\partial \ln Z^{(+)}}{\partial \beta}&=\sum_{p>0}\varepsilon(p)\left[\tanh\left(\frac{\beta \varepsilon(p)}{2}\right)\frac{1}{1+T^{(p)}}\right.\nonumber \\
&\left. + \coth\left(\frac{\beta \varepsilon(p)}{2}\right)\frac{T^{(p)})}{1+T^{(p)}}\right].
\end{align}
Similarly, we have
\begin{align}
\langle {\cal H}^{(-)} \rangle_T&=\frac{\partial \ln Z^{(-)}}{\partial \beta}=\frac{S^{(+)}+S^{(-)}T^{(q)}}{C^{(+)}+C^{(-)}T^{(q)}}\nonumber \\
&+\sum_q\varepsilon(q)\left[\tanh\left(\frac{\beta \varepsilon(q)}{2}\right)\frac{1}{1+T^{(q)}C^{(-)}/C^{(+)}}\right.\nonumber \\
&\left. + \coth\left(\frac{\beta \varepsilon(q)}{2}\right)\frac{T^{(q)}C^{(-)}/C^{(+)}}{1+T^{(q)}C^{(-)}/C^{(+)}}\right].
\end{align}
Here, we use the definitions
\begin{eqnarray}
T^{(p)}&=&\prod_{p>0} \tanh^2\left(\frac{\beta \varepsilon(p)}{2}\right),\nonumber\\ T^{(q)}&=&\prod_{0<q<\pi} \tanh^2\left(\frac{\beta \varepsilon(q)}{2}\right),
\end{eqnarray}
and
\be
S^{(\pm)}=\sinh(\beta)\pm h\sinh(\beta h).
\ee
%


\section{Proving the fact that $\rho_{23}\ge 0$}

\label{ap:rho230}

In this section, we show that 
\be
\rho_{23}=g_c/2\ge0.
\ee
This can be proved starting with the definition in Eq.~(\ref{par_corr}) in the thermodynamic limit
\begin{align}
g_c=\frac{2}{\pi} \int_0^{\pi}{\rm d}p \cos p (\cos p -h)t(p,T,h)\varepsilon^{-1}(p,h),
\end{align}
where the integral is separated in two halves, $\int_0^{\pi/2} + \int_{\pi/2}^{\pi}$, leading to
\begin{align}
g_c=\tilde{g}_c(h)+\tilde{g}_c(-h)
\end{align}
with
\begin{align}
\tilde{g}_c(\pm h) =\frac{2}{\pi} \int_0^{\pi/2}{\rm d}p \cos p (\cos p \pm h)t(p,T,\pm h)\varepsilon^{-1}(p,\pm h).
\end{align}
Bringing the two expressions below one integral, the integrand is of the form
\begin{align}
I(p,T,h)&=C(p,h)\left[(\cos p -h) \varepsilon(p,-h)t(p,T, h)\right.\nonumber\\
&\left.+(\cos p +h) \varepsilon(p,h)t(p,T, -h)\right],
\end{align}
where $C(p,h)=\pi^{-1}\cos p \varepsilon^{-1}(p, h)\varepsilon^{-1}(p,-h) \ge 0$ and thus $I(p,T,h) \ge 0$.

\bibliography{XYent8_arxiv4}

\begin{thebibliography}{95}%
\makeatletter
\providecommand \@ifxundefined [1]{%
 \@ifx{#1\undefined}
}%
\providecommand \@ifnum [1]{%
 \ifnum #1\expandafter \@firstoftwo
 \else \expandafter \@secondoftwo
 \fi
}%
\providecommand \@ifx [1]{%
 \ifx #1\expandafter \@firstoftwo
 \else \expandafter \@secondoftwo
 \fi
}%
\providecommand \natexlab [1]{#1}%
\providecommand \enquote  [1]{``#1''}%
\providecommand \bibnamefont  [1]{#1}%
\providecommand \bibfnamefont [1]{#1}%
\providecommand \citenamefont [1]{#1}%
\providecommand \href@noop [0]{\@secondoftwo}%
\providecommand \href [0]{\begingroup \@sanitize@url \@href}%
\providecommand \@href[1]{\@@startlink{#1}\@@href}%
\providecommand \@@href[1]{\endgroup#1\@@endlink}%
\providecommand \@sanitize@url [0]{\catcode `\\12\catcode `\$12\catcode
  `\&12\catcode `\#12\catcode `\^12\catcode `\_12\catcode `\%12\relax}%
\providecommand \@@startlink[1]{}%
\providecommand \@@endlink[0]{}%
\providecommand \url  [0]{\begingroup\@sanitize@url \@url }%
\providecommand \@url [1]{\endgroup\@href {#1}{\urlprefix }}%
\providecommand \urlprefix  [0]{URL }%
\providecommand \Eprint [0]{\href }%
\providecommand \doibase [0]{https://doi.org/}%
\providecommand \selectlanguage [0]{\@gobble}%
\providecommand \bibinfo  [0]{\@secondoftwo}%
\providecommand \bibfield  [0]{\@secondoftwo}%
\providecommand \translation [1]{[#1]}%
\providecommand \BibitemOpen [0]{}%
\providecommand \bibitemStop [0]{}%
\providecommand \bibitemNoStop [0]{.\EOS\space}%
\providecommand \EOS [0]{\spacefactor3000\relax}%
\providecommand \BibitemShut  [1]{\csname bibitem#1\endcsname}%
\let\auto@bib@innerbib\@empty
\bibitem [{\citenamefont {Horodecki}\ \emph {et~al.}(2009)\citenamefont
  {Horodecki}, \citenamefont {Horodecki}, \citenamefont {Horodecki},\ and\
  \citenamefont {Horodecki}}]{Horodecki2009Quantum}%
  \BibitemOpen
  \bibfield  {author} {\bibinfo {author} {\bibfnamefont {R.}~\bibnamefont
  {Horodecki}}, \bibinfo {author} {\bibfnamefont {P.}~\bibnamefont
  {Horodecki}}, \bibinfo {author} {\bibfnamefont {M.}~\bibnamefont
  {Horodecki}},\ and\ \bibinfo {author} {\bibfnamefont {K.}~\bibnamefont
  {Horodecki}},\ }\bibfield  {title} {\bibinfo {title} {Quantum entanglement},\
  }\href {https://doi.org/10.1103/RevModPhys.81.865} {\bibfield  {journal}
  {\bibinfo  {journal} {Rev. Mod. Phys.}\ }\textbf {\bibinfo {volume} {81}},\
  \bibinfo {pages} {865} (\bibinfo {year} {2009})}\BibitemShut {NoStop}%
\bibitem [{\citenamefont {G{\"u}hne}\ and\ \citenamefont
  {T{\'o}th}(2009)}]{Guhne2009Entanglement}%
  \BibitemOpen
  \bibfield  {author} {\bibinfo {author} {\bibfnamefont {O.}~\bibnamefont
  {G{\"u}hne}}\ and\ \bibinfo {author} {\bibfnamefont {G.}~\bibnamefont
  {T{\'o}th}},\ }\bibfield  {title} {\bibinfo {title} {Entanglement
  detection},\ }\href
  {https://doi.org/https://doi.org/10.1016/j.physrep.2009.02.004} {\bibfield
  {journal} {\bibinfo  {journal} {Phys. Rep.}\ }\textbf {\bibinfo {volume}
  {474}},\ \bibinfo {pages} {1} (\bibinfo {year} {2009})}\BibitemShut {NoStop}%
\bibitem [{\citenamefont {Friis}\ \emph {et~al.}(2019)\citenamefont {Friis},
  \citenamefont {Vitagliano}, \citenamefont {Malik},\ and\ \citenamefont
  {Huber}}]{Friis2019}%
  \BibitemOpen
  \bibfield  {author} {\bibinfo {author} {\bibfnamefont {N.}~\bibnamefont
  {Friis}}, \bibinfo {author} {\bibfnamefont {G.}~\bibnamefont {Vitagliano}},
  \bibinfo {author} {\bibfnamefont {M.}~\bibnamefont {Malik}},\ and\ \bibinfo
  {author} {\bibfnamefont {M.}~\bibnamefont {Huber}},\ }\bibfield  {title}
  {\bibinfo {title} {Entanglement certification from theory to experiment},\
  }\href {https://doi.org/10.1038/s42254-018-0003-5} {\bibfield  {journal}
  {\bibinfo  {journal} {Nat. Rev. Phys.}\ }\textbf {\bibinfo {volume} {1}},\
  \bibinfo {pages} {72} (\bibinfo {year} {2019})}\BibitemShut {NoStop}%
\bibitem [{\citenamefont {Peres}(1996)}]{Peres1996Separability}%
  \BibitemOpen
  \bibfield  {author} {\bibinfo {author} {\bibfnamefont {A.}~\bibnamefont
  {Peres}},\ }\bibfield  {title} {\bibinfo {title} {Separability criterion for
  density matrices},\ }\href {https://doi.org/10.1103/PhysRevLett.77.1413}
  {\bibfield  {journal} {\bibinfo  {journal} {Phys. Rev. Lett.}\ }\textbf
  {\bibinfo {volume} {77}},\ \bibinfo {pages} {1413} (\bibinfo {year}
  {1996})}\BibitemShut {NoStop}%
\bibitem [{\citenamefont {Horodecki}(1997)}]{Horodecki1997Separability}%
  \BibitemOpen
  \bibfield  {author} {\bibinfo {author} {\bibfnamefont {P.}~\bibnamefont
  {Horodecki}},\ }\bibfield  {title} {\bibinfo {title} {Separability criterion
  and inseparable mixed states with positive partial transposition},\ }\href
  {https://doi.org/10.1016/S0375-9601(97)00416-7} {\bibfield  {journal}
  {\bibinfo  {journal} {Phys. Lett. A}\ }\textbf {\bibinfo {volume} {232}},\
  \bibinfo {pages} {333 } (\bibinfo {year} {1997})}\BibitemShut {NoStop}%
\bibitem [{\citenamefont {Giedke}\ \emph {et~al.}(2001)\citenamefont {Giedke},
  \citenamefont {Kraus}, \citenamefont {Lewenstein},\ and\ \citenamefont
  {Cirac}}]{Giedke2001Entanglement}%
  \BibitemOpen
  \bibfield  {author} {\bibinfo {author} {\bibfnamefont {G.}~\bibnamefont
  {Giedke}}, \bibinfo {author} {\bibfnamefont {B.}~\bibnamefont {Kraus}},
  \bibinfo {author} {\bibfnamefont {M.}~\bibnamefont {Lewenstein}},\ and\
  \bibinfo {author} {\bibfnamefont {J.~I.}\ \bibnamefont {Cirac}},\ }\bibfield
  {title} {\bibinfo {title} {Entanglement criteria for all bipartite {Gaussian}
  states},\ }\href {https://doi.org/10.1103/PhysRevLett.87.167904} {\bibfield
  {journal} {\bibinfo  {journal} {Phys. Rev. Lett.}\ }\textbf {\bibinfo
  {volume} {87}},\ \bibinfo {pages} {167904} (\bibinfo {year}
  {2001})}\BibitemShut {NoStop}%
\bibitem [{\citenamefont {Horodecki}\ \emph {et~al.}(1996)\citenamefont
  {Horodecki}, \citenamefont {Horodecki},\ and\ \citenamefont
  {Horodecki}}]{Horodecki1996Separability}%
  \BibitemOpen
  \bibfield  {author} {\bibinfo {author} {\bibfnamefont {M.}~\bibnamefont
  {Horodecki}}, \bibinfo {author} {\bibfnamefont {P.}~\bibnamefont
  {Horodecki}},\ and\ \bibinfo {author} {\bibfnamefont {R.}~\bibnamefont
  {Horodecki}},\ }\bibfield  {title} {\bibinfo {title} {Separability of mixed
  states: necessary and sufficient conditions},\ }\href
  {https://doi.org/https://doi.org/10.1016/S0375-9601(96)00706-2} {\bibfield
  {journal} {\bibinfo  {journal} {Phys. Lett. A}\ }\textbf {\bibinfo {volume}
  {223}},\ \bibinfo {pages} {1 } (\bibinfo {year} {1996})}\BibitemShut
  {NoStop}%
\bibitem [{\citenamefont {Terhal}(2000)}]{Terhal2000Bell}%
  \BibitemOpen
  \bibfield  {author} {\bibinfo {author} {\bibfnamefont {B.~M.}\ \bibnamefont
  {Terhal}},\ }\bibfield  {title} {\bibinfo {title} {Bell inequalities and the
  separability criterion},\ }\href
  {https://doi.org/https://doi.org/10.1016/S0375-9601(00)00401-1} {\bibfield
  {journal} {\bibinfo  {journal} {Phys. Lett. A}\ }\textbf {\bibinfo {volume}
  {271}},\ \bibinfo {pages} {319} (\bibinfo {year} {2000})}\BibitemShut
  {NoStop}%
\bibitem [{\citenamefont {Lewenstein}\ \emph {et~al.}(2000)\citenamefont
  {Lewenstein}, \citenamefont {Kraus}, \citenamefont {Cirac},\ and\
  \citenamefont {Horodecki}}]{Lewenstein2000Optimization}%
  \BibitemOpen
  \bibfield  {author} {\bibinfo {author} {\bibfnamefont {M.}~\bibnamefont
  {Lewenstein}}, \bibinfo {author} {\bibfnamefont {B.}~\bibnamefont {Kraus}},
  \bibinfo {author} {\bibfnamefont {J.~I.}\ \bibnamefont {Cirac}},\ and\
  \bibinfo {author} {\bibfnamefont {P.}~\bibnamefont {Horodecki}},\ }\bibfield
  {title} {\bibinfo {title} {Optimization of entanglement witnesses},\ }\href
  {https://doi.org/10.1103/PhysRevA.62.052310} {\bibfield  {journal} {\bibinfo
  {journal} {Phys. Rev. A}\ }\textbf {\bibinfo {volume} {62}},\ \bibinfo
  {pages} {052310} (\bibinfo {year} {2000})}\BibitemShut {NoStop}%
\bibitem [{\citenamefont {Ac{\'{\i}}n}\ \emph {et~al.}(2001)\citenamefont
  {Ac{\'{\i}}n}, \citenamefont {Bru\ss{}}, \citenamefont {Lewenstein},\ and\
  \citenamefont {Sanpera}}]{Acin2001Classification}%
  \BibitemOpen
  \bibfield  {author} {\bibinfo {author} {\bibfnamefont {A.}~\bibnamefont
  {Ac{\'{\i}}n}}, \bibinfo {author} {\bibfnamefont {D.}~\bibnamefont
  {Bru\ss{}}}, \bibinfo {author} {\bibfnamefont {M.}~\bibnamefont
  {Lewenstein}},\ and\ \bibinfo {author} {\bibfnamefont {A.}~\bibnamefont
  {Sanpera}},\ }\bibfield  {title} {\bibinfo {title} {Classification of mixed
  three-qubit states},\ }\href {https://doi.org/10.1103/PhysRevLett.87.040401}
  {\bibfield  {journal} {\bibinfo  {journal} {Phys. Rev. Lett.}\ }\textbf
  {\bibinfo {volume} {87}},\ \bibinfo {pages} {040401} (\bibinfo {year}
  {2001})}\BibitemShut {NoStop}%
\bibitem [{\citenamefont {T\'oth}(2005)}]{Toth2005EntanglementWitnesses}%
  \BibitemOpen
  \bibfield  {author} {\bibinfo {author} {\bibfnamefont {G.}~\bibnamefont
  {T\'oth}},\ }\bibfield  {title} {\bibinfo {title} {Entanglement witnesses in
  spin models},\ }\href {https://doi.org/10.1103/PhysRevA.71.010301} {\bibfield
   {journal} {\bibinfo  {journal} {Phys. Rev. A}\ }\textbf {\bibinfo {volume}
  {71}},\ \bibinfo {pages} {010301(R)} (\bibinfo {year} {2005})}\BibitemShut
  {NoStop}%
\bibitem [{\citenamefont {T{\'o}th}\ and\ \citenamefont
  {G{\"u}hne}(2006)}]{Toth2006Detection}%
  \BibitemOpen
  \bibfield  {author} {\bibinfo {author} {\bibfnamefont {G.}~\bibnamefont
  {T{\'o}th}}\ and\ \bibinfo {author} {\bibfnamefont {O.}~\bibnamefont
  {G{\"u}hne}},\ }\bibfield  {title} {\bibinfo {title} {Detection of
  multipartite entanglement with two-body correlations},\ }\href
  {https://doi.org/10.1007/s00340-005-2057-1} {\bibfield  {journal} {\bibinfo
  {journal} {Appl. Phys. B}\ }\textbf {\bibinfo {volume} {82}},\ \bibinfo
  {pages} {237} (\bibinfo {year} {2006})}\BibitemShut {NoStop}%
\bibitem [{\citenamefont {G\"uhne}\ and\ \citenamefont
  {T\'oth}(2006)}]{Guhne2006Energy}%
  \BibitemOpen
  \bibfield  {author} {\bibinfo {author} {\bibfnamefont {O.}~\bibnamefont
  {G\"uhne}}\ and\ \bibinfo {author} {\bibfnamefont {G.}~\bibnamefont
  {T\'oth}},\ }\bibfield  {title} {\bibinfo {title} {Energy and multipartite
  entanglement in multidimensional and frustrated spin models},\ }\href
  {https://doi.org/10.1103/PhysRevA.73.052319} {\bibfield  {journal} {\bibinfo
  {journal} {Phys. Rev. A}\ }\textbf {\bibinfo {volume} {73}},\ \bibinfo
  {pages} {052319} (\bibinfo {year} {2006})}\BibitemShut {NoStop}%
\bibitem [{\citenamefont {G{\"u}hne}\ \emph {et~al.}(2005)\citenamefont
  {G{\"u}hne}, \citenamefont {T\'oth},\ and\ \citenamefont
  {Briegel}}]{Guhne2005Multipartite}%
  \BibitemOpen
  \bibfield  {author} {\bibinfo {author} {\bibfnamefont {O.}~\bibnamefont
  {G{\"u}hne}}, \bibinfo {author} {\bibfnamefont {G.}~\bibnamefont {T\'oth}},\
  and\ \bibinfo {author} {\bibfnamefont {H.~J.}\ \bibnamefont {Briegel}},\
  }\bibfield  {title} {\bibinfo {title} {Multipartite entanglement in spin
  chains},\ }\href {http://stacks.iop.org/1367-2630/7/i=1/a=229} {\bibfield
  {journal} {\bibinfo  {journal} {New J. Phys.}\ }\textbf {\bibinfo {volume}
  {7}},\ \bibinfo {pages} {229} (\bibinfo {year} {2005})}\BibitemShut {NoStop}%
\bibitem [{\citenamefont {{Brukner}}\ and\ \citenamefont
  {{Vedral}}(2004)}]{Brukner2004MacroscopicB}%
  \BibitemOpen
  \bibfield  {author} {\bibinfo {author} {\bibfnamefont {C.}~\bibnamefont
  {{Brukner}}}\ and\ \bibinfo {author} {\bibfnamefont {V.}~\bibnamefont
  {{Vedral}}},\ }\bibfield  {title} {\bibinfo {title} {{Macroscopic
  Thermodynamical Witnesses of Quantum Entanglement}},\ }\href
  {https://arxiv.org/abs/quant-ph/0406040} {\bibfield  {journal} {\bibinfo
  {journal} {arXiv: quant-ph/0406040}\ } (\bibinfo {year} {2004})}\BibitemShut
  {NoStop}%
\bibitem [{\citenamefont {Dowling}\ \emph {et~al.}(2004)\citenamefont
  {Dowling}, \citenamefont {Doherty},\ and\ \citenamefont
  {Bartlett}}]{Dowling2004Energy}%
  \BibitemOpen
  \bibfield  {author} {\bibinfo {author} {\bibfnamefont {M.~R.}\ \bibnamefont
  {Dowling}}, \bibinfo {author} {\bibfnamefont {A.~C.}\ \bibnamefont
  {Doherty}},\ and\ \bibinfo {author} {\bibfnamefont {S.~D.}\ \bibnamefont
  {Bartlett}},\ }\bibfield  {title} {\bibinfo {title} {Energy as an
  entanglement witness for quantum many-body systems},\ }\href
  {https://doi.org/10.1103/PhysRevA.70.062113} {\bibfield  {journal} {\bibinfo
  {journal} {Phys. Rev. A}\ }\textbf {\bibinfo {volume} {70}},\ \bibinfo
  {pages} {062113} (\bibinfo {year} {2004})}\BibitemShut {NoStop}%
\bibitem [{\citenamefont {Wu}\ \emph {et~al.}(2005)\citenamefont {Wu},
  \citenamefont {Bandyopadhyay}, \citenamefont {Sarandy},\ and\ \citenamefont
  {Lidar}}]{Wu2005Entanglement}%
  \BibitemOpen
  \bibfield  {author} {\bibinfo {author} {\bibfnamefont {L.-A.}\ \bibnamefont
  {Wu}}, \bibinfo {author} {\bibfnamefont {S.}~\bibnamefont {Bandyopadhyay}},
  \bibinfo {author} {\bibfnamefont {M.~S.}\ \bibnamefont {Sarandy}},\ and\
  \bibinfo {author} {\bibfnamefont {D.~A.}\ \bibnamefont {Lidar}},\ }\bibfield
  {title} {\bibinfo {title} {Entanglement observables and witnesses for
  interacting quantum spin systems},\ }\href
  {https://doi.org/10.1103/PhysRevA.72.032309} {\bibfield  {journal} {\bibinfo
  {journal} {Phys. Rev. A}\ }\textbf {\bibinfo {volume} {72}},\ \bibinfo
  {pages} {032309} (\bibinfo {year} {2005})}\BibitemShut {NoStop}%
\bibitem [{\citenamefont {Wang}(2002)}]{Wang2002Threshold}%
  \BibitemOpen
  \bibfield  {author} {\bibinfo {author} {\bibfnamefont {X.}~\bibnamefont
  {Wang}},\ }\bibfield  {title} {\bibinfo {title} {Threshold temperature for
  pairwise and many-particle thermal entanglement in the isotropic heisenberg
  model},\ }\href {https://doi.org/10.1103/PhysRevA.66.044305} {\bibfield
  {journal} {\bibinfo  {journal} {Phys. Rev. A}\ }\textbf {\bibinfo {volume}
  {66}},\ \bibinfo {pages} {044305} (\bibinfo {year} {2002})}\BibitemShut
  {NoStop}%
\bibitem [{\citenamefont {V\'ertesi}\ and\ \citenamefont
  {Bene}(2006)}]{Vertesi2006Thermal}%
  \BibitemOpen
  \bibfield  {author} {\bibinfo {author} {\bibfnamefont {T.}~\bibnamefont
  {V\'ertesi}}\ and\ \bibinfo {author} {\bibfnamefont {E.}~\bibnamefont
  {Bene}},\ }\bibfield  {title} {\bibinfo {title} {Thermal entanglement in the
  nanotubular system {${\mathrm{Na}}_{2}{\mathrm{V}}_{3}{\mathrm{O}}_{7}$}},\
  }\href {https://doi.org/10.1103/PhysRevB.73.134404} {\bibfield  {journal}
  {\bibinfo  {journal} {Phys. Rev. B}\ }\textbf {\bibinfo {volume} {73}},\
  \bibinfo {pages} {134404} (\bibinfo {year} {2006})}\BibitemShut {NoStop}%
\bibitem [{\citenamefont {Siloi}\ and\ \citenamefont
  {Troiani}(2012)}]{Sioli2012Towards}%
  \BibitemOpen
  \bibfield  {author} {\bibinfo {author} {\bibfnamefont {I.}~\bibnamefont
  {Siloi}}\ and\ \bibinfo {author} {\bibfnamefont {F.}~\bibnamefont
  {Troiani}},\ }\bibfield  {title} {\bibinfo {title} {Towards the chemical
  tuning of entanglement in molecular nanomagnets},\ }\href
  {https://doi.org/10.1103/PhysRevB.86.224404} {\bibfield  {journal} {\bibinfo
  {journal} {Phys. Rev. B}\ }\textbf {\bibinfo {volume} {86}},\ \bibinfo
  {pages} {224404} (\bibinfo {year} {2012})}\BibitemShut {NoStop}%
\bibitem [{\citenamefont {Siloi}\ and\ \citenamefont
  {Troiani}(2013)}]{Siloi2013Quantum}%
  \BibitemOpen
  \bibfield  {author} {\bibinfo {author} {\bibfnamefont {I.}~\bibnamefont
  {Siloi}}\ and\ \bibinfo {author} {\bibfnamefont {F.}~\bibnamefont
  {Troiani}},\ }\bibfield  {title} {\bibinfo {title} {Quantum entanglement in
  heterometallic wheels},\ }\href {https://doi.org/10.1140/epjb/e2012-30681-1}
  {\bibfield  {journal} {\bibinfo  {journal} {Eur. Phys. J. B}\ }\textbf
  {\bibinfo {volume} {86}},\ \bibinfo {pages} {71} (\bibinfo {year}
  {2013})}\BibitemShut {NoStop}%
\bibitem [{\citenamefont {Troiani}\ and\ \citenamefont
  {Siloi}(2012)}]{Troiani2012Energy}%
  \BibitemOpen
  \bibfield  {author} {\bibinfo {author} {\bibfnamefont {F.}~\bibnamefont
  {Troiani}}\ and\ \bibinfo {author} {\bibfnamefont {I.}~\bibnamefont
  {Siloi}},\ }\bibfield  {title} {\bibinfo {title} {Energy as a witness of
  multipartite entanglement in chains of arbitrary spins},\ }\href
  {https://doi.org/10.1103/PhysRevA.86.032330} {\bibfield  {journal} {\bibinfo
  {journal} {Phys. Rev. A}\ }\textbf {\bibinfo {volume} {86}},\ \bibinfo
  {pages} {032330} (\bibinfo {year} {2012})}\BibitemShut {NoStop}%
\bibitem [{\citenamefont {Homayoun}\ and\ \citenamefont
  {Aghayar}(2019)}]{Homayoun2019Energy}%
  \BibitemOpen
  \bibfield  {author} {\bibinfo {author} {\bibfnamefont {T.}~\bibnamefont
  {Homayoun}}\ and\ \bibinfo {author} {\bibfnamefont {K.}~\bibnamefont
  {Aghayar}},\ }\bibfield  {title} {\bibinfo {title} {Energy as an entanglement
  witnesses for one dimensional {XYZ} {Heisenberg} lattice: Optimization
  approach},\ }\href {https://doi.org/10.1007/s10955-019-02289-1} {\bibfield
  {journal} {\bibinfo  {journal} {J. Stat. Phys.}\ }\textbf {\bibinfo {volume}
  {176}},\ \bibinfo {pages} {85} (\bibinfo {year} {2019})}\BibitemShut
  {NoStop}%
\bibitem [{\citenamefont {Troiani}\ \emph {et~al.}(2013)\citenamefont
  {Troiani}, \citenamefont {Carretta},\ and\ \citenamefont
  {Santini}}]{Troiani2013Detection}%
  \BibitemOpen
  \bibfield  {author} {\bibinfo {author} {\bibfnamefont {F.}~\bibnamefont
  {Troiani}}, \bibinfo {author} {\bibfnamefont {S.}~\bibnamefont {Carretta}},\
  and\ \bibinfo {author} {\bibfnamefont {P.}~\bibnamefont {Santini}},\
  }\bibfield  {title} {\bibinfo {title} {Detection of entanglement between
  collective spins},\ }\href {https://doi.org/10.1103/PhysRevB.88.195421}
  {\bibfield  {journal} {\bibinfo  {journal} {Phys. Rev. B}\ }\textbf {\bibinfo
  {volume} {88}},\ \bibinfo {pages} {195421} (\bibinfo {year}
  {2013})}\BibitemShut {NoStop}%
\bibitem [{\citenamefont {Osterloh}\ \emph {et~al.}(2002)\citenamefont
  {Osterloh}, \citenamefont {Amico}, \citenamefont {Falci},\ and\ \citenamefont
  {Fazio}}]{Osterloh2002}%
  \BibitemOpen
  \bibfield  {author} {\bibinfo {author} {\bibfnamefont {A.}~\bibnamefont
  {Osterloh}}, \bibinfo {author} {\bibfnamefont {L.}~\bibnamefont {Amico}},
  \bibinfo {author} {\bibfnamefont {G.}~\bibnamefont {Falci}},\ and\ \bibinfo
  {author} {\bibfnamefont {R.}~\bibnamefont {Fazio}},\ }\bibfield  {title}
  {\bibinfo {title} {Scaling of entanglement close to a quantum phase
  transition},\ }\href {https://doi.org/10.1038/416608a} {\bibfield  {journal}
  {\bibinfo  {journal} {Nature (London)}\ }\textbf {\bibinfo {volume} {416}},\
  \bibinfo {pages} {608} (\bibinfo {year} {2002})}\BibitemShut {NoStop}%
\bibitem [{\citenamefont {Osborne}\ and\ \citenamefont
  {Nielsen}(2002)}]{PhysRevA.66.032110}%
  \BibitemOpen
  \bibfield  {author} {\bibinfo {author} {\bibfnamefont {T.~J.}\ \bibnamefont
  {Osborne}}\ and\ \bibinfo {author} {\bibfnamefont {M.~A.}\ \bibnamefont
  {Nielsen}},\ }\bibfield  {title} {\bibinfo {title} {Entanglement in a simple
  quantum phase transition},\ }\href
  {https://doi.org/10.1103/PhysRevA.66.032110} {\bibfield  {journal} {\bibinfo
  {journal} {Phys. Rev. A}\ }\textbf {\bibinfo {volume} {66}},\ \bibinfo
  {pages} {032110} (\bibinfo {year} {2002})}\BibitemShut {NoStop}%
\bibitem [{\citenamefont {Patan{\`{e}}}\ \emph {et~al.}(2007)\citenamefont
  {Patan{\`{e}}}, \citenamefont {Fazio},\ and\ \citenamefont
  {Amico}}]{Patan__2007}%
  \BibitemOpen
  \bibfield  {author} {\bibinfo {author} {\bibfnamefont {D.}~\bibnamefont
  {Patan{\`{e}}}}, \bibinfo {author} {\bibfnamefont {R.}~\bibnamefont
  {Fazio}},\ and\ \bibinfo {author} {\bibfnamefont {L.}~\bibnamefont {Amico}},\
  }\bibfield  {title} {\bibinfo {title} {Bound entanglement in the {XY}
  model},\ }\href {https://doi.org/10.1088/1367-2630/9/9/322} {\bibfield
  {journal} {\bibinfo  {journal} {New J. Phys.}\ }\textbf {\bibinfo {volume}
  {9}},\ \bibinfo {pages} {322} (\bibinfo {year} {2007})}\BibitemShut {NoStop}%
\bibitem [{\citenamefont {Hofmann}\ \emph {et~al.}(2014)\citenamefont
  {Hofmann}, \citenamefont {Osterloh},\ and\ \citenamefont
  {G\"uhne}}]{PhysRevB.89.134101}%
  \BibitemOpen
  \bibfield  {author} {\bibinfo {author} {\bibfnamefont {M.}~\bibnamefont
  {Hofmann}}, \bibinfo {author} {\bibfnamefont {A.}~\bibnamefont {Osterloh}},\
  and\ \bibinfo {author} {\bibfnamefont {O.}~\bibnamefont {G\"uhne}},\
  }\bibfield  {title} {\bibinfo {title} {Scaling of genuine multiparticle
  entanglement close to a quantum phase transition},\ }\href
  {https://doi.org/10.1103/PhysRevB.89.134101} {\bibfield  {journal} {\bibinfo
  {journal} {Phys. Rev. B}\ }\textbf {\bibinfo {volume} {89}},\ \bibinfo
  {pages} {134101} (\bibinfo {year} {2014})}\BibitemShut {NoStop}%
\bibitem [{\citenamefont {Giampaolo}\ and\ \citenamefont
  {Hiesmayr}(2013)}]{PhysRevA.88.052305}%
  \BibitemOpen
  \bibfield  {author} {\bibinfo {author} {\bibfnamefont {S.~M.}\ \bibnamefont
  {Giampaolo}}\ and\ \bibinfo {author} {\bibfnamefont {B.~C.}\ \bibnamefont
  {Hiesmayr}},\ }\bibfield  {title} {\bibinfo {title} {Genuine multipartite
  entanglement in the $xy$ model},\ }\href
  {https://doi.org/10.1103/PhysRevA.88.052305} {\bibfield  {journal} {\bibinfo
  {journal} {Phys. Rev. A}\ }\textbf {\bibinfo {volume} {88}},\ \bibinfo
  {pages} {052305} (\bibinfo {year} {2013})}\BibitemShut {NoStop}%
\bibitem [{\citenamefont {Vidal}\ and\ \citenamefont
  {Werner}(2002)}]{Vidal2002Computable}%
  \BibitemOpen
  \bibfield  {author} {\bibinfo {author} {\bibfnamefont {G.}~\bibnamefont
  {Vidal}}\ and\ \bibinfo {author} {\bibfnamefont {R.~F.}\ \bibnamefont
  {Werner}},\ }\bibfield  {title} {\bibinfo {title} {Computable measure of
  entanglement},\ }\href {https://doi.org/10.1103/PhysRevA.65.032314}
  {\bibfield  {journal} {\bibinfo  {journal} {Phys. Rev. A}\ }\textbf {\bibinfo
  {volume} {65}},\ \bibinfo {pages} {032314} (\bibinfo {year}
  {2002})}\BibitemShut {NoStop}%
\bibitem [{\citenamefont {Barouch}\ \emph {et~al.}(1970)\citenamefont
  {Barouch}, \citenamefont {McCoy},\ and\ \citenamefont
  {Dresden}}]{PhysRevA.2.1075}%
  \BibitemOpen
  \bibfield  {author} {\bibinfo {author} {\bibfnamefont {E.}~\bibnamefont
  {Barouch}}, \bibinfo {author} {\bibfnamefont {B.~M.}\ \bibnamefont {McCoy}},\
  and\ \bibinfo {author} {\bibfnamefont {M.}~\bibnamefont {Dresden}},\
  }\bibfield  {title} {\bibinfo {title} {Statistical mechanics of the
  $\mathrm{XY}$ model. {I}},\ }\href {https://doi.org/10.1103/PhysRevA.2.1075}
  {\bibfield  {journal} {\bibinfo  {journal} {Phys. Rev. A}\ }\textbf {\bibinfo
  {volume} {2}},\ \bibinfo {pages} {1075} (\bibinfo {year} {1970})}\BibitemShut
  {NoStop}%
\bibitem [{\citenamefont {Barouch}\ and\ \citenamefont
  {McCoy}(1971)}]{PhysRevA.3.2137}%
  \BibitemOpen
  \bibfield  {author} {\bibinfo {author} {\bibfnamefont {E.}~\bibnamefont
  {Barouch}}\ and\ \bibinfo {author} {\bibfnamefont {B.~M.}\ \bibnamefont
  {McCoy}},\ }\bibfield  {title} {\bibinfo {title} {Statistical mechanics of
  the $\mathrm{XY}$ model. {III}},\ }\href
  {https://doi.org/10.1103/PhysRevA.3.2137} {\bibfield  {journal} {\bibinfo
  {journal} {Phys. Rev. A}\ }\textbf {\bibinfo {volume} {3}},\ \bibinfo {pages}
  {2137} (\bibinfo {year} {1971})}\BibitemShut {NoStop}%
\bibitem [{\citenamefont {Igl\'oi}\ and\ \citenamefont
  {Rieger}(2000)}]{PhysRevLett.85.3233}%
  \BibitemOpen
  \bibfield  {author} {\bibinfo {author} {\bibfnamefont {F.}~\bibnamefont
  {Igl\'oi}}\ and\ \bibinfo {author} {\bibfnamefont {H.}~\bibnamefont
  {Rieger}},\ }\bibfield  {title} {\bibinfo {title} {Long-range correlations in
  the nonequilibrium quantum relaxation of a spin chain},\ }\href
  {https://doi.org/10.1103/PhysRevLett.85.3233} {\bibfield  {journal} {\bibinfo
   {journal} {Phys. Rev. Lett.}\ }\textbf {\bibinfo {volume} {85}},\ \bibinfo
  {pages} {3233} (\bibinfo {year} {2000})}\BibitemShut {NoStop}%
\bibitem [{\citenamefont {Sengupta}\ \emph {et~al.}(2004)\citenamefont
  {Sengupta}, \citenamefont {Powell},\ and\ \citenamefont
  {Sachdev}}]{PhysRevA.69.053616}%
  \BibitemOpen
  \bibfield  {author} {\bibinfo {author} {\bibfnamefont {K.}~\bibnamefont
  {Sengupta}}, \bibinfo {author} {\bibfnamefont {S.}~\bibnamefont {Powell}},\
  and\ \bibinfo {author} {\bibfnamefont {S.}~\bibnamefont {Sachdev}},\
  }\bibfield  {title} {\bibinfo {title} {Quench dynamics across quantum
  critical points},\ }\href {https://doi.org/10.1103/PhysRevA.69.053616}
  {\bibfield  {journal} {\bibinfo  {journal} {Phys. Rev. A}\ }\textbf {\bibinfo
  {volume} {69}},\ \bibinfo {pages} {053616} (\bibinfo {year}
  {2004})}\BibitemShut {NoStop}%
\bibitem [{\citenamefont {Polkovnikov}\ \emph {et~al.}(2011)\citenamefont
  {Polkovnikov}, \citenamefont {Sengupta}, \citenamefont {Silva},\ and\
  \citenamefont {Vengalattore}}]{RevModPhys.83.863}%
  \BibitemOpen
  \bibfield  {author} {\bibinfo {author} {\bibfnamefont {A.}~\bibnamefont
  {Polkovnikov}}, \bibinfo {author} {\bibfnamefont {K.}~\bibnamefont
  {Sengupta}}, \bibinfo {author} {\bibfnamefont {A.}~\bibnamefont {Silva}},\
  and\ \bibinfo {author} {\bibfnamefont {M.}~\bibnamefont {Vengalattore}},\
  }\bibfield  {title} {\bibinfo {title} {Colloquium: Nonequilibrium dynamics of
  closed interacting quantum systems},\ }\href
  {https://doi.org/10.1103/RevModPhys.83.863} {\bibfield  {journal} {\bibinfo
  {journal} {Rev. Mod. Phys.}\ }\textbf {\bibinfo {volume} {83}},\ \bibinfo
  {pages} {863} (\bibinfo {year} {2011})}\BibitemShut {NoStop}%
\bibitem [{\citenamefont {Sotiriadis}\ \emph {et~al.}(2012)\citenamefont
  {Sotiriadis}, \citenamefont {Fioretto},\ and\ \citenamefont
  {Mussardo}}]{Sotiriadis_2012}%
  \BibitemOpen
  \bibfield  {author} {\bibinfo {author} {\bibfnamefont {S.}~\bibnamefont
  {Sotiriadis}}, \bibinfo {author} {\bibfnamefont {D.}~\bibnamefont
  {Fioretto}},\ and\ \bibinfo {author} {\bibfnamefont {G.}~\bibnamefont
  {Mussardo}},\ }\bibfield  {title} {\bibinfo {title}
  {Zamolodchikov{\textendash}{Faddeev} algebra and quantum quenches in
  integrable field theories},\ }\href
  {https://doi.org/10.1088/1742-5468/2012/02/p02017} {\bibfield  {journal}
  {\bibinfo  {journal} {J. Stat. Mech.: Theory Exp.}\ }\textbf {\bibinfo
  {volume} {2012}}\bibinfo  {number} { (02)},\ \bibinfo {pages}
  {P02017}}\BibitemShut {NoStop}%
\bibitem [{\citenamefont {Roux}(2009)}]{PhysRevA.79.021608}%
  \BibitemOpen
\bibfield  {number} {  }\bibfield  {author} {\bibinfo {author} {\bibfnamefont
  {G.}~\bibnamefont {Roux}},\ }\bibfield  {title} {\bibinfo {title} {Quenches
  in quantum many-body systems: One-dimensional bose-hubbard model
  reexamined},\ }\href {https://doi.org/10.1103/PhysRevA.79.021608} {\bibfield
  {journal} {\bibinfo  {journal} {Phys. Rev. A}\ }\textbf {\bibinfo {volume}
  {79}},\ \bibinfo {pages} {021608} (\bibinfo {year} {2009})}\BibitemShut
  {NoStop}%
\bibitem [{\citenamefont {Sotiriadis}\ \emph {et~al.}(2009)\citenamefont
  {Sotiriadis}, \citenamefont {Calabrese},\ and\ \citenamefont
  {Cardy}}]{Sotiriadis_2009}%
  \BibitemOpen
  \bibfield  {author} {\bibinfo {author} {\bibfnamefont {S.}~\bibnamefont
  {Sotiriadis}}, \bibinfo {author} {\bibfnamefont {P.}~\bibnamefont
  {Calabrese}},\ and\ \bibinfo {author} {\bibfnamefont {J.}~\bibnamefont
  {Cardy}},\ }\bibfield  {title} {\bibinfo {title} {Quantum quench from a
  thermal initial state},\ }\href {https://doi.org/10.1209/0295-5075/87/20002}
  {\bibfield  {journal} {\bibinfo  {journal} {{EPL} (Europhysics Letters)}\
  }\textbf {\bibinfo {volume} {87}},\ \bibinfo {pages} {20002} (\bibinfo {year}
  {2009})}\BibitemShut {NoStop}%
\bibitem [{\citenamefont {Kollar}\ and\ \citenamefont
  {Eckstein}(2008)}]{PhysRevA.78.013626}%
  \BibitemOpen
  \bibfield  {author} {\bibinfo {author} {\bibfnamefont {M.}~\bibnamefont
  {Kollar}}\ and\ \bibinfo {author} {\bibfnamefont {M.}~\bibnamefont
  {Eckstein}},\ }\bibfield  {title} {\bibinfo {title} {Relaxation of a
  one-dimensional mott insulator after an interaction quench},\ }\href
  {https://doi.org/10.1103/PhysRevA.78.013626} {\bibfield  {journal} {\bibinfo
  {journal} {Phys. Rev. A}\ }\textbf {\bibinfo {volume} {78}},\ \bibinfo
  {pages} {013626} (\bibinfo {year} {2008})}\BibitemShut {NoStop}%
\bibitem [{\citenamefont {Barthel}\ and\ \citenamefont
  {Schollw\"ock}(2008)}]{PhysRevLett.100.100601}%
  \BibitemOpen
  \bibfield  {author} {\bibinfo {author} {\bibfnamefont {T.}~\bibnamefont
  {Barthel}}\ and\ \bibinfo {author} {\bibfnamefont {U.}~\bibnamefont
  {Schollw\"ock}},\ }\bibfield  {title} {\bibinfo {title} {Dephasing and the
  steady state in quantum many-particle systems},\ }\href
  {https://doi.org/10.1103/PhysRevLett.100.100601} {\bibfield  {journal}
  {\bibinfo  {journal} {Phys. Rev. Lett.}\ }\textbf {\bibinfo {volume} {100}},\
  \bibinfo {pages} {100601} (\bibinfo {year} {2008})}\BibitemShut {NoStop}%
\bibitem [{\citenamefont {Cramer}\ \emph
  {et~al.}(2008{\natexlab{a}})\citenamefont {Cramer}, \citenamefont {Flesch},
  \citenamefont {McCulloch}, \citenamefont {Schollw\"ock},\ and\ \citenamefont
  {Eisert}}]{PhysRevLett.101.063001}%
  \BibitemOpen
  \bibfield  {author} {\bibinfo {author} {\bibfnamefont {M.}~\bibnamefont
  {Cramer}}, \bibinfo {author} {\bibfnamefont {A.}~\bibnamefont {Flesch}},
  \bibinfo {author} {\bibfnamefont {I.~P.}\ \bibnamefont {McCulloch}}, \bibinfo
  {author} {\bibfnamefont {U.}~\bibnamefont {Schollw\"ock}},\ and\ \bibinfo
  {author} {\bibfnamefont {J.}~\bibnamefont {Eisert}},\ }\bibfield  {title}
  {\bibinfo {title} {Exploring local quantum many-body relaxation by atoms in
  optical superlattices},\ }\href
  {https://doi.org/10.1103/PhysRevLett.101.063001} {\bibfield  {journal}
  {\bibinfo  {journal} {Phys. Rev. Lett.}\ }\textbf {\bibinfo {volume} {101}},\
  \bibinfo {pages} {063001} (\bibinfo {year} {2008}{\natexlab{a}})}\BibitemShut
  {NoStop}%
\bibitem [{\citenamefont {Cramer}\ \emph
  {et~al.}(2008{\natexlab{b}})\citenamefont {Cramer}, \citenamefont {Dawson},
  \citenamefont {Eisert},\ and\ \citenamefont
  {Osborne}}]{PhysRevLett.100.030602}%
  \BibitemOpen
  \bibfield  {author} {\bibinfo {author} {\bibfnamefont {M.}~\bibnamefont
  {Cramer}}, \bibinfo {author} {\bibfnamefont {C.~M.}\ \bibnamefont {Dawson}},
  \bibinfo {author} {\bibfnamefont {J.}~\bibnamefont {Eisert}},\ and\ \bibinfo
  {author} {\bibfnamefont {T.~J.}\ \bibnamefont {Osborne}},\ }\bibfield
  {title} {\bibinfo {title} {Exact relaxation in a class of nonequilibrium
  quantum lattice systems},\ }\href
  {https://doi.org/10.1103/PhysRevLett.100.030602} {\bibfield  {journal}
  {\bibinfo  {journal} {Phys. Rev. Lett.}\ }\textbf {\bibinfo {volume} {100}},\
  \bibinfo {pages} {030602} (\bibinfo {year} {2008}{\natexlab{b}})}\BibitemShut
  {NoStop}%
\bibitem [{\citenamefont {Manmana}\ \emph {et~al.}(2007)\citenamefont
  {Manmana}, \citenamefont {Wessel}, \citenamefont {Noack},\ and\ \citenamefont
  {Muramatsu}}]{PhysRevLett.98.210405}%
  \BibitemOpen
  \bibfield  {author} {\bibinfo {author} {\bibfnamefont {S.~R.}\ \bibnamefont
  {Manmana}}, \bibinfo {author} {\bibfnamefont {S.}~\bibnamefont {Wessel}},
  \bibinfo {author} {\bibfnamefont {R.~M.}\ \bibnamefont {Noack}},\ and\
  \bibinfo {author} {\bibfnamefont {A.}~\bibnamefont {Muramatsu}},\ }\bibfield
  {title} {\bibinfo {title} {Strongly correlated fermions after a quantum
  quench},\ }\href {https://doi.org/10.1103/PhysRevLett.98.210405} {\bibfield
  {journal} {\bibinfo  {journal} {Phys. Rev. Lett.}\ }\textbf {\bibinfo
  {volume} {98}},\ \bibinfo {pages} {210405} (\bibinfo {year}
  {2007})}\BibitemShut {NoStop}%
\bibitem [{\citenamefont {Cazalilla}(2006)}]{PhysRevLett.97.156403}%
  \BibitemOpen
  \bibfield  {author} {\bibinfo {author} {\bibfnamefont {M.~A.}\ \bibnamefont
  {Cazalilla}},\ }\bibfield  {title} {\bibinfo {title} {Effect of suddenly
  turning on interactions in the {Luttinger} model},\ }\href
  {https://doi.org/10.1103/PhysRevLett.97.156403} {\bibfield  {journal}
  {\bibinfo  {journal} {Phys. Rev. Lett.}\ }\textbf {\bibinfo {volume} {97}},\
  \bibinfo {pages} {156403} (\bibinfo {year} {2006})}\BibitemShut {NoStop}%
\bibitem [{\citenamefont {Calabrese}\ and\ \citenamefont
  {Cardy}(2006)}]{PhysRevLett.96.136801}%
  \BibitemOpen
  \bibfield  {author} {\bibinfo {author} {\bibfnamefont {P.}~\bibnamefont
  {Calabrese}}\ and\ \bibinfo {author} {\bibfnamefont {J.}~\bibnamefont
  {Cardy}},\ }\bibfield  {title} {\bibinfo {title} {Time dependence of
  correlation functions following a quantum quench},\ }\href
  {https://doi.org/10.1103/PhysRevLett.96.136801} {\bibfield  {journal}
  {\bibinfo  {journal} {Phys. Rev. Lett.}\ }\textbf {\bibinfo {volume} {96}},\
  \bibinfo {pages} {136801} (\bibinfo {year} {2006})}\BibitemShut {NoStop}%
\bibitem [{\citenamefont {Rigol}\ \emph {et~al.}(2007)\citenamefont {Rigol},
  \citenamefont {Dunjko}, \citenamefont {Yurovsky},\ and\ \citenamefont
  {Olshanii}}]{PhysRevLett.98.050405}%
  \BibitemOpen
  \bibfield  {author} {\bibinfo {author} {\bibfnamefont {M.}~\bibnamefont
  {Rigol}}, \bibinfo {author} {\bibfnamefont {V.}~\bibnamefont {Dunjko}},
  \bibinfo {author} {\bibfnamefont {V.}~\bibnamefont {Yurovsky}},\ and\
  \bibinfo {author} {\bibfnamefont {M.}~\bibnamefont {Olshanii}},\ }\bibfield
  {title} {\bibinfo {title} {Relaxation in a completely integrable many-body
  quantum system: An ab initio study of the dynamics of the highly excited
  states of {1D} lattice hard-core bosons},\ }\href
  {https://doi.org/10.1103/PhysRevLett.98.050405} {\bibfield  {journal}
  {\bibinfo  {journal} {Phys. Rev. Lett.}\ }\textbf {\bibinfo {volume} {98}},\
  \bibinfo {pages} {050405} (\bibinfo {year} {2007})}\BibitemShut {NoStop}%
\bibitem [{\citenamefont {Hamazaki}\ \emph {et~al.}(2016)\citenamefont
  {Hamazaki}, \citenamefont {Ikeda},\ and\ \citenamefont
  {Ueda}}]{PhysRevE.93.032116}%
  \BibitemOpen
  \bibfield  {author} {\bibinfo {author} {\bibfnamefont {R.}~\bibnamefont
  {Hamazaki}}, \bibinfo {author} {\bibfnamefont {T.~N.}\ \bibnamefont
  {Ikeda}},\ and\ \bibinfo {author} {\bibfnamefont {M.}~\bibnamefont {Ueda}},\
  }\bibfield  {title} {\bibinfo {title} {Generalized {Gibbs} ensemble in a
  nonintegrable system with an extensive number of local symmetries},\ }\href
  {https://doi.org/10.1103/PhysRevE.93.032116} {\bibfield  {journal} {\bibinfo
  {journal} {Phys. Rev. E}\ }\textbf {\bibinfo {volume} {93}},\ \bibinfo
  {pages} {032116} (\bibinfo {year} {2016})}\BibitemShut {NoStop}%
\bibitem [{\citenamefont {Larson}(2013)}]{Larson_2013}%
  \BibitemOpen
  \bibfield  {author} {\bibinfo {author} {\bibfnamefont {J.}~\bibnamefont
  {Larson}},\ }\bibfield  {title} {\bibinfo {title} {Integrability versus
  quantum thermalization},\ }\href
  {https://doi.org/10.1088/0953-4075/46/22/224016} {\bibfield  {journal}
  {\bibinfo  {journal} {J. Phys. B: At. Mol. Opt. Phys.}\ }\textbf {\bibinfo
  {volume} {46}},\ \bibinfo {pages} {224016} (\bibinfo {year}
  {2013})}\BibitemShut {NoStop}%
\bibitem [{\citenamefont {Yurovsky}\ and\ \citenamefont
  {Olshanii}(2011)}]{PhysRevLett.106.025303}%
  \BibitemOpen
  \bibfield  {author} {\bibinfo {author} {\bibfnamefont {V.~A.}\ \bibnamefont
  {Yurovsky}}\ and\ \bibinfo {author} {\bibfnamefont {M.}~\bibnamefont
  {Olshanii}},\ }\bibfield  {title} {\bibinfo {title} {Memory of the initial
  conditions in an incompletely chaotic quantum system: Universal predictions
  with application to cold atoms},\ }\href
  {https://doi.org/10.1103/PhysRevLett.106.025303} {\bibfield  {journal}
  {\bibinfo  {journal} {Phys. Rev. Lett.}\ }\textbf {\bibinfo {volume} {106}},\
  \bibinfo {pages} {025303} (\bibinfo {year} {2011})}\BibitemShut {NoStop}%
\bibitem [{\citenamefont {Olshanii}\ \emph {et~al.}(2012)\citenamefont
  {Olshanii}, \citenamefont {Jacobs}, \citenamefont {Rigol}, \citenamefont
  {Dunjko}, \citenamefont {Kennard},\ and\ \citenamefont
  {Yurovsky}}]{Olshanii2012}%
  \BibitemOpen
  \bibfield  {author} {\bibinfo {author} {\bibfnamefont {M.}~\bibnamefont
  {Olshanii}}, \bibinfo {author} {\bibfnamefont {K.}~\bibnamefont {Jacobs}},
  \bibinfo {author} {\bibfnamefont {M.}~\bibnamefont {Rigol}}, \bibinfo
  {author} {\bibfnamefont {V.}~\bibnamefont {Dunjko}}, \bibinfo {author}
  {\bibfnamefont {H.}~\bibnamefont {Kennard}},\ and\ \bibinfo {author}
  {\bibfnamefont {V.~A.}\ \bibnamefont {Yurovsky}},\ }\bibfield  {title}
  {\bibinfo {title} {An exactly solvable model for the integrability--chaos
  transition in rough quantum billiards},\ }\href
  {https://doi.org/10.1038/ncomms1653} {\bibfield  {journal} {\bibinfo
  {journal} {Nat. Commun.}\ }\textbf {\bibinfo {volume} {3}},\ \bibinfo {pages}
  {641} (\bibinfo {year} {2012})}\BibitemShut {NoStop}%
\bibitem [{\citenamefont {Vidmar}\ and\ \citenamefont
  {Rigol}(2016)}]{Vidmar_2016}%
  \BibitemOpen
  \bibfield  {author} {\bibinfo {author} {\bibfnamefont {L.}~\bibnamefont
  {Vidmar}}\ and\ \bibinfo {author} {\bibfnamefont {M.}~\bibnamefont {Rigol}},\
  }\bibfield  {title} {\bibinfo {title} {Generalized {Gibbs} ensemble in
  integrable lattice models},\ }\href
  {https://doi.org/10.1088/1742-5468/2016/06/064007} {\bibfield  {journal}
  {\bibinfo  {journal} {J. Stat. Mech.: Theory Exp.}\ }\textbf {\bibinfo
  {volume} {2016}}\bibinfo  {number} { (6)},\ \bibinfo {pages}
  {064007}}\BibitemShut {NoStop}%
\bibitem [{\citenamefont {Ilievski}\ \emph {et~al.}(2016)\citenamefont
  {Ilievski}, \citenamefont {Medenjak}, \citenamefont {Prosen},\ and\
  \citenamefont {Zadnik}}]{Ilievski_2016}%
  \BibitemOpen
\bibfield  {number} {  }\bibfield  {author} {\bibinfo {author} {\bibfnamefont
  {E.}~\bibnamefont {Ilievski}}, \bibinfo {author} {\bibfnamefont
  {M.}~\bibnamefont {Medenjak}}, \bibinfo {author} {\bibfnamefont
  {T.}~\bibnamefont {Prosen}},\ and\ \bibinfo {author} {\bibfnamefont
  {L.}~\bibnamefont {Zadnik}},\ }\bibfield  {title} {\bibinfo {title}
  {Quasilocal charges in integrable lattice systems},\ }\href
  {https://doi.org/10.1088/1742-5468/2016/06/064008} {\bibfield  {journal}
  {\bibinfo  {journal} {J. Stat. Mech.: Theory Exp.}\ }\textbf {\bibinfo
  {volume} {2016}}\bibinfo  {number} { (6)},\ \bibinfo {pages}
  {064008}}\BibitemShut {NoStop}%
\bibitem [{\citenamefont {Ilievski}\ \emph {et~al.}(2015)\citenamefont
  {Ilievski}, \citenamefont {De~Nardis}, \citenamefont {Wouters}, \citenamefont
  {Caux}, \citenamefont {Essler},\ and\ \citenamefont
  {Prosen}}]{PhysRevLett.115.157201}%
  \BibitemOpen
\bibfield  {number} {  }\bibfield  {author} {\bibinfo {author} {\bibfnamefont
  {E.}~\bibnamefont {Ilievski}}, \bibinfo {author} {\bibfnamefont
  {J.}~\bibnamefont {De~Nardis}}, \bibinfo {author} {\bibfnamefont
  {B.}~\bibnamefont {Wouters}}, \bibinfo {author} {\bibfnamefont {J.-S.}\
  \bibnamefont {Caux}}, \bibinfo {author} {\bibfnamefont {F.~H.~L.}\
  \bibnamefont {Essler}},\ and\ \bibinfo {author} {\bibfnamefont
  {T.}~\bibnamefont {Prosen}},\ }\bibfield  {title} {\bibinfo {title} {Complete
  generalized {Gibbs} ensembles in an interacting theory},\ }\href
  {https://doi.org/10.1103/PhysRevLett.115.157201} {\bibfield  {journal}
  {\bibinfo  {journal} {Phys. Rev. Lett.}\ }\textbf {\bibinfo {volume} {115}},\
  \bibinfo {pages} {157201} (\bibinfo {year} {2015})}\BibitemShut {NoStop}%
\bibitem [{\citenamefont {Essler}\ \emph {et~al.}(2015)\citenamefont {Essler},
  \citenamefont {Mussardo},\ and\ \citenamefont {Panfil}}]{PhysRevA.91.051602}%
  \BibitemOpen
  \bibfield  {author} {\bibinfo {author} {\bibfnamefont {F.~H.~L.}\
  \bibnamefont {Essler}}, \bibinfo {author} {\bibfnamefont {G.}~\bibnamefont
  {Mussardo}},\ and\ \bibinfo {author} {\bibfnamefont {M.}~\bibnamefont
  {Panfil}},\ }\bibfield  {title} {\bibinfo {title} {Generalized {Gibbs}
  ensembles for quantum field theories},\ }\href
  {https://doi.org/10.1103/PhysRevA.91.051602} {\bibfield  {journal} {\bibinfo
  {journal} {Phys. Rev. A}\ }\textbf {\bibinfo {volume} {91}},\ \bibinfo
  {pages} {051602} (\bibinfo {year} {2015})}\BibitemShut {NoStop}%
\bibitem [{\citenamefont {Pozsgay}(2014{\natexlab{a}})}]{Pozsgay_2014a}%
  \BibitemOpen
  \bibfield  {author} {\bibinfo {author} {\bibfnamefont {B.}~\bibnamefont
  {Pozsgay}},\ }\bibfield  {title} {\bibinfo {title} {Quantum quenches and
  generalized {Gibbs} ensemble in a bethe ansatz solvable lattice model of
  interacting bosons},\ }\href
  {https://doi.org/10.1088/1742-5468/2014/10/p10045} {\bibfield  {journal}
  {\bibinfo  {journal} {J. Stat. Mech.: Theory Exp.}\ }\textbf {\bibinfo
  {volume} {2014}}\bibinfo  {number} { (10)},\ \bibinfo {pages}
  {P10045}}\BibitemShut {NoStop}%
\bibitem [{\citenamefont {Pozsgay}(2014{\natexlab{b}})}]{Pozsgay_2014}%
  \BibitemOpen
\bibfield  {number} {  }\bibfield  {author} {\bibinfo {author} {\bibfnamefont
  {B.}~\bibnamefont {Pozsgay}},\ }\bibfield  {title} {\bibinfo {title} {Failure
  of the generalized eigenstate thermalization hypothesis in integrable models
  with multiple particle species},\ }\href
  {https://doi.org/10.1088/1742-5468/2014/09/p09026} {\bibfield  {journal}
  {\bibinfo  {journal} {J. Stat. Mech.: Theory Exp.}\ }\textbf {\bibinfo
  {volume} {2014}}\bibinfo  {number} { (9)},\ \bibinfo {pages}
  {P09026}}\BibitemShut {NoStop}%
\bibitem [{\citenamefont {Goldstein}\ and\ \citenamefont
  {Andrei}(2014)}]{PhysRevA.90.043625}%
  \BibitemOpen
\bibfield  {number} {  }\bibfield  {author} {\bibinfo {author} {\bibfnamefont
  {G.}~\bibnamefont {Goldstein}}\ and\ \bibinfo {author} {\bibfnamefont
  {N.}~\bibnamefont {Andrei}},\ }\bibfield  {title} {\bibinfo {title} {Failure
  of the local generalized {Gibbs} ensemble for integrable models with bound
  states},\ }\href {https://doi.org/10.1103/PhysRevA.90.043625} {\bibfield
  {journal} {\bibinfo  {journal} {Phys. Rev. A}\ }\textbf {\bibinfo {volume}
  {90}},\ \bibinfo {pages} {043625} (\bibinfo {year} {2014})}\BibitemShut
  {NoStop}%
\bibitem [{\citenamefont {Pozsgay}\ \emph {et~al.}(2014)\citenamefont
  {Pozsgay}, \citenamefont {Mesty\'an}, \citenamefont {Werner}, \citenamefont
  {Kormos}, \citenamefont {Zar\'and},\ and\ \citenamefont
  {Tak\'acs}}]{PhysRevLett.113.117203}%
  \BibitemOpen
  \bibfield  {author} {\bibinfo {author} {\bibfnamefont {B.}~\bibnamefont
  {Pozsgay}}, \bibinfo {author} {\bibfnamefont {M.}~\bibnamefont {Mesty\'an}},
  \bibinfo {author} {\bibfnamefont {M.~A.}\ \bibnamefont {Werner}}, \bibinfo
  {author} {\bibfnamefont {M.}~\bibnamefont {Kormos}}, \bibinfo {author}
  {\bibfnamefont {G.}~\bibnamefont {Zar\'and}},\ and\ \bibinfo {author}
  {\bibfnamefont {G.}~\bibnamefont {Tak\'acs}},\ }\bibfield  {title} {\bibinfo
  {title} {Correlations after quantum quenches in the {$XXZ$} spin chain:
  Failure of the generalized {Gibbs} ensemble},\ }\href
  {https://doi.org/10.1103/PhysRevLett.113.117203} {\bibfield  {journal}
  {\bibinfo  {journal} {Phys. Rev. Lett.}\ }\textbf {\bibinfo {volume} {113}},\
  \bibinfo {pages} {117203} (\bibinfo {year} {2014})}\BibitemShut {NoStop}%
\bibitem [{\citenamefont {Wouters}\ \emph {et~al.}(2014)\citenamefont
  {Wouters}, \citenamefont {De~Nardis}, \citenamefont {Brockmann},
  \citenamefont {Fioretto}, \citenamefont {Rigol},\ and\ \citenamefont
  {Caux}}]{PhysRevLett.113.117202}%
  \BibitemOpen
  \bibfield  {author} {\bibinfo {author} {\bibfnamefont {B.}~\bibnamefont
  {Wouters}}, \bibinfo {author} {\bibfnamefont {J.}~\bibnamefont {De~Nardis}},
  \bibinfo {author} {\bibfnamefont {M.}~\bibnamefont {Brockmann}}, \bibinfo
  {author} {\bibfnamefont {D.}~\bibnamefont {Fioretto}}, \bibinfo {author}
  {\bibfnamefont {M.}~\bibnamefont {Rigol}},\ and\ \bibinfo {author}
  {\bibfnamefont {J.-S.}\ \bibnamefont {Caux}},\ }\bibfield  {title} {\bibinfo
  {title} {Quenching the anisotropic heisenberg chain: Exact solution and
  generalized {Gibbs} ensemble predictions},\ }\href
  {https://doi.org/10.1103/PhysRevLett.113.117202} {\bibfield  {journal}
  {\bibinfo  {journal} {Phys. Rev. Lett.}\ }\textbf {\bibinfo {volume} {113}},\
  \bibinfo {pages} {117202} (\bibinfo {year} {2014})}\BibitemShut {NoStop}%
\bibitem [{\citenamefont {Calabrese}\ \emph {et~al.}(2012)\citenamefont
  {Calabrese}, \citenamefont {Essler},\ and\ \citenamefont
  {Fagotti}}]{Calabrese_2012}%
  \BibitemOpen
  \bibfield  {author} {\bibinfo {author} {\bibfnamefont {P.}~\bibnamefont
  {Calabrese}}, \bibinfo {author} {\bibfnamefont {F.~H.~L.}\ \bibnamefont
  {Essler}},\ and\ \bibinfo {author} {\bibfnamefont {M.}~\bibnamefont
  {Fagotti}},\ }\bibfield  {title} {\bibinfo {title} {Quantum quench in the
  transverse field ising chain: I. time evolution of order parameter
  correlators},\ }\href {https://doi.org/10.1088/1742-5468/2012/07/p07016}
  {\bibfield  {journal} {\bibinfo  {journal} {J. Stat. Mech.: Theory Exp.}\
  }\textbf {\bibinfo {volume} {2012}}\bibinfo  {number} { (07)},\ \bibinfo
  {pages} {P07016}}\BibitemShut {NoStop}%
\bibitem [{\citenamefont {Blass}\ \emph {et~al.}(2012)\citenamefont {Blass},
  \citenamefont {Rieger},\ and\ \citenamefont {Igl{\'{o}}i}}]{Blass_2012}%
  \BibitemOpen
\bibfield  {number} {  }\bibfield  {author} {\bibinfo {author} {\bibfnamefont
  {B.}~\bibnamefont {Blass}}, \bibinfo {author} {\bibfnamefont
  {H.}~\bibnamefont {Rieger}},\ and\ \bibinfo {author} {\bibfnamefont
  {F.}~\bibnamefont {Igl{\'{o}}i}},\ }\bibfield  {title} {\bibinfo {title}
  {Quantum relaxation and finite-size effects in the {XY} chain in a transverse
  field after global quenches},\ }\href
  {https://doi.org/10.1209/0295-5075/99/30004} {\bibfield  {journal} {\bibinfo
  {journal} {{EPL} (Europhysics Letters)}\ }\textbf {\bibinfo {volume} {99}},\
  \bibinfo {pages} {30004} (\bibinfo {year} {2012})}\BibitemShut {NoStop}%
\bibitem [{\citenamefont {Lieb}\ \emph {et~al.}(1961)\citenamefont {Lieb},
  \citenamefont {Schultz},\ and\ \citenamefont {Mattis}}]{Lieb1961Two}%
  \BibitemOpen
  \bibfield  {author} {\bibinfo {author} {\bibfnamefont {E.}~\bibnamefont
  {Lieb}}, \bibinfo {author} {\bibfnamefont {T.}~\bibnamefont {Schultz}},\ and\
  \bibinfo {author} {\bibfnamefont {D.}~\bibnamefont {Mattis}},\ }\bibfield
  {title} {\bibinfo {title} {Two soluble models of an antiferromagnetic
  chain},\ }\href
  {https://doi.org/https://doi.org/10.1016/0003-4916(61)90115-4} {\bibfield
  {journal} {\bibinfo  {journal} {Ann. Phys}\ }\textbf {\bibinfo {volume}
  {16}},\ \bibinfo {pages} {407} (\bibinfo {year} {1961})}\BibitemShut
  {NoStop}%
\bibitem [{\citenamefont {Pfeuty}(1970)}]{Pfeuty1970The}%
  \BibitemOpen
  \bibfield  {author} {\bibinfo {author} {\bibfnamefont {P.}~\bibnamefont
  {Pfeuty}},\ }\bibfield  {title} {\bibinfo {title} {The one-dimensional ising
  model with a transverse field},\ }\href
  {https://doi.org/10.1016/0003-4916(70)90270-8} {\bibfield  {journal}
  {\bibinfo  {journal} {Ann. Phys.}\ }\textbf {\bibinfo {volume} {57}},\
  \bibinfo {pages} {79 } (\bibinfo {year} {1970})}\BibitemShut {NoStop}%
\bibitem [{\citenamefont {Lewenstein}\ \emph {et~al.}(2001)\citenamefont
  {Lewenstein}, \citenamefont {Kraus}, \citenamefont {Horodecki},\ and\
  \citenamefont {Cirac}}]{Lewenstein2001Characterization}%
  \BibitemOpen
  \bibfield  {author} {\bibinfo {author} {\bibfnamefont {M.}~\bibnamefont
  {Lewenstein}}, \bibinfo {author} {\bibfnamefont {B.}~\bibnamefont {Kraus}},
  \bibinfo {author} {\bibfnamefont {P.}~\bibnamefont {Horodecki}},\ and\
  \bibinfo {author} {\bibfnamefont {J.~I.}\ \bibnamefont {Cirac}},\ }\bibfield
  {title} {\bibinfo {title} {Characterization of separable states and
  entanglement witnesses},\ }\href {https://doi.org/10.1103/PhysRevA.63.044304}
  {\bibfield  {journal} {\bibinfo  {journal} {Phys. Rev. A}\ }\textbf {\bibinfo
  {volume} {63}},\ \bibinfo {pages} {044304} (\bibinfo {year}
  {2001})}\BibitemShut {NoStop}%
\bibitem [{\citenamefont {Terhal}(2002)}]{Terhal2002Detecting}%
  \BibitemOpen
  \bibfield  {author} {\bibinfo {author} {\bibfnamefont {B.~M.}\ \bibnamefont
  {Terhal}},\ }\bibfield  {title} {\bibinfo {title} {Detecting quantum
  entanglement},\ }\href {https://doi.org/10.1016/S0304-3975(02)00139-1}
  {\bibfield  {journal} {\bibinfo  {journal} {Theor. Comput. Sci.}\ }\textbf
  {\bibinfo {volume} {287}},\ \bibinfo {pages} {313 } (\bibinfo {year}
  {2002})}\BibitemShut {NoStop}%
\bibitem [{\citenamefont {Bourennane}\ \emph {et~al.}(2004)\citenamefont
  {Bourennane}, \citenamefont {Eibl}, \citenamefont {Kurtsiefer}, \citenamefont
  {Gaertner}, \citenamefont {Weinfurter}, \citenamefont {G\"uhne},
  \citenamefont {Hyllus}, \citenamefont {Bru\ss{}}, \citenamefont
  {Lewenstein},\ and\ \citenamefont {Sanpera}}]{Bourennane2004Experimental}%
  \BibitemOpen
  \bibfield  {author} {\bibinfo {author} {\bibfnamefont {M.}~\bibnamefont
  {Bourennane}}, \bibinfo {author} {\bibfnamefont {M.}~\bibnamefont {Eibl}},
  \bibinfo {author} {\bibfnamefont {C.}~\bibnamefont {Kurtsiefer}}, \bibinfo
  {author} {\bibfnamefont {S.}~\bibnamefont {Gaertner}}, \bibinfo {author}
  {\bibfnamefont {H.}~\bibnamefont {Weinfurter}}, \bibinfo {author}
  {\bibfnamefont {O.}~\bibnamefont {G\"uhne}}, \bibinfo {author} {\bibfnamefont
  {P.}~\bibnamefont {Hyllus}}, \bibinfo {author} {\bibfnamefont
  {D.}~\bibnamefont {Bru\ss{}}}, \bibinfo {author} {\bibfnamefont
  {M.}~\bibnamefont {Lewenstein}},\ and\ \bibinfo {author} {\bibfnamefont
  {A.}~\bibnamefont {Sanpera}},\ }\bibfield  {title} {\bibinfo {title}
  {Experimental detection of multipartite entanglement using witness
  operators},\ }\href {https://doi.org/10.1103/PhysRevLett.92.087902}
  {\bibfield  {journal} {\bibinfo  {journal} {Phys. Rev. Lett.}\ }\textbf
  {\bibinfo {volume} {92}},\ \bibinfo {pages} {087902} (\bibinfo {year}
  {2004})}\BibitemShut {NoStop}%
\bibitem [{\citenamefont {{Walther}}\ \emph {et~al.}(2005)\citenamefont
  {{Walther}}, \citenamefont {{Resch}}, \citenamefont {{Rudolph}},
  \citenamefont {{Schenck}}, \citenamefont {{Weinfurter}}, \citenamefont
  {{Vedral}}, \citenamefont {{Aspelmeyer}},\ and\ \citenamefont
  {{Zeilinger}}}]{Walther2005ExperimentalOneWay}%
  \BibitemOpen
  \bibfield  {author} {\bibinfo {author} {\bibfnamefont {P.}~\bibnamefont
  {{Walther}}}, \bibinfo {author} {\bibfnamefont {K.~J.}\ \bibnamefont
  {{Resch}}}, \bibinfo {author} {\bibfnamefont {T.}~\bibnamefont {{Rudolph}}},
  \bibinfo {author} {\bibfnamefont {E.}~\bibnamefont {{Schenck}}}, \bibinfo
  {author} {\bibfnamefont {H.}~\bibnamefont {{Weinfurter}}}, \bibinfo {author}
  {\bibfnamefont {V.}~\bibnamefont {{Vedral}}}, \bibinfo {author}
  {\bibfnamefont {M.}~\bibnamefont {{Aspelmeyer}}},\ and\ \bibinfo {author}
  {\bibfnamefont {A.}~\bibnamefont {{Zeilinger}}},\ }\bibfield  {title}
  {\bibinfo {title} {{Experimental one-way quantum computing}},\ }\href
  {https://doi.org/10.1038/nature03347} {\bibfield  {journal} {\bibinfo
  {journal} {\nat}\ }\textbf {\bibinfo {volume} {434}},\ \bibinfo {pages} {169}
  (\bibinfo {year} {2005})}\BibitemShut {NoStop}%
\bibitem [{\citenamefont {Kiesel}\ \emph {et~al.}(2005)\citenamefont {Kiesel},
  \citenamefont {Schmid}, \citenamefont {Weber}, \citenamefont {T\'oth},
  \citenamefont {G\"uhne}, \citenamefont {Ursin},\ and\ \citenamefont
  {Weinfurter}}]{Kiesel2005Experimental}%
  \BibitemOpen
  \bibfield  {author} {\bibinfo {author} {\bibfnamefont {N.}~\bibnamefont
  {Kiesel}}, \bibinfo {author} {\bibfnamefont {C.}~\bibnamefont {Schmid}},
  \bibinfo {author} {\bibfnamefont {U.}~\bibnamefont {Weber}}, \bibinfo
  {author} {\bibfnamefont {G.}~\bibnamefont {T\'oth}}, \bibinfo {author}
  {\bibfnamefont {O.}~\bibnamefont {G\"uhne}}, \bibinfo {author} {\bibfnamefont
  {R.}~\bibnamefont {Ursin}},\ and\ \bibinfo {author} {\bibfnamefont
  {H.}~\bibnamefont {Weinfurter}},\ }\bibfield  {title} {\bibinfo {title}
  {Experimental analysis of a four-qubit photon cluster state},\ }\href
  {https://doi.org/10.1103/PhysRevLett.95.210502} {\bibfield  {journal}
  {\bibinfo  {journal} {Phys. Rev. Lett.}\ }\textbf {\bibinfo {volume} {95}},\
  \bibinfo {pages} {210502} (\bibinfo {year} {2005})}\BibitemShut {NoStop}%
\bibitem [{\citenamefont {Wieczorek}\ \emph {et~al.}(2009)\citenamefont
  {Wieczorek}, \citenamefont {Krischek}, \citenamefont {Kiesel}, \citenamefont
  {Michelberger}, \citenamefont {T\'oth},\ and\ \citenamefont
  {Weinfurter}}]{Wieczorek2009Experimental}%
  \BibitemOpen
  \bibfield  {author} {\bibinfo {author} {\bibfnamefont {W.}~\bibnamefont
  {Wieczorek}}, \bibinfo {author} {\bibfnamefont {R.}~\bibnamefont {Krischek}},
  \bibinfo {author} {\bibfnamefont {N.}~\bibnamefont {Kiesel}}, \bibinfo
  {author} {\bibfnamefont {P.}~\bibnamefont {Michelberger}}, \bibinfo {author}
  {\bibfnamefont {G.}~\bibnamefont {T\'oth}},\ and\ \bibinfo {author}
  {\bibfnamefont {H.}~\bibnamefont {Weinfurter}},\ }\bibfield  {title}
  {\bibinfo {title} {Experimental entanglement of a six-photon symmetric dicke
  state},\ }\href {https://doi.org/10.1103/PhysRevLett.103.020504} {\bibfield
  {journal} {\bibinfo  {journal} {Phys. Rev. Lett.}\ }\textbf {\bibinfo
  {volume} {103}},\ \bibinfo {pages} {020504} (\bibinfo {year}
  {2009})}\BibitemShut {NoStop}%
\bibitem [{\citenamefont {Prevedel}\ \emph {et~al.}(2009)\citenamefont
  {Prevedel}, \citenamefont {Cronenberg}, \citenamefont {Tame}, \citenamefont
  {Paternostro}, \citenamefont {Walther}, \citenamefont {Kim},\ and\
  \citenamefont {Zeilinger}}]{Prevedel2009Experimental}%
  \BibitemOpen
  \bibfield  {author} {\bibinfo {author} {\bibfnamefont {R.}~\bibnamefont
  {Prevedel}}, \bibinfo {author} {\bibfnamefont {G.}~\bibnamefont
  {Cronenberg}}, \bibinfo {author} {\bibfnamefont {M.~S.}\ \bibnamefont
  {Tame}}, \bibinfo {author} {\bibfnamefont {M.}~\bibnamefont {Paternostro}},
  \bibinfo {author} {\bibfnamefont {P.}~\bibnamefont {Walther}}, \bibinfo
  {author} {\bibfnamefont {M.~S.}\ \bibnamefont {Kim}},\ and\ \bibinfo {author}
  {\bibfnamefont {A.}~\bibnamefont {Zeilinger}},\ }\bibfield  {title} {\bibinfo
  {title} {Experimental realization of dicke states of up to six qubits for
  multiparty quantum networking},\ }\href
  {https://doi.org/10.1103/PhysRevLett.103.020503} {\bibfield  {journal}
  {\bibinfo  {journal} {Phys. Rev. Lett.}\ }\textbf {\bibinfo {volume} {103}},\
  \bibinfo {pages} {020503} (\bibinfo {year} {2009})}\BibitemShut {NoStop}%
\bibitem [{\citenamefont {Gao}\ \emph {et~al.}(2010)\citenamefont {Gao},
  \citenamefont {Lu}, \citenamefont {Yao}, \citenamefont {Xu}, \citenamefont
  {G{\"u}hne}, \citenamefont {Goebel}, \citenamefont {Chen}, \citenamefont
  {Peng}, \citenamefont {Chen},\ and\ \citenamefont
  {Pan}}]{Gao2010Experimental}%
  \BibitemOpen
  \bibfield  {author} {\bibinfo {author} {\bibfnamefont {W.-B.}\ \bibnamefont
  {Gao}}, \bibinfo {author} {\bibfnamefont {C.-Y.}\ \bibnamefont {Lu}},
  \bibinfo {author} {\bibfnamefont {X.-C.}\ \bibnamefont {Yao}}, \bibinfo
  {author} {\bibfnamefont {P.}~\bibnamefont {Xu}}, \bibinfo {author}
  {\bibfnamefont {O.}~\bibnamefont {G{\"u}hne}}, \bibinfo {author}
  {\bibfnamefont {A.}~\bibnamefont {Goebel}}, \bibinfo {author} {\bibfnamefont
  {Y.-A.}\ \bibnamefont {Chen}}, \bibinfo {author} {\bibfnamefont {C.-Z.}\
  \bibnamefont {Peng}}, \bibinfo {author} {\bibfnamefont {Z.-B.}\ \bibnamefont
  {Chen}},\ and\ \bibinfo {author} {\bibfnamefont {J.-W.}\ \bibnamefont
  {Pan}},\ }\bibfield  {title} {\bibinfo {title} {Experimental demonstration of
  a hyper-entangled ten-qubit schr{\"o}dinger cat state},\ }\href
  {https://doi.org/10.1038/nphys1603} {\bibfield  {journal} {\bibinfo
  {journal} {Nat. Phys.}\ }\textbf {\bibinfo {volume} {6}},\ \bibinfo {pages}
  {331} (\bibinfo {year} {2010})}\BibitemShut {NoStop}%
\bibitem [{\citenamefont {Gong}\ \emph {et~al.}(2019)\citenamefont {Gong},
  \citenamefont {Chen}, \citenamefont {Zheng}, \citenamefont {Wang},
  \citenamefont {Zha}, \citenamefont {Deng}, \citenamefont {Yan}, \citenamefont
  {Rong}, \citenamefont {Wu}, \citenamefont {Li}, \citenamefont {Chen},
  \citenamefont {Zhao}, \citenamefont {Liang}, \citenamefont {Lin},
  \citenamefont {Xu}, \citenamefont {Guo}, \citenamefont {Sun}, \citenamefont
  {Castellano}, \citenamefont {Wang}, \citenamefont {Peng}, \citenamefont {Lu},
  \citenamefont {Zhu},\ and\ \citenamefont {Pan}}]{Genuine2019Gong}%
  \BibitemOpen
  \bibfield  {author} {\bibinfo {author} {\bibfnamefont {M.}~\bibnamefont
  {Gong}}, \bibinfo {author} {\bibfnamefont {M.-C.}\ \bibnamefont {Chen}},
  \bibinfo {author} {\bibfnamefont {Y.}~\bibnamefont {Zheng}}, \bibinfo
  {author} {\bibfnamefont {S.}~\bibnamefont {Wang}}, \bibinfo {author}
  {\bibfnamefont {C.}~\bibnamefont {Zha}}, \bibinfo {author} {\bibfnamefont
  {H.}~\bibnamefont {Deng}}, \bibinfo {author} {\bibfnamefont {Z.}~\bibnamefont
  {Yan}}, \bibinfo {author} {\bibfnamefont {H.}~\bibnamefont {Rong}}, \bibinfo
  {author} {\bibfnamefont {Y.}~\bibnamefont {Wu}}, \bibinfo {author}
  {\bibfnamefont {S.}~\bibnamefont {Li}}, \bibinfo {author} {\bibfnamefont
  {F.}~\bibnamefont {Chen}}, \bibinfo {author} {\bibfnamefont {Y.}~\bibnamefont
  {Zhao}}, \bibinfo {author} {\bibfnamefont {F.}~\bibnamefont {Liang}},
  \bibinfo {author} {\bibfnamefont {J.}~\bibnamefont {Lin}}, \bibinfo {author}
  {\bibfnamefont {Y.}~\bibnamefont {Xu}}, \bibinfo {author} {\bibfnamefont
  {C.}~\bibnamefont {Guo}}, \bibinfo {author} {\bibfnamefont {L.}~\bibnamefont
  {Sun}}, \bibinfo {author} {\bibfnamefont {A.~D.}\ \bibnamefont {Castellano}},
  \bibinfo {author} {\bibfnamefont {H.}~\bibnamefont {Wang}}, \bibinfo {author}
  {\bibfnamefont {C.}~\bibnamefont {Peng}}, \bibinfo {author} {\bibfnamefont
  {C.-Y.}\ \bibnamefont {Lu}}, \bibinfo {author} {\bibfnamefont
  {X.}~\bibnamefont {Zhu}},\ and\ \bibinfo {author} {\bibfnamefont {J.-W.}\
  \bibnamefont {Pan}},\ }\bibfield  {title} {\bibinfo {title} {Genuine 12-qubit
  entanglement on a superconducting quantum processor},\ }\href
  {https://doi.org/10.1103/PhysRevLett.122.110501} {\bibfield  {journal}
  {\bibinfo  {journal} {Phys. Rev. Lett.}\ }\textbf {\bibinfo {volume} {122}},\
  \bibinfo {pages} {110501} (\bibinfo {year} {2019})}\BibitemShut {NoStop}%
\bibitem [{\citenamefont {H{\"a}ffner}\ \emph {et~al.}(2005)\citenamefont
  {H{\"a}ffner}, \citenamefont {H{\"a}nsel}, \citenamefont {Roos},
  \citenamefont {Benhelm}, \citenamefont {Chwalla}, \citenamefont {K{\"o}rber},
  \citenamefont {Rapol}, \citenamefont {Riebe}, \citenamefont {Schmidt},
  \citenamefont {Becher}, \citenamefont {G\"uhne}, \citenamefont {D\"ur},\ and\
  \citenamefont {Blatt}}]{Haffner2005Scalable}%
  \BibitemOpen
  \bibfield  {author} {\bibinfo {author} {\bibfnamefont {H.}~\bibnamefont
  {H{\"a}ffner}}, \bibinfo {author} {\bibfnamefont {W.}~\bibnamefont
  {H{\"a}nsel}}, \bibinfo {author} {\bibfnamefont {C.}~\bibnamefont {Roos}},
  \bibinfo {author} {\bibfnamefont {J.}~\bibnamefont {Benhelm}}, \bibinfo
  {author} {\bibfnamefont {M.}~\bibnamefont {Chwalla}}, \bibinfo {author}
  {\bibfnamefont {T.}~\bibnamefont {K{\"o}rber}}, \bibinfo {author}
  {\bibfnamefont {U.}~\bibnamefont {Rapol}}, \bibinfo {author} {\bibfnamefont
  {M.}~\bibnamefont {Riebe}}, \bibinfo {author} {\bibfnamefont
  {P.}~\bibnamefont {Schmidt}}, \bibinfo {author} {\bibfnamefont
  {C.}~\bibnamefont {Becher}}, \bibinfo {author} {\bibfnamefont
  {O.}~\bibnamefont {G\"uhne}}, \bibinfo {author} {\bibfnamefont
  {W.}~\bibnamefont {D\"ur}},\ and\ \bibinfo {author} {\bibfnamefont
  {R.}~\bibnamefont {Blatt}},\ }\bibfield  {title} {\bibinfo {title} {Scalable
  multiparticle entanglement of trapped ions},\ }\href
  {https://doi.org/10.1038/nature04279} {\bibfield  {journal} {\bibinfo
  {journal} {Nature (London)}\ }\textbf {\bibinfo {volume} {438}},\ \bibinfo
  {pages} {643} (\bibinfo {year} {2005})}\BibitemShut {NoStop}%
\bibitem [{\citenamefont {Monz}\ \emph {et~al.}(2011)\citenamefont {Monz},
  \citenamefont {Schindler}, \citenamefont {Barreiro}, \citenamefont {Chwalla},
  \citenamefont {Nigg}, \citenamefont {Coish}, \citenamefont {Harlander},
  \citenamefont {H\"ansel}, \citenamefont {Hennrich},\ and\ \citenamefont
  {Blatt}}]{14qubit2011Monz}%
  \BibitemOpen
  \bibfield  {author} {\bibinfo {author} {\bibfnamefont {T.}~\bibnamefont
  {Monz}}, \bibinfo {author} {\bibfnamefont {P.}~\bibnamefont {Schindler}},
  \bibinfo {author} {\bibfnamefont {J.~T.}\ \bibnamefont {Barreiro}}, \bibinfo
  {author} {\bibfnamefont {M.}~\bibnamefont {Chwalla}}, \bibinfo {author}
  {\bibfnamefont {D.}~\bibnamefont {Nigg}}, \bibinfo {author} {\bibfnamefont
  {W.~A.}\ \bibnamefont {Coish}}, \bibinfo {author} {\bibfnamefont
  {M.}~\bibnamefont {Harlander}}, \bibinfo {author} {\bibfnamefont
  {W.}~\bibnamefont {H\"ansel}}, \bibinfo {author} {\bibfnamefont
  {M.}~\bibnamefont {Hennrich}},\ and\ \bibinfo {author} {\bibfnamefont
  {R.}~\bibnamefont {Blatt}},\ }\bibfield  {title} {\bibinfo {title} {14-qubit
  entanglement: Creation and coherence},\ }\href
  {https://doi.org/10.1103/PhysRevLett.106.130506} {\bibfield  {journal}
  {\bibinfo  {journal} {Phys. Rev. Lett.}\ }\textbf {\bibinfo {volume} {106}},\
  \bibinfo {pages} {130506} (\bibinfo {year} {2011})}\BibitemShut {NoStop}%
\bibitem [{\citenamefont {Fel'dman}\ and\ \citenamefont
  {Pyrkov}(2008)}]{Feldman2008}%
  \BibitemOpen
  \bibfield  {author} {\bibinfo {author} {\bibfnamefont {E.~B.}\ \bibnamefont
  {Fel'dman}}\ and\ \bibinfo {author} {\bibfnamefont {A.~N.}\ \bibnamefont
  {Pyrkov}},\ }\bibfield  {title} {\bibinfo {title} {Evolution of spin
  entanglement and an entanglement witness in multiple-quantum nmr
  experiments},\ }\href {https://doi.org/10.1134/S0021364008180124} {\bibfield
  {journal} {\bibinfo  {journal} {JETP Letters}\ }\textbf {\bibinfo {volume}
  {88}},\ \bibinfo {pages} {398} (\bibinfo {year} {2008})}\BibitemShut
  {NoStop}%
\bibitem [{\citenamefont {G\"arttner}\ \emph {et~al.}(2018)\citenamefont
  {G\"arttner}, \citenamefont {Hauke},\ and\ \citenamefont
  {Rey}}]{Garttner2018Relating}%
  \BibitemOpen
  \bibfield  {author} {\bibinfo {author} {\bibfnamefont {M.}~\bibnamefont
  {G\"arttner}}, \bibinfo {author} {\bibfnamefont {P.}~\bibnamefont {Hauke}},\
  and\ \bibinfo {author} {\bibfnamefont {A.~M.}\ \bibnamefont {Rey}},\
  }\bibfield  {title} {\bibinfo {title} {Relating out-of-time-order
  correlations to entanglement via multiple-quantum coherences},\ }\href
  {https://doi.org/10.1103/PhysRevLett.120.040402} {\bibfield  {journal}
  {\bibinfo  {journal} {Phys. Rev. Lett.}\ }\textbf {\bibinfo {volume} {120}},\
  \bibinfo {pages} {040402} (\bibinfo {year} {2018})}\BibitemShut {NoStop}%
\bibitem [{\citenamefont {Werner}(1989)}]{Werner1989Quantum}%
  \BibitemOpen
  \bibfield  {author} {\bibinfo {author} {\bibfnamefont {R.~F.}\ \bibnamefont
  {Werner}},\ }\bibfield  {title} {\bibinfo {title} {Quantum states with
  {Einstein-Podolsky-Rosen} correlations admitting a hidden-variable model},\
  }\href {https://doi.org/10.1103/PhysRevA.40.4277} {\bibfield  {journal}
  {\bibinfo  {journal} {Phys. Rev. A}\ }\textbf {\bibinfo {volume} {40}},\
  \bibinfo {pages} {4277} (\bibinfo {year} {1989})}\BibitemShut {NoStop}%
\bibitem [{\citenamefont {Horodecki}\ \emph {et~al.}(1998)\citenamefont
  {Horodecki}, \citenamefont {Horodecki},\ and\ \citenamefont
  {Horodecki}}]{Horodecki1998Mixed-State}%
  \BibitemOpen
  \bibfield  {author} {\bibinfo {author} {\bibfnamefont {M.}~\bibnamefont
  {Horodecki}}, \bibinfo {author} {\bibfnamefont {P.}~\bibnamefont
  {Horodecki}},\ and\ \bibinfo {author} {\bibfnamefont {R.}~\bibnamefont
  {Horodecki}},\ }\bibfield  {title} {\bibinfo {title} {Mixed-state
  entanglement and distillation: Is there a ``bound'' entanglement in
  nature?},\ }\href {https://doi.org/10.1103/PhysRevLett.80.5239} {\bibfield
  {journal} {\bibinfo  {journal} {Phys. Rev. Lett.}\ }\textbf {\bibinfo
  {volume} {80}},\ \bibinfo {pages} {5239} (\bibinfo {year}
  {1998})}\BibitemShut {NoStop}%
\bibitem [{\citenamefont {Sanpera}\ \emph {et~al.}(1998)\citenamefont
  {Sanpera}, \citenamefont {Tarrach},\ and\ \citenamefont
  {Vidal}}]{PhysRevA.58.826}%
  \BibitemOpen
  \bibfield  {author} {\bibinfo {author} {\bibfnamefont {A.}~\bibnamefont
  {Sanpera}}, \bibinfo {author} {\bibfnamefont {R.}~\bibnamefont {Tarrach}},\
  and\ \bibinfo {author} {\bibfnamefont {G.}~\bibnamefont {Vidal}},\ }\bibfield
   {title} {\bibinfo {title} {Local description of quantum inseparability},\
  }\href {https://doi.org/10.1103/PhysRevA.58.826} {\bibfield  {journal}
  {\bibinfo  {journal} {Phys. Rev. A}\ }\textbf {\bibinfo {volume} {58}},\
  \bibinfo {pages} {826} (\bibinfo {year} {1998})}\BibitemShut {NoStop}%
\bibitem [{\citenamefont {Rana}(2013)}]{PhysRevA.87.054301}%
  \BibitemOpen
  \bibfield  {author} {\bibinfo {author} {\bibfnamefont {S.}~\bibnamefont
  {Rana}},\ }\bibfield  {title} {\bibinfo {title} {Negative eigenvalues of
  partial transposition of arbitrary bipartite states},\ }\href
  {https://doi.org/10.1103/PhysRevA.87.054301} {\bibfield  {journal} {\bibinfo
  {journal} {Phys. Rev. A}\ }\textbf {\bibinfo {volume} {87}},\ \bibinfo
  {pages} {054301} (\bibinfo {year} {2013})}\BibitemShut {NoStop}%
\bibitem [{\citenamefont {Dutta}\ \emph {et~al.}(2015)\citenamefont {Dutta},
  \citenamefont {Aeppli}, \citenamefont {Chakrabarti}, \citenamefont
  {Divakaran}, \citenamefont {Rosenbaum},\ and\ \citenamefont
  {Sen}}]{dutta_aeppli_chakrabarti_divakaran_rosenbaum_sen_2015}%
  \BibitemOpen
  \bibfield  {author} {\bibinfo {author} {\bibfnamefont {A.}~\bibnamefont
  {Dutta}}, \bibinfo {author} {\bibfnamefont {G.}~\bibnamefont {Aeppli}},
  \bibinfo {author} {\bibfnamefont {B.~K.}\ \bibnamefont {Chakrabarti}},
  \bibinfo {author} {\bibfnamefont {U.}~\bibnamefont {Divakaran}}, \bibinfo
  {author} {\bibfnamefont {T.~F.}\ \bibnamefont {Rosenbaum}},\ and\ \bibinfo
  {author} {\bibfnamefont {D.}~\bibnamefont {Sen}},\ }\href
  {https://doi.org/10.1017/CBO9781107706057} {\emph {\bibinfo {title} {Quantum
  Phase Transitions in Transverse Field Spin Models: From Statistical Physics
  to Quantum Information}}}\ (\bibinfo  {publisher} {Cambridge University
  Press},\ \bibinfo {year} {2015})\BibitemShut {NoStop}%
\bibitem [{\citenamefont {Greiner}\ \emph {et~al.}(2002)\citenamefont
  {Greiner}, \citenamefont {Mandel}, \citenamefont {H{\"a}nsch},\ and\
  \citenamefont {Bloch}}]{Greiner2002}%
  \BibitemOpen
  \bibfield  {author} {\bibinfo {author} {\bibfnamefont {M.}~\bibnamefont
  {Greiner}}, \bibinfo {author} {\bibfnamefont {O.}~\bibnamefont {Mandel}},
  \bibinfo {author} {\bibfnamefont {T.~W.}\ \bibnamefont {H{\"a}nsch}},\ and\
  \bibinfo {author} {\bibfnamefont {I.}~\bibnamefont {Bloch}},\ }\bibfield
  {title} {\bibinfo {title} {Collapse and revival of the matter wave field of a
  bose--einstein condensate},\ }\href {https://doi.org/10.1038/nature00968}
  {\bibfield  {journal} {\bibinfo  {journal} {Nature (London)}\ }\textbf
  {\bibinfo {volume} {419}},\ \bibinfo {pages} {51} (\bibinfo {year}
  {2002})}\BibitemShut {NoStop}%
\bibitem [{\citenamefont {Paredes}\ \emph {et~al.}(2004)\citenamefont
  {Paredes}, \citenamefont {Widera}, \citenamefont {Murg}, \citenamefont
  {Mandel}, \citenamefont {F{\"o}lling}, \citenamefont {Cirac}, \citenamefont
  {Shlyapnikov}, \citenamefont {H{\"a}nsch},\ and\ \citenamefont
  {Bloch}}]{Paredes2004}%
  \BibitemOpen
  \bibfield  {author} {\bibinfo {author} {\bibfnamefont {B.}~\bibnamefont
  {Paredes}}, \bibinfo {author} {\bibfnamefont {A.}~\bibnamefont {Widera}},
  \bibinfo {author} {\bibfnamefont {V.}~\bibnamefont {Murg}}, \bibinfo {author}
  {\bibfnamefont {O.}~\bibnamefont {Mandel}}, \bibinfo {author} {\bibfnamefont
  {S.}~\bibnamefont {F{\"o}lling}}, \bibinfo {author} {\bibfnamefont
  {I.}~\bibnamefont {Cirac}}, \bibinfo {author} {\bibfnamefont {G.~V.}\
  \bibnamefont {Shlyapnikov}}, \bibinfo {author} {\bibfnamefont {T.~W.}\
  \bibnamefont {H{\"a}nsch}},\ and\ \bibinfo {author} {\bibfnamefont
  {I.}~\bibnamefont {Bloch}},\ }\bibfield  {title} {\bibinfo {title}
  {Tonks--{Girardeau} gas of ultracold atoms in an optical lattice},\ }\href
  {https://doi.org/10.1038/nature02530} {\bibfield  {journal} {\bibinfo
  {journal} {Nature (London)}\ }\textbf {\bibinfo {volume} {429}},\ \bibinfo
  {pages} {277} (\bibinfo {year} {2004})}\BibitemShut {NoStop}%
\bibitem [{\citenamefont {Sadler}\ \emph {et~al.}(2006)\citenamefont {Sadler},
  \citenamefont {Higbie}, \citenamefont {Leslie}, \citenamefont
  {Vengalattore},\ and\ \citenamefont {Stamper-Kurn}}]{Sadler2006}%
  \BibitemOpen
  \bibfield  {author} {\bibinfo {author} {\bibfnamefont {L.~E.}\ \bibnamefont
  {Sadler}}, \bibinfo {author} {\bibfnamefont {J.~M.}\ \bibnamefont {Higbie}},
  \bibinfo {author} {\bibfnamefont {S.~R.}\ \bibnamefont {Leslie}}, \bibinfo
  {author} {\bibfnamefont {M.}~\bibnamefont {Vengalattore}},\ and\ \bibinfo
  {author} {\bibfnamefont {D.~M.}\ \bibnamefont {Stamper-Kurn}},\ }\bibfield
  {title} {\bibinfo {title} {Spontaneous symmetry breaking in a quenched
  ferromagnetic spinor bose--einstein condensate},\ }\href
  {https://doi.org/10.1038/nature05094} {\bibfield  {journal} {\bibinfo
  {journal} {Nature (London)}\ }\textbf {\bibinfo {volume} {443}},\ \bibinfo
  {pages} {312} (\bibinfo {year} {2006})}\BibitemShut {NoStop}%
\bibitem [{\citenamefont {Lamacraft}(2007)}]{PhysRevLett.98.160404}%
  \BibitemOpen
  \bibfield  {author} {\bibinfo {author} {\bibfnamefont {A.}~\bibnamefont
  {Lamacraft}},\ }\bibfield  {title} {\bibinfo {title} {Quantum quenches in a
  spinor condensate},\ }\href {https://doi.org/10.1103/PhysRevLett.98.160404}
  {\bibfield  {journal} {\bibinfo  {journal} {Phys. Rev. Lett.}\ }\textbf
  {\bibinfo {volume} {98}},\ \bibinfo {pages} {160404} (\bibinfo {year}
  {2007})}\BibitemShut {NoStop}%
\bibitem [{\citenamefont {Kinoshita}\ \emph {et~al.}(2006)\citenamefont
  {Kinoshita}, \citenamefont {Wenger},\ and\ \citenamefont
  {Weiss}}]{Kinoshita2006}%
  \BibitemOpen
  \bibfield  {author} {\bibinfo {author} {\bibfnamefont {T.}~\bibnamefont
  {Kinoshita}}, \bibinfo {author} {\bibfnamefont {T.}~\bibnamefont {Wenger}},\
  and\ \bibinfo {author} {\bibfnamefont {D.~S.}\ \bibnamefont {Weiss}},\
  }\bibfield  {title} {\bibinfo {title} {A quantum newton's cradle},\ }\href
  {https://doi.org/10.1038/nature04693} {\bibfield  {journal} {\bibinfo
  {journal} {Nature (London)}\ }\textbf {\bibinfo {volume} {440}},\ \bibinfo
  {pages} {900} (\bibinfo {year} {2006})}\BibitemShut {NoStop}%
\bibitem [{\citenamefont {Hofferberth}\ \emph {et~al.}(2007)\citenamefont
  {Hofferberth}, \citenamefont {Lesanovsky}, \citenamefont {Fischer},
  \citenamefont {Schumm},\ and\ \citenamefont
  {Schmiedmayer}}]{Hofferberth2007}%
  \BibitemOpen
  \bibfield  {author} {\bibinfo {author} {\bibfnamefont {S.}~\bibnamefont
  {Hofferberth}}, \bibinfo {author} {\bibfnamefont {I.}~\bibnamefont
  {Lesanovsky}}, \bibinfo {author} {\bibfnamefont {B.}~\bibnamefont {Fischer}},
  \bibinfo {author} {\bibfnamefont {T.}~\bibnamefont {Schumm}},\ and\ \bibinfo
  {author} {\bibfnamefont {J.}~\bibnamefont {Schmiedmayer}},\ }\bibfield
  {title} {\bibinfo {title} {Non-equilibrium coherence dynamics in
  one-dimensional {Bose} gases},\ }\href {https://doi.org/10.1038/nature06149}
  {\bibfield  {journal} {\bibinfo  {journal} {Nature (London)}\ }\textbf
  {\bibinfo {volume} {449}},\ \bibinfo {pages} {324} (\bibinfo {year}
  {2007})}\BibitemShut {NoStop}%
\bibitem [{\citenamefont {Bloch}\ \emph {et~al.}(2008)\citenamefont {Bloch},
  \citenamefont {Dalibard},\ and\ \citenamefont {Zwerger}}]{RevModPhys.80.885}%
  \BibitemOpen
  \bibfield  {author} {\bibinfo {author} {\bibfnamefont {I.}~\bibnamefont
  {Bloch}}, \bibinfo {author} {\bibfnamefont {J.}~\bibnamefont {Dalibard}},\
  and\ \bibinfo {author} {\bibfnamefont {W.}~\bibnamefont {Zwerger}},\
  }\bibfield  {title} {\bibinfo {title} {Many-body physics with ultracold
  gases},\ }\href {https://doi.org/10.1103/RevModPhys.80.885} {\bibfield
  {journal} {\bibinfo  {journal} {Rev. Mod. Phys.}\ }\textbf {\bibinfo {volume}
  {80}},\ \bibinfo {pages} {885} (\bibinfo {year} {2008})}\BibitemShut
  {NoStop}%
\bibitem [{\citenamefont {Trotzky}\ \emph {et~al.}(2012)\citenamefont
  {Trotzky}, \citenamefont {Chen}, \citenamefont {Flesch}, \citenamefont
  {McCulloch}, \citenamefont {Schollw{\"o}ck}, \citenamefont {Eisert},\ and\
  \citenamefont {Bloch}}]{Trotzky2012}%
  \BibitemOpen
  \bibfield  {author} {\bibinfo {author} {\bibfnamefont {S.}~\bibnamefont
  {Trotzky}}, \bibinfo {author} {\bibfnamefont {Y.-A.}\ \bibnamefont {Chen}},
  \bibinfo {author} {\bibfnamefont {A.}~\bibnamefont {Flesch}}, \bibinfo
  {author} {\bibfnamefont {I.~P.}\ \bibnamefont {McCulloch}}, \bibinfo {author}
  {\bibfnamefont {U.}~\bibnamefont {Schollw{\"o}ck}}, \bibinfo {author}
  {\bibfnamefont {J.}~\bibnamefont {Eisert}},\ and\ \bibinfo {author}
  {\bibfnamefont {I.}~\bibnamefont {Bloch}},\ }\bibfield  {title} {\bibinfo
  {title} {Probing the relaxation towards equilibrium in an isolated strongly
  correlated one-dimensional bose gas},\ }\href
  {https://doi.org/10.1038/nphys2232} {\bibfield  {journal} {\bibinfo
  {journal} {Nat. Phys.}\ }\textbf {\bibinfo {volume} {8}},\ \bibinfo {pages}
  {325} (\bibinfo {year} {2012})}\BibitemShut {NoStop}%
\bibitem [{\citenamefont {Cheneau}\ \emph {et~al.}(2012)\citenamefont
  {Cheneau}, \citenamefont {Barmettler}, \citenamefont {Poletti}, \citenamefont
  {Endres}, \citenamefont {Schau{\ss}}, \citenamefont {Fukuhara}, \citenamefont
  {Gross}, \citenamefont {Bloch}, \citenamefont {Kollath},\ and\ \citenamefont
  {Kuhr}}]{Cheneau2012}%
  \BibitemOpen
  \bibfield  {author} {\bibinfo {author} {\bibfnamefont {M.}~\bibnamefont
  {Cheneau}}, \bibinfo {author} {\bibfnamefont {P.}~\bibnamefont {Barmettler}},
  \bibinfo {author} {\bibfnamefont {D.}~\bibnamefont {Poletti}}, \bibinfo
  {author} {\bibfnamefont {M.}~\bibnamefont {Endres}}, \bibinfo {author}
  {\bibfnamefont {P.}~\bibnamefont {Schau{\ss}}}, \bibinfo {author}
  {\bibfnamefont {T.}~\bibnamefont {Fukuhara}}, \bibinfo {author}
  {\bibfnamefont {C.}~\bibnamefont {Gross}}, \bibinfo {author} {\bibfnamefont
  {I.}~\bibnamefont {Bloch}}, \bibinfo {author} {\bibfnamefont
  {C.}~\bibnamefont {Kollath}},\ and\ \bibinfo {author} {\bibfnamefont
  {S.}~\bibnamefont {Kuhr}},\ }\bibfield  {title} {\bibinfo {title}
  {Light-cone-like spreading of correlations in a quantum many-body system},\
  }\href {https://doi.org/10.1038/nature10748} {\bibfield  {journal} {\bibinfo
  {journal} {Nature (London)}\ }\textbf {\bibinfo {volume} {481}},\ \bibinfo
  {pages} {484} (\bibinfo {year} {2012})}\BibitemShut {NoStop}%
\bibitem [{\citenamefont {Paul}\ \emph {et~al.}(2022)\citenamefont {Paul},
  \citenamefont {Titum},\ and\ \citenamefont {Maghrebi}}]{Paul2022Hidden}%
  \BibitemOpen
  \bibfield  {author} {\bibinfo {author} {\bibfnamefont {S.}~\bibnamefont
  {Paul}}, \bibinfo {author} {\bibfnamefont {P.}~\bibnamefont {Titum}},\ and\
  \bibinfo {author} {\bibfnamefont {M.~F.}\ \bibnamefont {Maghrebi}},\
  }\bibfield  {title} {\bibinfo {title} {Hidden quantum criticality and
  entanglement in quench dynamics},\ }\href {https://arxiv.org/abs/2202.04654}
  {\bibfield  {journal} {\bibinfo  {journal} {arXiv:2202.04654}\ } (\bibinfo
  {year} {2022})}\BibitemShut {NoStop}%
\bibitem [{\citenamefont {Eisler}\ and\ \citenamefont
  {Zimbor{\'{a}}s}(2015)}]{Eisler_2015}%
  \BibitemOpen
  \bibfield  {author} {\bibinfo {author} {\bibfnamefont {V.}~\bibnamefont
  {Eisler}}\ and\ \bibinfo {author} {\bibfnamefont {Z.}~\bibnamefont
  {Zimbor{\'{a}}s}},\ }\bibfield  {title} {\bibinfo {title} {On the partial
  transpose of fermionic gaussian states},\ }\href
  {https://doi.org/10.1088/1367-2630/17/5/053048} {\bibfield  {journal}
  {\bibinfo  {journal} {New J. Phys.}\ }\textbf {\bibinfo {volume} {17}},\
  \bibinfo {pages} {053048} (\bibinfo {year} {2015})}\BibitemShut {NoStop}%
\bibitem [{\citenamefont {Eisert}\ \emph {et~al.}(2018)\citenamefont {Eisert},
  \citenamefont {Eisler},\ and\ \citenamefont
  {Zimbor\'as}}]{PhysRevB.97.165123}%
  \BibitemOpen
  \bibfield  {author} {\bibinfo {author} {\bibfnamefont {J.}~\bibnamefont
  {Eisert}}, \bibinfo {author} {\bibfnamefont {V.}~\bibnamefont {Eisler}},\
  and\ \bibinfo {author} {\bibfnamefont {Z.}~\bibnamefont {Zimbor\'as}},\
  }\bibfield  {title} {\bibinfo {title} {Entanglement negativity bounds for
  fermionic gaussian states},\ }\href
  {https://doi.org/10.1103/PhysRevB.97.165123} {\bibfield  {journal} {\bibinfo
  {journal} {Phys. Rev. B}\ }\textbf {\bibinfo {volume} {97}},\ \bibinfo
  {pages} {165123} (\bibinfo {year} {2018})}\BibitemShut {NoStop}%
\bibitem [{\citenamefont {Burkhardt}\ and\ \citenamefont
  {Guim}(1985)}]{Burkhardt_1985}%
  \BibitemOpen
  \bibfield  {author} {\bibinfo {author} {\bibfnamefont {T.~W.}\ \bibnamefont
  {Burkhardt}}\ and\ \bibinfo {author} {\bibfnamefont {I.}~\bibnamefont
  {Guim}},\ }\bibfield  {title} {\bibinfo {title} {Finite-size scaling of the
  quantum ising chain with periodic, free, and antiperiodic boundary
  conditions},\ }\href {https://doi.org/10.1088/0305-4470/18/1/006} {\bibfield
  {journal} {\bibinfo  {journal} {J. Phys. A: Math. Gen.}\ }\textbf {\bibinfo
  {volume} {18}},\ \bibinfo {pages} {L33} (\bibinfo {year} {1985})}\BibitemShut
  {NoStop}%
\bibitem [{\citenamefont {Cabrera}\ and\ \citenamefont
  {Jullien}(1987)}]{PhysRevB.35.7062}%
  \BibitemOpen
  \bibfield  {author} {\bibinfo {author} {\bibfnamefont {G.~G.}\ \bibnamefont
  {Cabrera}}\ and\ \bibinfo {author} {\bibfnamefont {R.}~\bibnamefont
  {Jullien}},\ }\bibfield  {title} {\bibinfo {title} {Role of boundary
  conditions in the finite-size ising model},\ }\href
  {https://doi.org/10.1103/PhysRevB.35.7062} {\bibfield  {journal} {\bibinfo
  {journal} {Phys. Rev. B}\ }\textbf {\bibinfo {volume} {35}},\ \bibinfo
  {pages} {7062} (\bibinfo {year} {1987})}\BibitemShut {NoStop}%
\end{thebibliography}%

\end{document}